\documentclass[pdftex,twocolumn,epjc3]{svjour3}

\RequirePackage[T1]{fontenc}

\smartqed

\RequirePackage{graphicx}
\RequirePackage{mathptmx}
\RequirePackage{eurosym}
\RequirePackage[colorlinks,citecolor=blue,urlcolor=blue,linkcolor=blue]{hyperref}

\journalname{Eur. Phys. J. C}

\usepackage{lineno,hyperref,ifthen}
\modulolinenumbers[5]
\usepackage{upgreek}
\usepackage{amsmath,amssymb}
\usepackage[printonlyused]{acronym}
\usepackage{todonotes} 
\setboolean{AC@nolist}{true} 
\makeatletter
\AtBeginDocument{
	\ifAC@nolist
		\renewcommand*\AC@hyperlink{\expandafter\@secondoftwo}
	\fi
}
\makeatother

\graphicspath{{./}{images/}}

\begin{document}
\sloppy
\title{An innovative silicon photomultiplier digitizing camera for gamma-ray astronomy
}
\author{
M.~Heller\thanksref{e2,addr1},
E.jr~Schioppa\thanksref{e1,addr1},
A.~Porcelli\thanksref{e3,addr1},
I.~Troyano Pujadas\thanksref{addr1},
K.~Zi{\c e}tara\thanksref{addr5},
D.~della Volpe\thanksref{addr1},
T.~Montaruli\thanksref{addr1},
F.~Cadoux\thanksref{addr1},
Y.~Favre\thanksref{addr1},
J.~A.~Aguilar\thanksref{addr1,addr13},
A.~Christov\thanksref{addr1},
E.~Prandini\thanksref{addr2},
P.~Rajda\thanksref{addr9},
M.~Rameez\thanksref{addr1},
W.~Bilnik\thanksref{addr9},
J.~B\l{}ocki\thanksref{addr3},
L.~Bogacz\thanksref{addr12},
J. Borkowski\thanksref{addr9},
T.~Bulik\thanksref{addr4},
A.~Frankowski\thanksref{addr7},
M.~Grudzi{\'n}ska\thanksref{addr4},
B.~Id{\'z}kowski\thanksref{addr5},
M.~Jamrozy\thanksref{addr5},
M.~Janiak\thanksref{addr7},
J.~Kasperek\thanksref{addr9},
K.~Lalik\thanksref{addr9},
E.~Lyard\thanksref{addr2},
E.~Mach\thanksref{addr3},
D.~Mandat\thanksref{addr10},
A.~Marsza{\l}ek\thanksref{addr3,addr5},
L.~D.~Medina~Miranda\thanksref{addr1},
J.~Micha{\l}owski\thanksref{addr3},
R.~Moderski\thanksref{addr7},
A.~Neronov\thanksref{addr2},
J.~Niemiec\thanksref{addr3},
M.~Ostrowski\thanksref{addr5},
P.~Pa{\'s}ko\thanksref{addr6},
M.~Pech\thanksref{addr10},
P.~Schovanek\thanksref{addr10},
K.~Seweryn\thanksref{addr6},
V.~Sliusar\thanksref{addr2,addr8},
K.~Skowron\thanksref{addr3},
{\L}.~Stawarz\thanksref{addr5},
M.~Stodulska\thanksref{addr3,addr5},
M.~Stodulski\thanksref{addr3},
R.~Walter\thanksref{addr2},
M.~Wi{\c e}cek\thanksref{addr9},
A.~Zagda\'{n}ski\thanksref{addr5}.
}
\thankstext{e2}{e-mail: \href{mailto:matthieu.heller@cern.ch}{matthieu.heller@cern.ch}}
\thankstext{e1}{e-mail: \href{mailto:enrico.junior.schioppa@cern.ch}{enrico.junior.schioppa@cern.ch}}
\thankstext{e3}{e-mail: \href{mailto:alessio.porcelli@unige.ch}{aporcell@uni-mainz.de}}

\institute{DPNC - Universit\'e de Gen\`eve, 24 Quai Ernest Ansermet, Gen\`eve, Switzerland \label{addr1}
	\and
	D\'epartment of Astronomy - Universit\'e de Gen\'eve, 16 Chemin de Ecogia, Gen\`eve, Switzerland \label{addr2}
	\and
	Instytut Fizyki J\c{a}drowej im. H. Niewodnicza\'nskiego Polskiej Akademii Nauk, ul. Radzikowskiego 152, 31-342 Krak\'ow, Poland \label{addr3}
	\and
	Astronomical Observatory, University of Warsaw, al. Ujazdowskie 4, 00-478 Warsaw, Poland \label{addr4}
	\and
	Astronomical Observatory, Jagellonian University, ul. Orla 171, 30-244 Krak\'ow, Poland \label{addr5}
	\and
	Centrum Bada\'n Kosmicznych Polskiej Akademii Nauk, Warsaw, Poland \label{addr6}
	\and
	Nicolaus Copernicus Astronomical Center, Polish Academy of Science, Warsaw, Poland \label{addr7}
	\and
	Astronomical Observatory, Taras Shevchenko National University of Kyiv, Observatorna str., 3, Kyiv, Ukraine \label{addr8}
	\and
	AGH University of Science and Technology, al.Mickiewicza 30, Krak\'ow, Poland \label{addr9}
	\and
	Institute of Physics of the Czech Academy of Sciences, 17. listopadu 50, Olomouc \& Na Slovance 2, Prague, Czech Republic \label{addr10}
	\and
	Vrije Universiteit Brussels, Pleinlaan 2 1050 Brussels, Belgium \label{addr11}
        \and
        Department of Information Technologies, Jagiellonian University, ul.\ prof.\ Stanis{\l}awa {\L}ojasiewicza 11, 30-348 Krak\'ow, Poland \label{addr12}
        \and
        Universit\'e Libre Bruxelles, Facult\'e des Sciences, Avenue Franklin Roosevelt 50, 1050 Brussels, Belgium \label{addr13}
}

\date{Received: date / Accepted: date}

\maketitle


\begin{abstract}
The \ac{SST-1M} is one of the three proposed designs for the \acp{SST} of the \ac{CTA} project. The \ac{SST-1M} will be equipped with a 4~m-diameter segmented mirror dish and an innovative fully digital camera based on \acp{SiPM}. Since the \ac{SST} sub-array will consist of up to 70 telescopes, the challenge is not only to build a telescope with excellent performance, but also to design it so that its components can be commissioned, assembled and tested by industry.
\newline
In this paper we review the basic steps that led to the design concepts for the \ac{SST-1M} camera and the ongoing realization of the first prototype, 
with focus on the innovative solutions adopted for the photodetector plane and the readout and trigger parts of the camera. 
In addition, we report on results of laboratory measurements on real scale elements that validate the camera design and show that it is capable of matching the \ac{CTA} requirements of operating up to high-moon-light background conditions.
\end{abstract}



\section{Introduction}
\label{sec:intro}
The \ac{CTA}, the next generation very high energy gamma-ray observatory, is a project to build two arrays of over 100 \acp{IACT} placed in two sites in the northern and southern hemispheres.
The array will consist of three types of telescopes: \acp{LST}, with $\sim$24~m mirror diameter, \acp{MST}, with $\sim$12~m mirror diameter and small size telescopes (SSTs), with $\sim$4~m mirror diameter\footnote{In this paper, we will call ``mirror'' the full reflective surface of the telescope, which, in our case, is composed of 18 hexagonal facets.}. About 70 small size telescopes will be installed in the southern site, which offers the best view of the galactic plane, and will be spaced at inter-telescope distances between 200-300~m to cover an air shower collecting surface of several km$^2$. This surface allows for observation of gamma-rays with energy between about 3~TeV and 300~TeV~\cite{CTAconcept}. Different \ac{SST} designs are being proposed, among which a single mirror Davies-Cotton telescope (SST-1M) based on \acl{SiPM} (\acs{SiPM}) photodetectors, whose camera is described in this paper. The other two projects~\cite{ASTRI,GCT} are dual mirror telescopes of Schwarzschild-Couder design.
\newline
\begin{figure*}[htp!]
	\centering
	\includegraphics[width=0.7\textwidth]{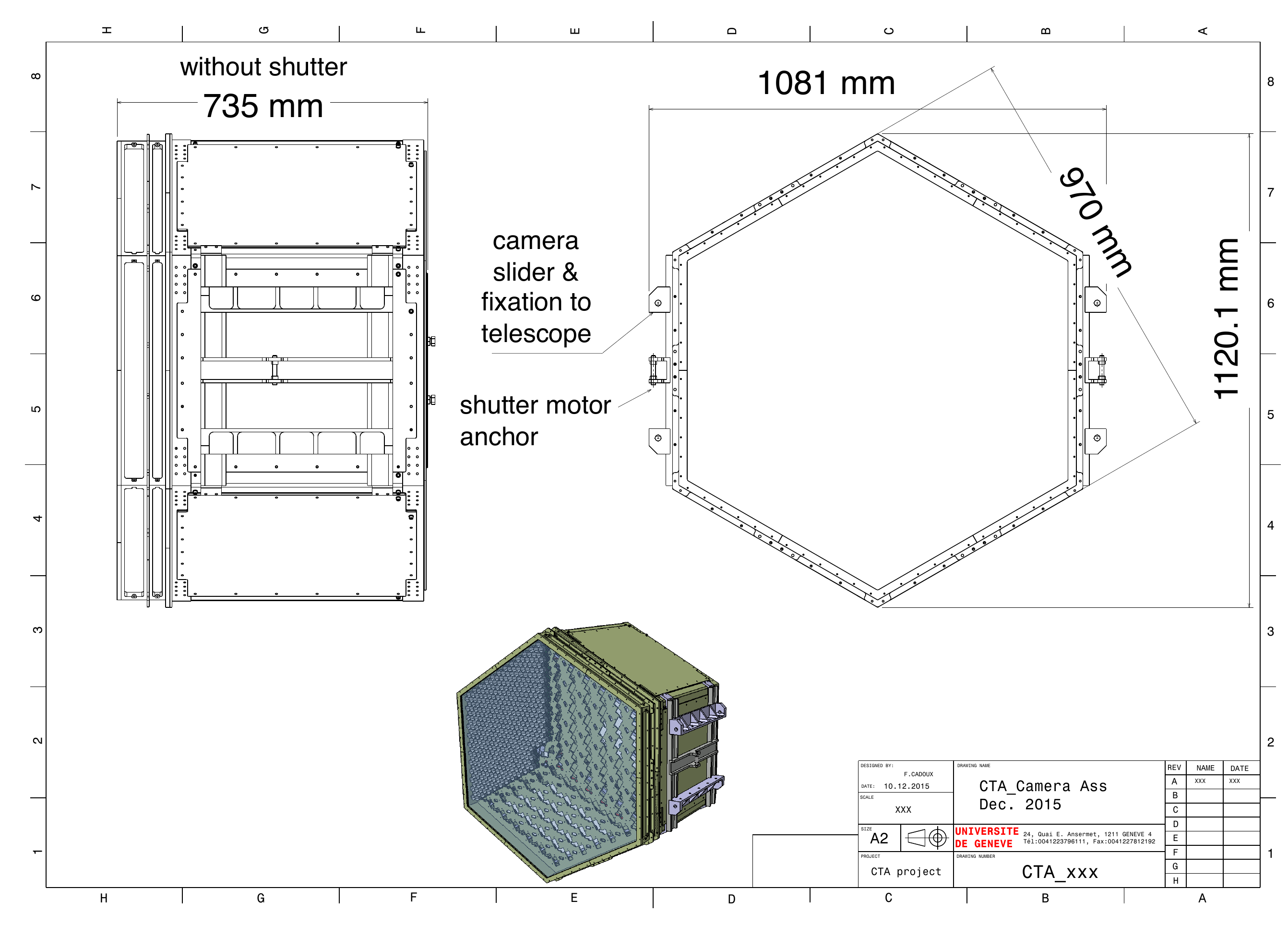}
	\caption{CAD drawing indicating the dimensions of the camera.}
	\label{fig:camera_dim}
\end{figure*}
The camera is a critical element of the proposed \ac{SST-1M} telescope, and has been designed to address the \ac{CTA} specifications on the sensitivity of the array, its angular resolution, the charge resolution and dynamic range of single cameras, the \ac{FoV} of at least $9^\circ$ for \acp{SST}, the uniformity of the response, as well as on the maintenance time and availability, while keeping the cost of single telescopes reasonable.
The SST-1M camera has been designed to achieve the best cost over performance ratio while satisfying the stringent \ac{CTA} requirements. Its components are made with standard industrial techniques, which make them reproducible and suited for large scale production. For these reasons, the camera features a few innovative strategies in both the optical system of the \ac{PDP} and the fully digital readout and trigger system, called DigiCam.
\newline
A camera prototype is being produced by the University of Geneva - UniGE (in charge of the PDP and its front-end electronics, the cooling system, the mechanics including the shutter, and the system for the integration on the telescope structure), the Jagellonian University and the AGH University of Science and Technology in Krak\'ow (in charge of the development of the readout and trigger system). This prototype not only serves to prove that the overall concept can meet the expected performance, but also serves as a test-bench to validate the production and assembly phases in view of the production of twenty SST-1M telescopes.

This paper is structured as follows: the general concept of the camera is described in Sec.~\ref{sec:design}, while Sec.s~\ref{sec:PDP} and \ref{sec:DigiCam} are dedicated to more details on the design of the PDP and of DigiCam, respectively.
Sec.~\ref{sec:cooling} describes the cooling system and Sec.~\ref{sec:safety} the housekeeping system. Sec.s~\ref{sec:tests} and \ref{sec:validation} 
are devoted to the description of the camera tests and validation of its performance estimated with the simulation described in Sec.~\ref{sec:performances}. Sec.~\ref{sec:calibration} describes initial plans on the calibration strategy during operation. In Sec.~\ref{sec:concl}, we draw the conclusions of the results and the plans for future operation and developments.


\section{Overview on the SST-1M camera design}
\label{sec:design}

\subsection{Camera structure}
\begin{figure}
	\centering
	\includegraphics[width=0.5\textwidth]{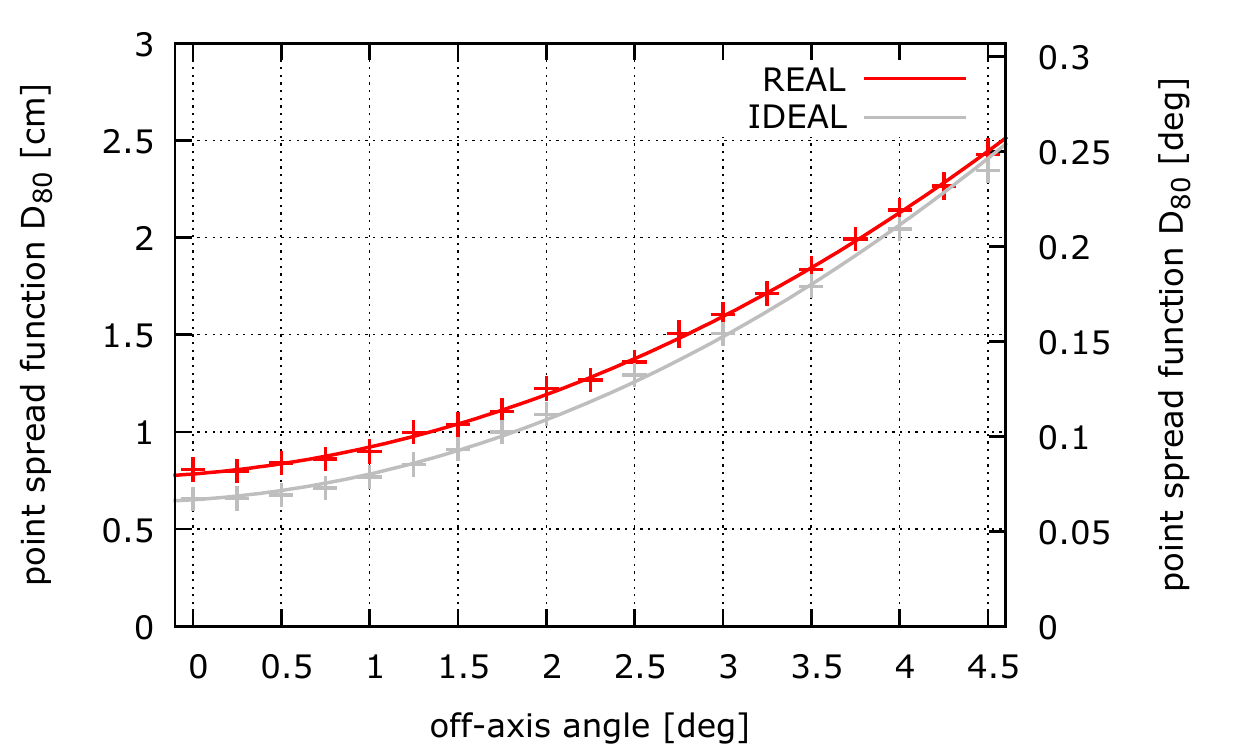}
	\caption{PSF of the optical system of the SST-1M telescope as a function of the off-axis angle for real mirror facets (measured focal and spot-size) and for ideal ones from ray-tracing simulation. The PSF is defined as the diameter of the region containing 80\% of the photons.}
	\label{fig:PSF}
\end{figure}

\begin{figure}[bt]
	\centering
	\includegraphics[width=0.5\textwidth]{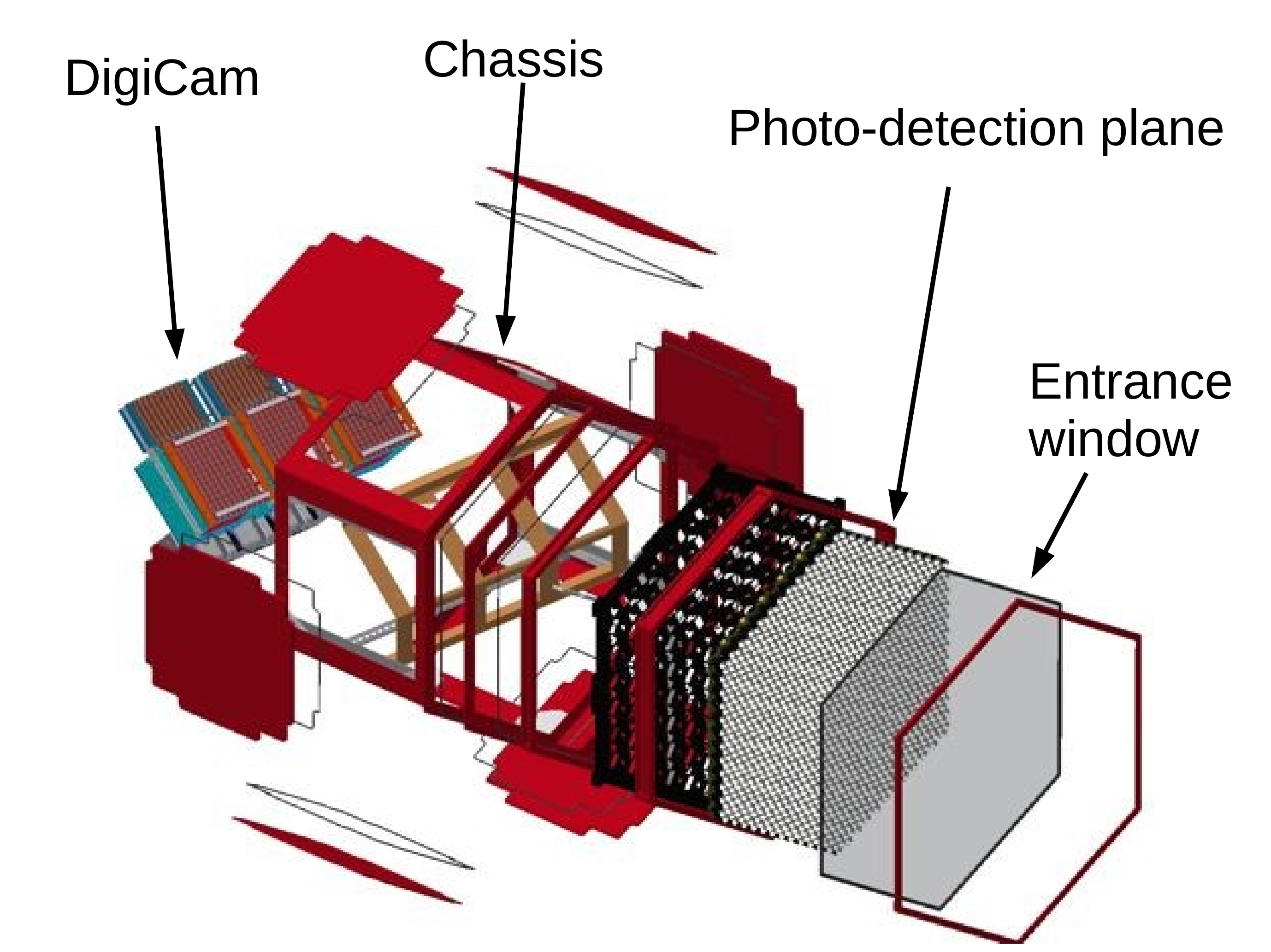}
	\caption{CAD drawing of the SST-1M camera, exploded view. 
	}
	\label{fig:cameraCAD}
\end{figure}

\begin{figure}
	\centering
        \includegraphics[width=0.45\textwidth]{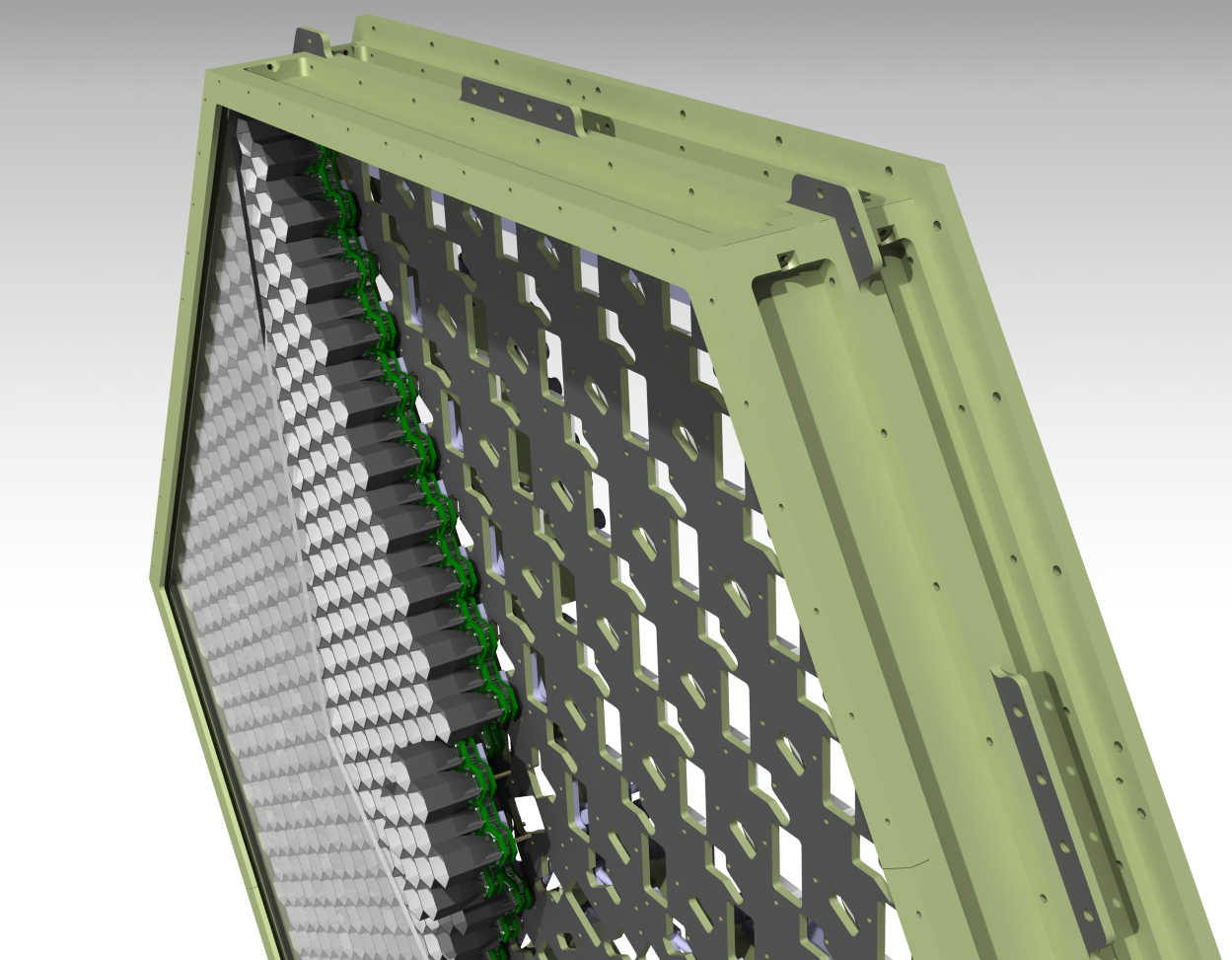}
	\caption{A CAD drawing showing the main features of the PDP. The 12 pixels modules (cones + pixels + front-end electronics) are mounted on the aluminium backplate, and the Borofloat window is fixed to the frame.}
	\label{fig:PDP}
\end{figure}

\begin{figure}[bt]
	\centering
	\includegraphics[width=0.5\textwidth]{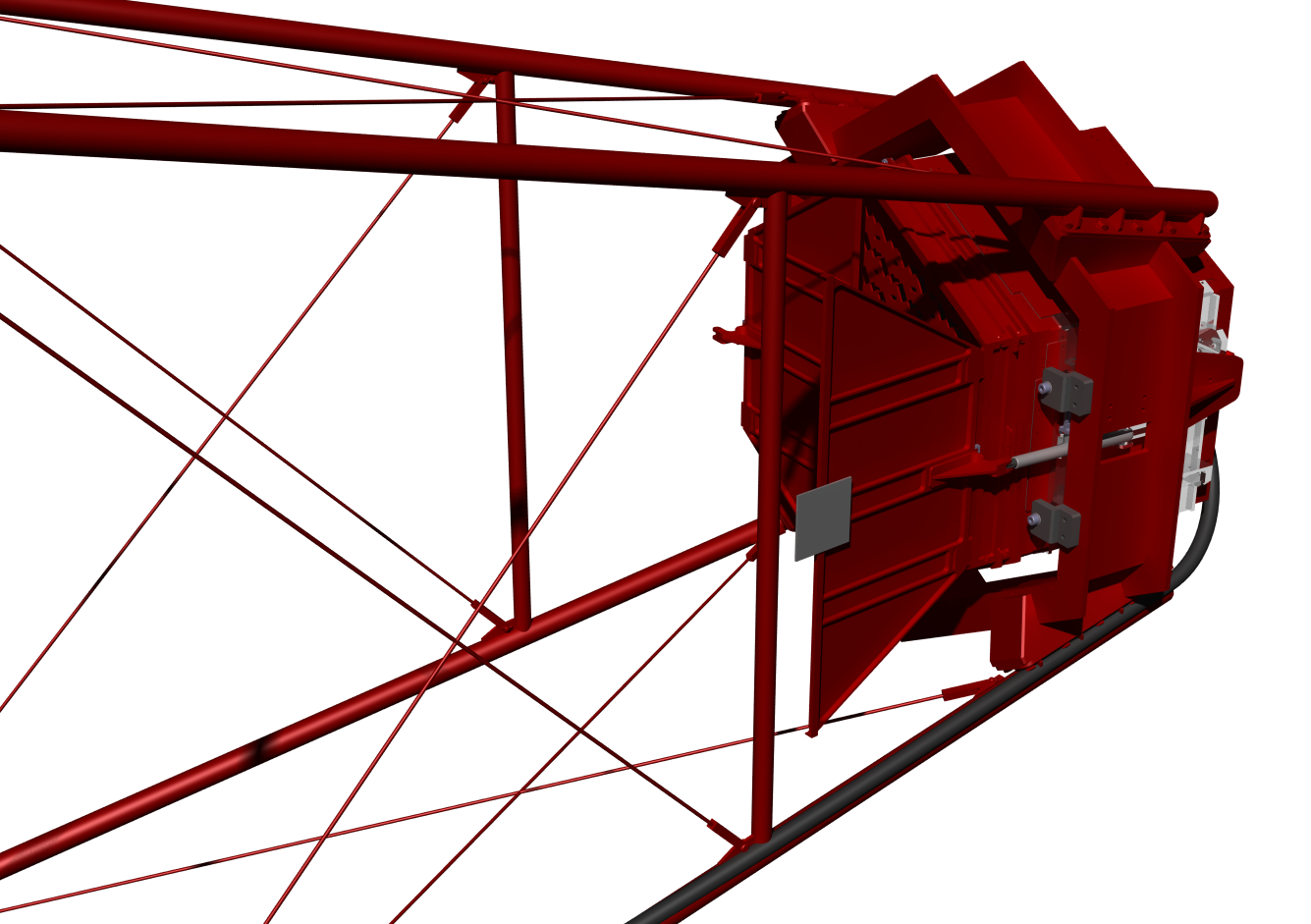}
	\caption{Drawing of the camera, including the shutter, installed on the telescope structure.}
	\label{fig:shutter}
\end{figure}

The geometry of the the \ac{SST-1M} camera is determined by the optical properties and geometry of the telescope, as was discussed in~\cite{CONES}. 
As can be seen in Fig.~\ref{fig:camera_dim}, the camera has a hexagonal shape with the vertex-to-vertex length of 1120~mm and a height of 735~mm. It weighs less than 200~kg.
According to the \ac{CTA} requirements, the SST optical \ac{PSF} shall not exceed 0.25$^\circ$ at $4^\circ$ off-axis and the telescope must focus a parallel beam of light (over 80\% of the required camera \acs{FoV} of 9$^\circ$) with a rms time spread of less than $1.5$~ns. 
To achieve the required \ac{PSF} with a Davies-Cotton design, a focal ratio of 1.4 is adopted for a telescope with effective mirror diameter of 4~m and
consequently focal length of 5.6~m. These dimensions fix the linear pixel size to 23.2~mm flat-to-flat for hexagonal pixels and a cut-off angle to 24$^\circ$, that can be achieved using light concentrators. The optical \ac{PSF}, shown in Fig.~\ref{fig:PSF}, is obtained with ray tracing including the mirror facet geometry and the measured spot size and the used focal length. To obtain the angular resolution of the telescope, the PSF has to be convolved with the precision coming from the camera and its pixel size. 
Simulations  indicate that for 80\% of the FoV, which corresponds to within $4^\circ$ off-axis, the largest time spread is 0.244~ns for on-axis rays.
\newline
A CAD drawing of the camera decomposed in its elements is shown in Fig.~\ref{fig:cameraCAD}. 
The mechanics features an entrance window that protects the PDP (see Fig.~\ref{fig:PDP}) and a shutter (see Fig.~\ref{fig:shutter}) that provides a light-tight cover when the telescope is in parking position and also protects the camera from environmental conditions. The camera mechanics guarantees protection from water and dust of at least IP65 level\footnote{International Protection Marking according to the IEC standard 60529.}. 

\subsection{General concept of the camera architecture}
The camera is composed of two main parts. The \ac{PDP} (described in Sec.~\ref{sec:PDP}), based on \ac{SiPM} sensors, and the trigger and readout system, DigiCam (see Sec.~\ref{sec:DigiCam}). DigiCam uses an innovative fully digital approach in gamma-ray astronomy. Another example of this kind in CTA is FlashCam, the camera for the mid-size telescopes \cite{ICRC_FlashCam}. The general idea behind such camera architecture is to have a continuous digitization of the signals issued by the PDP, and to use the same data to produce a trigger decision. 

The SST-1M camera takes snapshots of all pixels every 4~ns and stores them in ring buffers. As explained in Sec.~\ref{sec:DigiCam}, the trigger system applies the selection criteria on a lower resolution copy of these data. If an event passes the selection, the full resolution data are sent to a camera server via a 10~Gbps link (this bandwidth can be shared among events of different types) \footnote{The maximum camera throughput (throughput = trigger rate $\times$ data size) is $\sim$32~kHz for a readout window of 80~ns.}. 
The camera server filters the data reducing the event rate down to the CTA target of 600~Hz for the commissioning (300~Hz for normal operation). It also acts as the interface between the camera and the central array system. It not only ships the data to the array data stream, but it also transmits information and commands to and from \ac{ACTL} and handles the array trigger requests. 

The use of ring buffers in DigiCam allows the system to keep taking data while analyzing previous images for the trigger decision, providing a dead time free operation. 
Latest generation of FPGAs\footnote{Xilinx Virtex 7 family.} are used to achieve the high data throughput needed to aggregate the huge amount of data exchanged within the DigiCam hardware components (see Sec.~\ref{sec:DigiCam}), to have resources to guarantee low latency and high performance of the trigger algorithms and keep the flexibility for further evolution of the system.

The trigger logic is based on pattern matching algorithms which guarantee flexibility as different types of events (gamma, protons, muons, calibration events, etc.) produce different patterns, and the data can be triggered and flagged accordingly. 

\subsection{SiPM sensors in the SST-1M camera}
\label{SiPM section}
The use of \ac{SiPM} technology is quite recent in the field of gamma-ray astrophysics and it is an important innovative feature of the \acs{SST-1M} camera. Currently, FACT is the only telescope successfully operating the first SiPM-based camera on field~\cite{FACT}. It is very similar in dimensions to the SST-1M telescope but with half its \acs{FoV}.  \ac{SiPM}s offer many advantages with respect to the traditional \acp{PMT}, such as negligible ageing, insensitivity to magnetic fields, cost effectiveness, robustness against intense light, considerably lower voltage. For the case of the CTA SSTs, the capability of \acp{SiPM} of operating at high levels of light without any ageing, implies that data can also be taken with intense moonlight, increasing the telescope duty cycle, hence improving the discovery potential and sensitivity in the high-energy domain. The SST-1M camera will use an improved SiPM technology compared to the one used in FACT that reduces dramatically the cross-talk while the fill factor is not much affected (see Sec.~\ref{sec:sensors}). 
 
A relevant feature of the SST-1M camera design is that the sensors are DC coupled to the front-end electronics while the other SST solutions\cite{ASTRI,GCT} are AC coupled. With DC coupling, shifts in the baseline due to changes of the intensity of the \ac{NSB} and of the moon light,
can be measured and used to monitor such noise pixel-by-pixel. This information can be used by the entire array to monitor the stray light environmental noise.

Another innovative feature of the camera concerns the stabilization of the \ac{SiPM} working point. The breakdown voltage of the sensor depends strongly on temperature. For the sensors used in the SST-1M camera prototype, the breakdown voltage varies with temperature with a coefficient of typically 54~mV/$^\circ$C.
If no counter measures are taken, the sensors within the PDP operate at different gains, that is the conversion factor of  charge into the number of \acp{p.e.}. The gain can change due to temperature variations in time and can be different between pixels due to temperature gradients within the PDP (see Sec. \ref{cooling}). 
These effects would lead to a non-uniformity of the trigger efficiency in time and across the camera affecting the reconstruction of the gamma-ray events. 
Since the sensors are operated at a 2.8~V over-voltage (the difference between the bias voltage and the breakdown voltage $V_{ov} = V_{bias}-V_{break}$), this would imply a gain variation of 2\%/$^\circ$C\footnote{The gain $g$ is directly proportional to the over-voltage $V_{ov}$, hence $\frac{\Delta G}{G} = \frac{\Delta V_{ov}}{V_{ov}}$.}. 
A stabilization of the sensor working point has therefore been developed and is described in Sec.~\ref{electronics section}.

\section{The design and production of the PDP}
\label{sec:PDP}
The PDP (see Fig.~\ref{fig:PDP}) has 1296 pixels, distributed in 108 modules of 12 pixels. The PDP has a hexagonal sensitive area of 87.7 cm side-to-side and weighs about 35 kg.
Its mechanical stability is provided by an aluminium backplate, to which the modules are screwed. The backplate also serves as a heat sink for the \ac{PDP} cooling system (see Sec.~\ref{sec:cooling}).
 \begin{figure}
	\centering
        \includegraphics[width=0.45\textwidth]{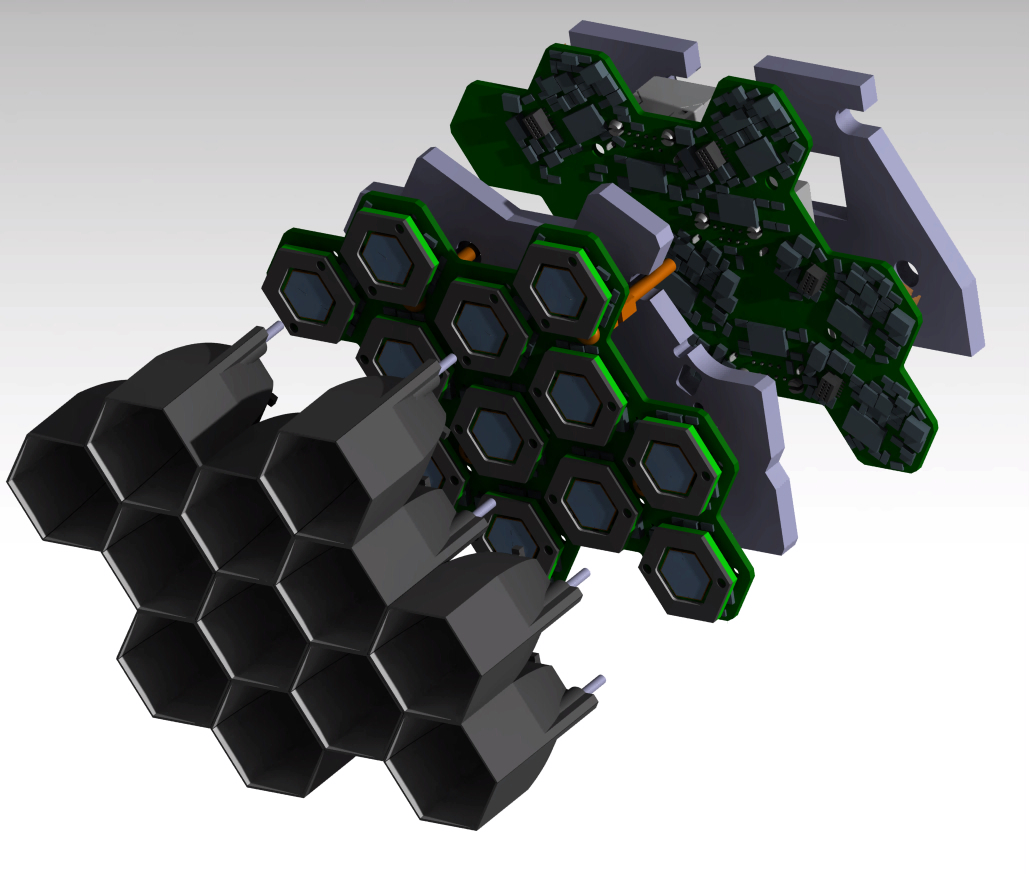}
                \includegraphics[width=0.45\textwidth]{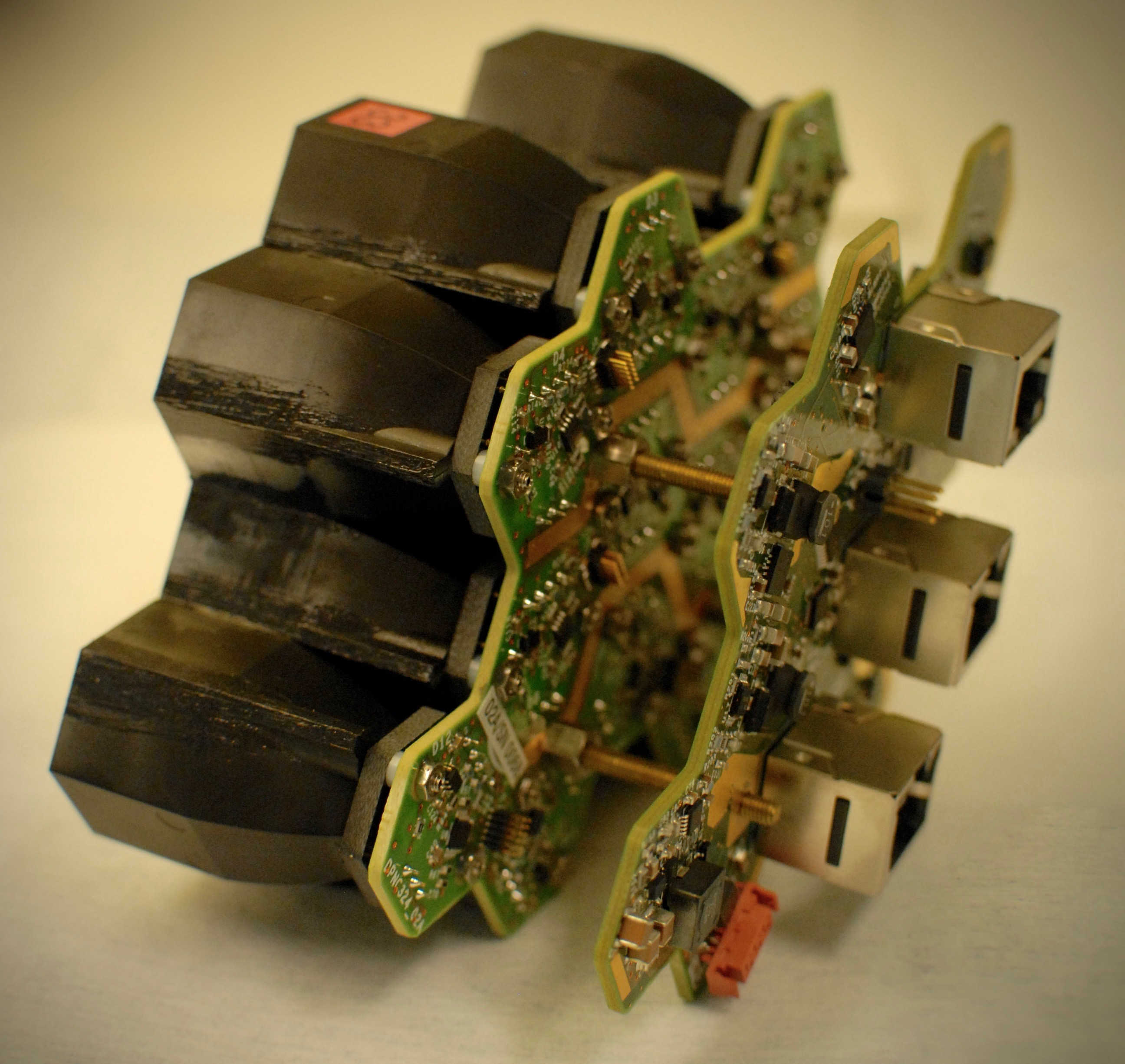}
        \caption{Top: a single 12 pixels module (drawing decomposed into the cones, preamplifier board with sensors and the slow control board, including the two layers of thermal foam). Bottom: a photo of a module prior to its final assembly.}
	\label{fig:module}
\end{figure}
A drawing and a photograph of a single module are shown in Fig.~\ref{fig:module}. 

\begin{figure}
	\centering
        \includegraphics[width=0.35\textwidth]{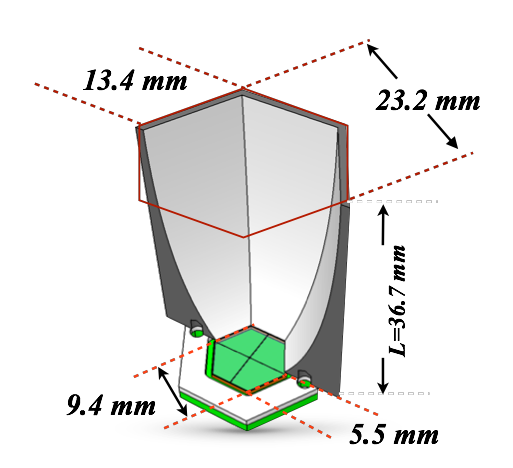}
	\caption{A drawing of a single pixel, composed of a light funnel (cut-out view) coupled to a sensor.}
	\label{fig:cone}
\end{figure}
The pixels are formed by a hexagonal hollow light-funnel with a compression factor of about 6 coupled to a large area, hexagonal \ac{SiPM} sensor~\cite{CONES}. A pixel design is shown in Fig.~\ref{fig:cone}.
The sensor has been designed in collaboration with Hamamatsu \footnote{http://www.hamamatsu.com/us/en/index.html} to reach the desired size. The choice of a hexagonal shape is important for trigger purposes. 
For an easy implementation of selection algorithms based on the recognition of circular, elliptical or ring-shaped patterns (these latter peculiar of muon events), it is desirable that the trigger can operate in fully symmetrical conditions. Both circular and hexagonal pixels provide such a feature since the center of each pixel is at the same distance from the centers of all its neighbors. The hexagonal shape was chosen since it minimizes the dead space between pixels.

The PDP also includes the front-end electronics which, due to space constraints, is implemented in two separate \acp{PCB} in each module. The front-end electronics boards - the preamplifier board and the \ac{SCB} - are introduced in Sec.~\ref{electronics section} and are described in detail in Ref.~\cite{electronics_paper}. The former has been specifically realized to handle the signals arising from the large area (hence large capacitance) sensors, the latter serves to manage the slow control parameters of each sensor (such as the bias voltage and the temperature) and to stabilize its operational point. The design of the two boards has been driven by the need of having a low noise, high-bandwidth and low power front-end electronics. Cost minimization has also been accounted for reaching $\sim$100\euro\ (including the cost of the sensors) per pixel in the production phase of 20 telescopes.
In what follows the different components of the PDP are described in detail.


\subsection{The entrance window}
The main protection of the \ac{PDP} against water and dust is provided by an entrance window made of 3.3~mm thick Borofloat llayer. Borofloat was chosen against PMMA due to its better mechanical rigidity. PMMA has good UV transmittance down to 280~nm (310~nm for Borofloat). Nonetheless, a rigid enough PMMA window would require a 6-8~mm thickness, hence absorbing too much incoming light, while \ac{FEA} studies indicate that for Borofloat 3.3~mm is sufficient. Given that the photo-detection efficiency of the sensors significantly degrades for wavelengths below 310~nm, it was decided to adopt the Borofloat solution.

The outer of the entrance window is coated with an anti-reflective layer to reduce Fresnel losses. The inner side is coated with a dichroic filter cutting off wavelengths above 540~nm. 
The coating of the window is a delicate procedure given its large surface. In order to obtain a uniform result, a large enough coating chamber is required. The only company offering such a possibility, among those we explored, is Thin Film Physics\footnote{http://www.tfp-thinfilms.com} (TFP). 
\begin{figure}
	\centering
        \includegraphics[width=0.45\textwidth]{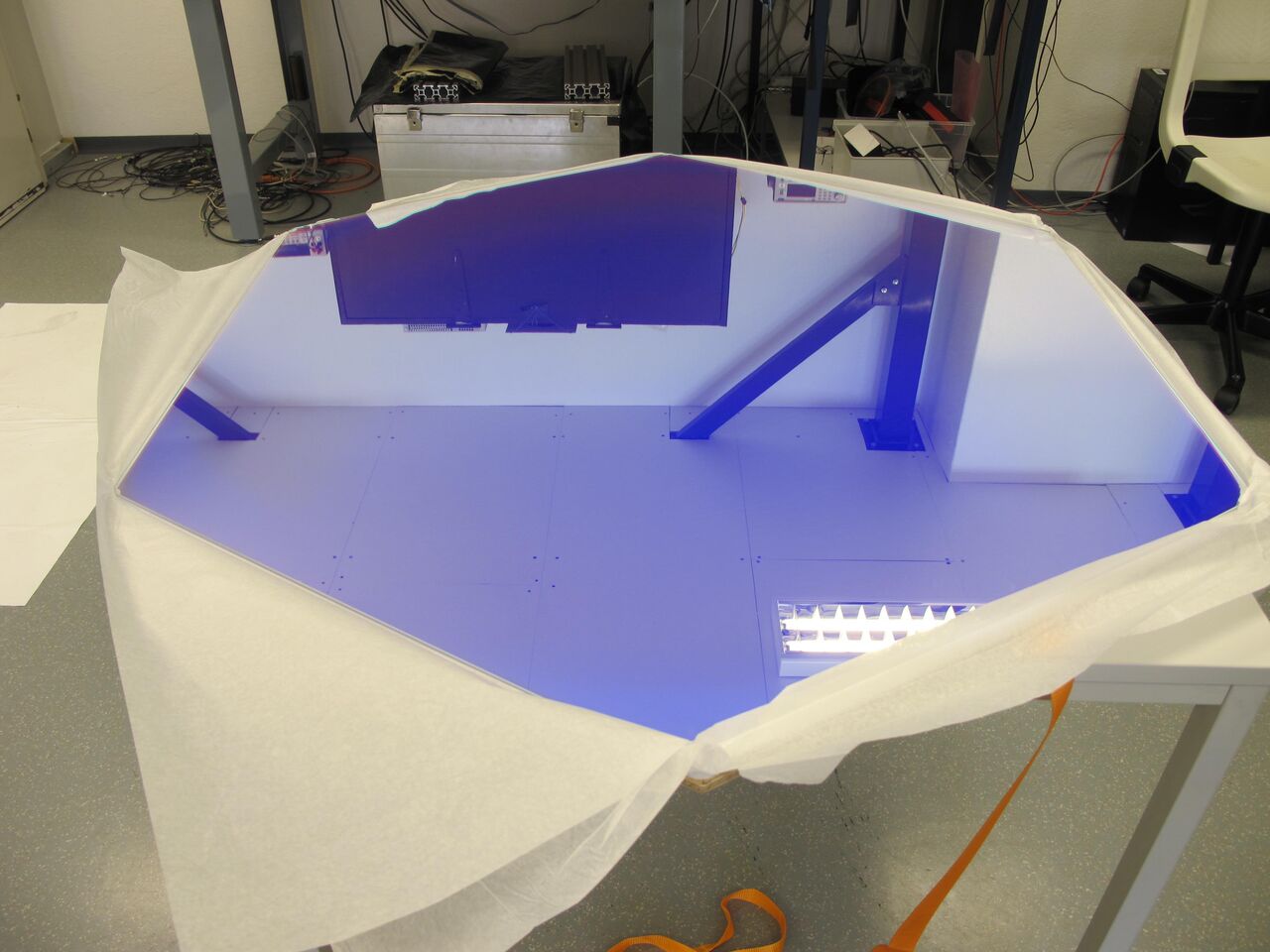}
	\caption{The coated Borofloat entrance window of the camera.}
	\label{fig:entrance}
\end{figure}
The first produced window is shown in Fig.~\ref{fig:entrance}. 
\begin{figure}
	\centering
        \includegraphics[width=0.45\textwidth]{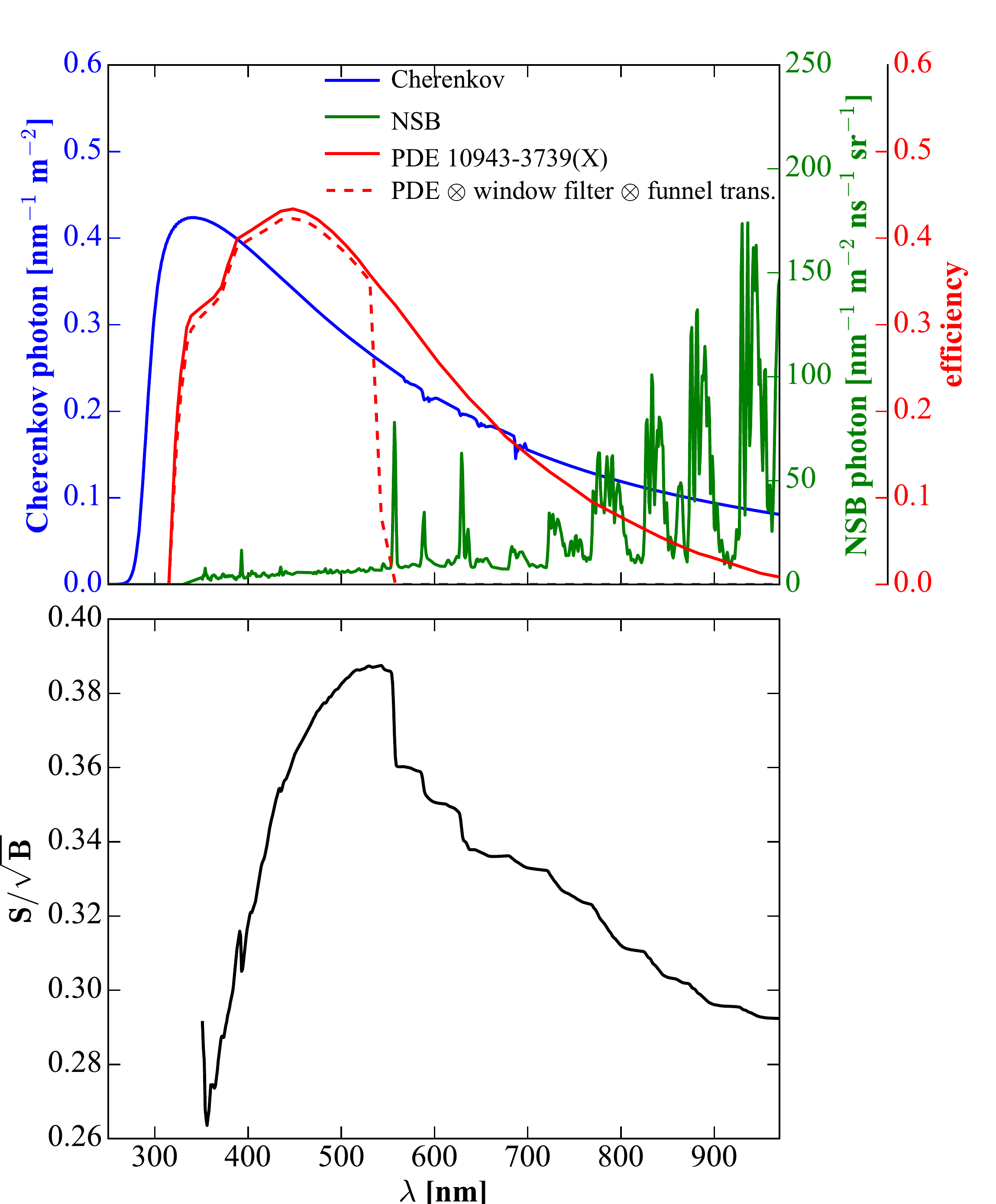}
	\caption{Top: the Cherenkov light spectrum (blue solid line) and the CTA reference \acl{NSB} spectrum (green solid line) for dark nights. For comparison, the photo-detection efficiency (\ac{PDE}) of the sensors is indicated by the red solid line, and its convolution with the wavelength filter due to the window and the light concentrator transmittance is shown by the dashed red line. Bottom: Signal-to-noise ratio as a function of the wavelength, showing a maximum at about 540~nm.}
	\label{SNRwindow}
\end{figure}
As a reference, at the top of Fig.~\ref{SNRwindow} the Cherenkov spectrum (blue line, calculated for showers at 20$^\circ$ zenith angle and detected at 2000~m above see level) and the CTA reference NSB (green line) spectrum\footnote{The peaks in the NSB spectrum correspond to absorption lines from the molecules in the atmosphere. Although the NSB partially depends on the diffused light due to the moon at each night, the spectral shape during dark nights at the final CTA site should not change dramatically compared to the reference curve. Also, the normalization of the Cherenkov spectrum may slightly change with inclination, energy of the showers and altitude of the detector \cite{NSB_1,NSB_2,NSB_3}.} are compared to the \ac{PDE} of the sensors (red line) and its convolution with the wavelength filter (red dashed line) on the window and the light concentrator transmittance (see next section); in the bottom panel of the figure, the SNR of the window is shown. Cutting the long wavelengths is more important for \acp{SiPM} than for \acp{PMT} since the \acp{SiPM} have higher sensitivity in the red and near infrared where the \ac{NSB} is larger.  The intense \ac{NSB} peaks at wavelengths larger than 540 nm are cut away by the filter layer on the window as shown by the red dashed line in Fig.~\ref{SNRwindow} (top).


\subsection{The hollow light concentrators}
Light guides are often used in gamma-ray telescope cameras to focus light from the pixel surface onto sensors of smaller area with good efficiency, and to reduce the contamination by stray light coming from the \ac{NSB} and from reflections of light on the terrain, snow, etc. The light guide design has to be the closest possible to the ideal Winston cone, whose efficiency is maximum up to a sharp cut-off angle which depends on the f/D ratio of the telescope ($f$ and $D$ being the focal length and the mirror diameter, respectively) and on the \ac{FoV}~\cite{CONES}.

The light funnel design was optimized by using the Zemax optical and illumination design software~\cite{zemax}. To maximize the collection efficiency of the cone (that is, the amount of outgoing light with respect to the amount of incoming light) the design of the funnel inner surface has been optimized using a cubic B\`ezier curve~\cite{Okumura}. 

Two possible light guide designs have been investigated: full cones and hollow cones. The calculations show that full cones in a material of the same optical properties as the camera window would provide a better compression factor (14.1, compared to 6.1 for the hollow cones). However, they would be more elongated (53.3~mm, while the hollow cones are 36.7~mm long), and would therefore absorb most of the light. A solution would be to reduce the pixel size, so that the length of the full cone would reduce accordingly. However, this would increase considerably the number of channels (and hence the cost) of the camera. Moreover, since the PSF is fixed by the telescope optics, there is no advantage in reducing the pixel size.

The adopted solution has been therefore the hollow light concentrator. A drawing of the light guide is shown in figure~\ref{fig:cone}. The light is collected from an entrance hexagonal surface of 23.2 mm side-to-side linear size and focused onto an exit hexagonal surface of 9.4 mm side-to-side linear size. The cut-off angle is around 24$^\circ$. 

In line with the overall camera design philosophy, the production strategy of the cones has been conceived for being cheap, reproducible and scalable, at the same time delivering a high quality product that could be tested on a subset of samples prior to assembly on the camera structure. The cones substrate is produced by Viaoptic GmbH \footnote{http://www.viaoptic.de/de/inhalt/landingpage.html} in black polycarbonate (MAKROLON 2405 SW 901510) using plastic injection molding, a well established mass production technique, followed by cleaning and coating. The Bezier shape eases the manufacturing process since computer-numerical-control machines are typically using this format. While the stringent precision on the geometry (shape and size with tolerance $< 40$~$\mu$m) of the cones substrate was met, the requirements on the roughness ($<50$~nm) of the inner surface where obtained with a dedicated optimization and polishing of the injection mould. This part is critical, since the overall reflectivity of the cone is driven by the smoothness of the reflective surfaces \cite{RoughDavies}, while the coating's role is to modify and/or enhance it. Also the coating technique, similar to sputtering, required some development in collaboration with TFP and BTE\footnote{http://www.bte-born.de}. Methods for the deposition of reflective layers on plastic are well established for flat surfaces, but in the case of the highly curved surfaces of the light concentrators such techniques are more difficult and required a dedicated optimization. 

\begin{figure}
	\centering
        \includegraphics[width=0.5\textwidth]{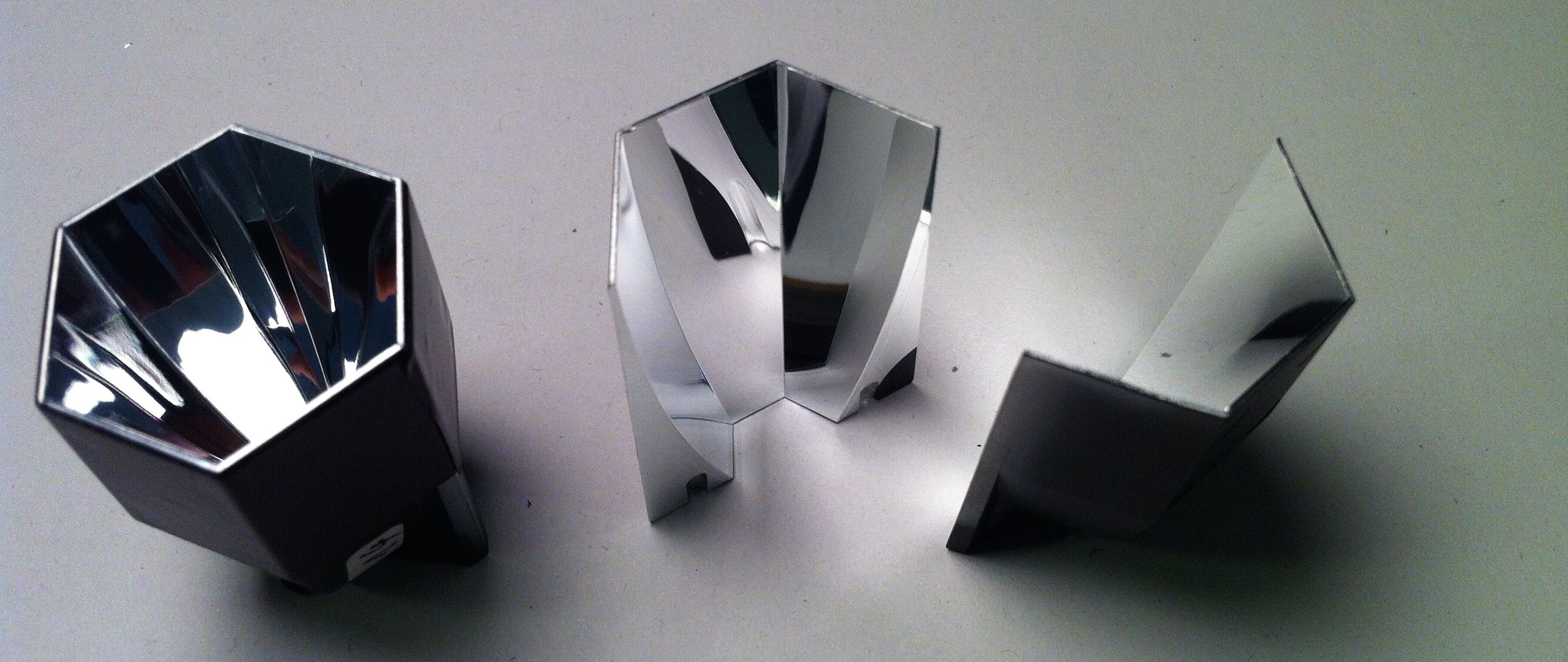} 
        \includegraphics[width=0.5\textwidth]{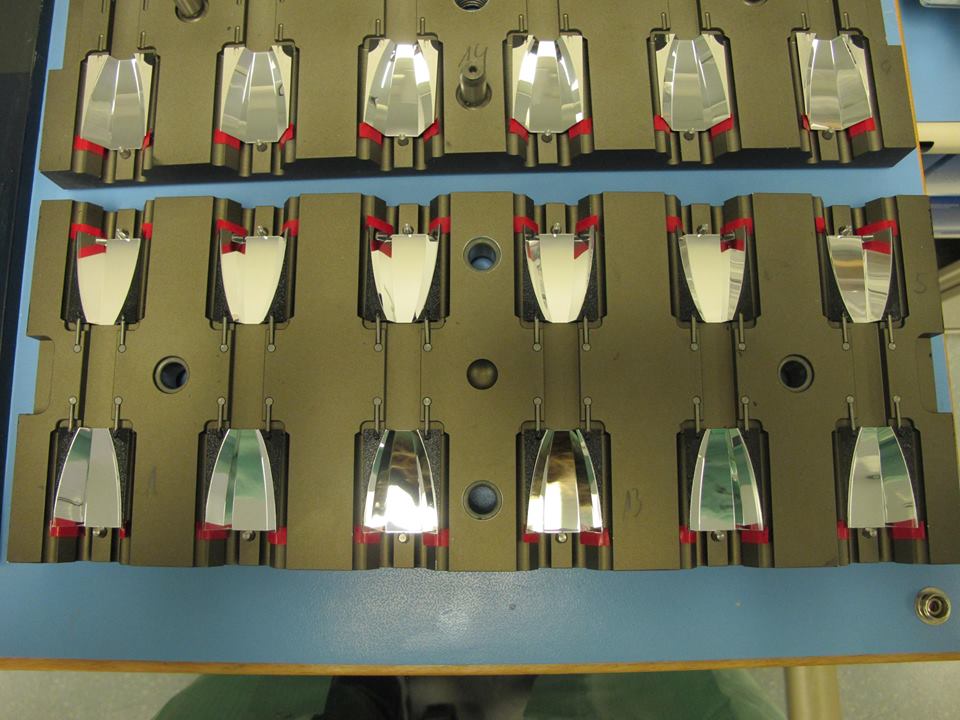}
        \caption{Top: A picture of an assembled cone (left) and of the two halves before being glued together. Bottom:  photo of the jig used to glue 24 half-cones together.}
	\label{cone picture}
\end{figure}

The cones are produced in two halves, that are coated separately and later on glued together (see Fig.~\ref{cone picture}) in jigs shown at the bottom of the figure. The assembly time of the camera elements is affected by the drying time of the glue of about 6 hours. To make the assembly process faster the number of jigs can be increased.

The overall optimization of the cones design required multiple production campaigns followed by dedicated measurements of the cone transmittance versus incident angle of light. For each production batch, ellipsometric evaluation of the coating on flat samples at the factory are followed by laboratory measurements at UniGE on a set of assembled cones with a dedicated test setup that allows measuring the reflection efficiency of a single cone in about half an hour time and to compare it with the one expected from simulations. Since individual testing of all the cones is too time consuming, assessments can be made only on samples of the produced cones. The high uniformity of the substrates is guaranteed by the producer thanks to the coating in chambers large enough to contain all cones and with high uniformity of sputtering.

\begin{figure}
	\centering
        \includegraphics[width=0.5\textwidth]{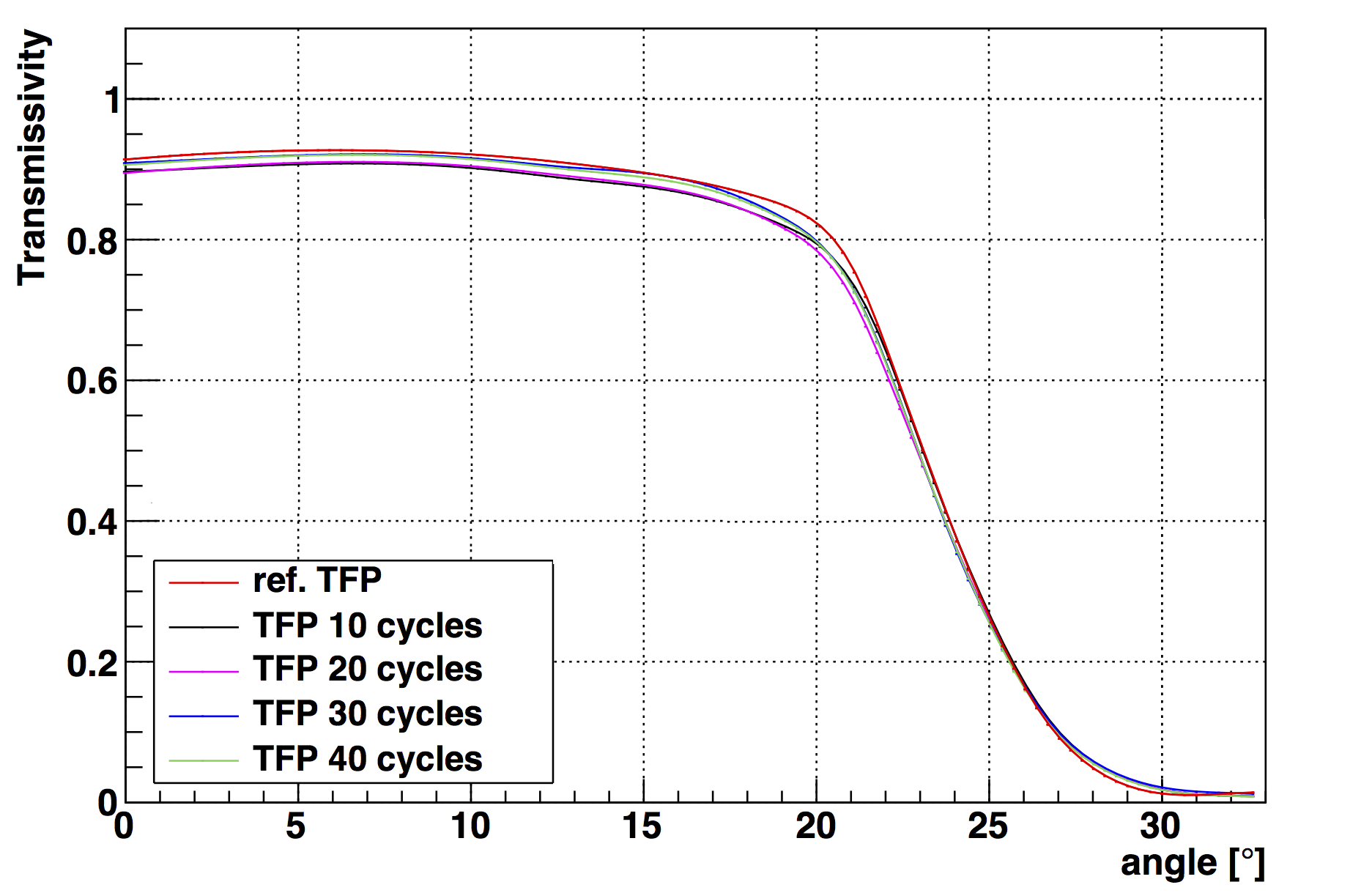} 
                \includegraphics[width=0.5\textwidth]{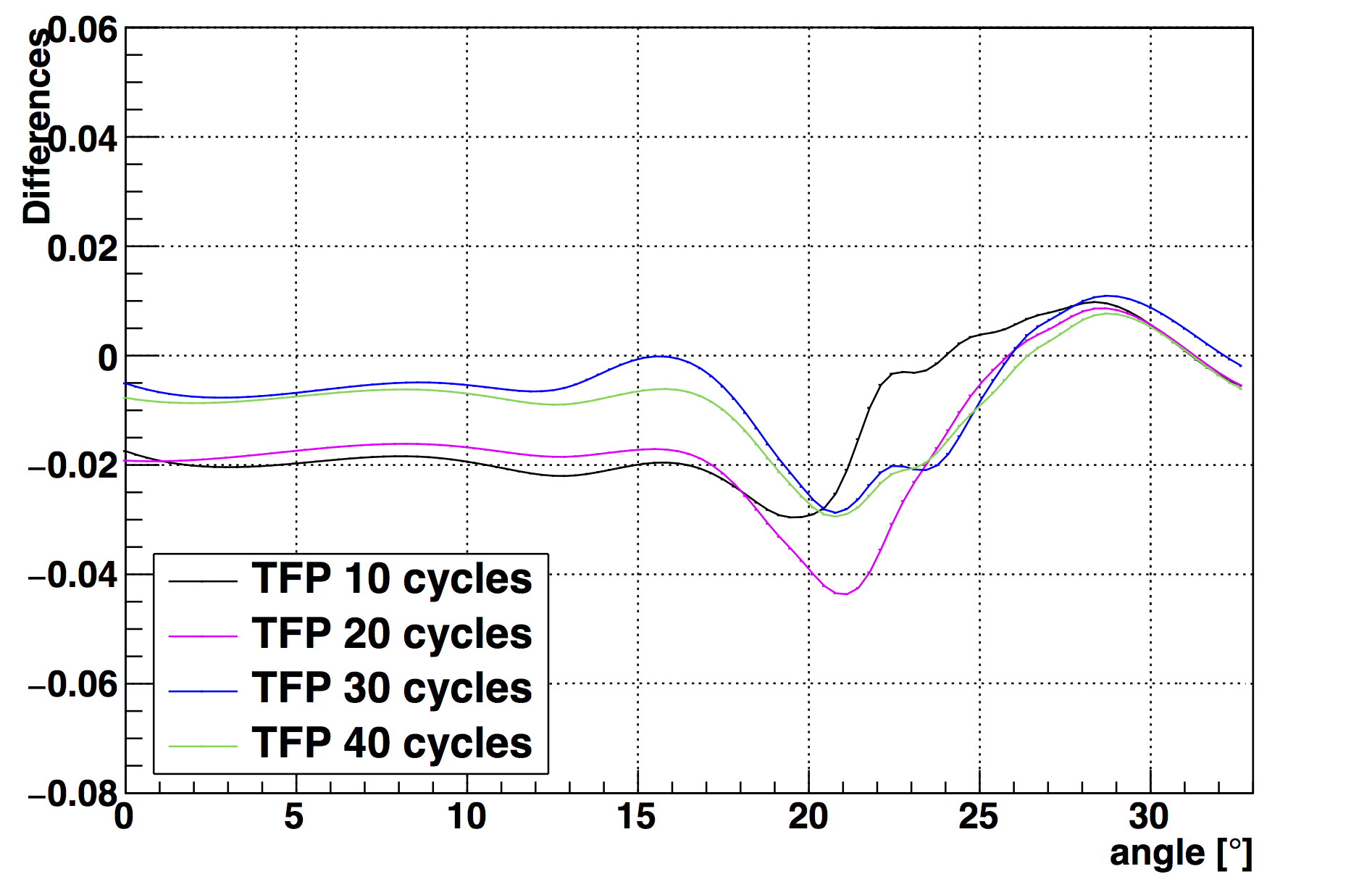} 
       \caption{Top: The transmission curve versus incidence angle on the light guide entrance surface for the reference TFP curve (red line) and for cones which underwent 10, 20, 30 and 40 thermal cycles between the thermal range of $-15^{\circ}$ to $35^{\circ}$ at constant (low) humidity levels. The reference TFP curve is obtained averaging 42 cones, chosen randomly, of the production of 1300 cones. Differences are smaller than the precision of the measurement set up. Bottom: difference between the measured transmittance of cones after temperature cycles and the reference TFP curve in the upper plot.}
	\label{cone measurement}
\end{figure}

The measuring set up and the Zemax simulation are fully described in~\cite{CONES}, where an agreement of the order of 2\%, comparable to the estimated systematic error of the measurement, is shown. The transmittance of the measured cones for an angle of incidence of $16^{\circ}$ on the entrance surface (which corresponds to the incidence angle that produces the maximum of the distribution of the light reflected on the telescope mirror) is about 88-90\%. This value does not include the absorption by the entrance window of the camera. BTE and TFP cones have shown negligible performance degradation after thermal cycles (within the 2\% measurement systematic errors). In Fig.~\ref{cone measurement} the average of the transmittance function of 42 cones is shown and compared to the transmittance of cones which underwent different numbers of thermal cycles.

\subsection{The \ac{SiPM} sensors}
\label{sec:sensors}
\begin{figure}
	\centering
        \includegraphics[width=0.45\textwidth]{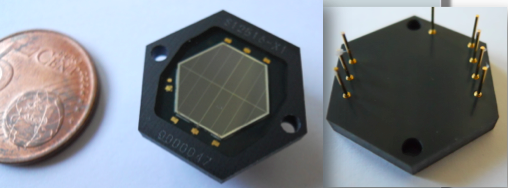}
	\caption{A picture of the custom-made large area hexagonal sensor. The side-to-side dimension of the sensitive area is 9.4~mm. There are 8 pins: 2 common cathodes, 2 \ac{NTC} pins, and 4 anodes.}
	\label{sensor}
\end{figure}
A picture of the hexagonal sensor is shown in Fig.~\ref{sensor}. With its 93.56~mm$^2$ sensitive surface, this device is one of the largest monolithic \ac{SiPM} produced. 
Since such large area hexagonal shaped sensors were not yet commercially available, the device was designed and produced in collaboration with Hamamatsu. 
The first version was named S12516(X)-050, followed by the version, called S10943-3739(X), which is used for the camera, which employs the Hamamatsu low cross-talk technology\footnote{Optical cross-talk is limited by trenches introduced between the micro-cells.}. 
This allows for an operation at higher over-voltage than the with the S12516(X)-050 sensors, translating into a higher \ac{PDE} and improved \ac{SNR}, especially relevant for the detection of few photons.

\begin{figure}
	\centering
        \includegraphics[width=0.4\textwidth]{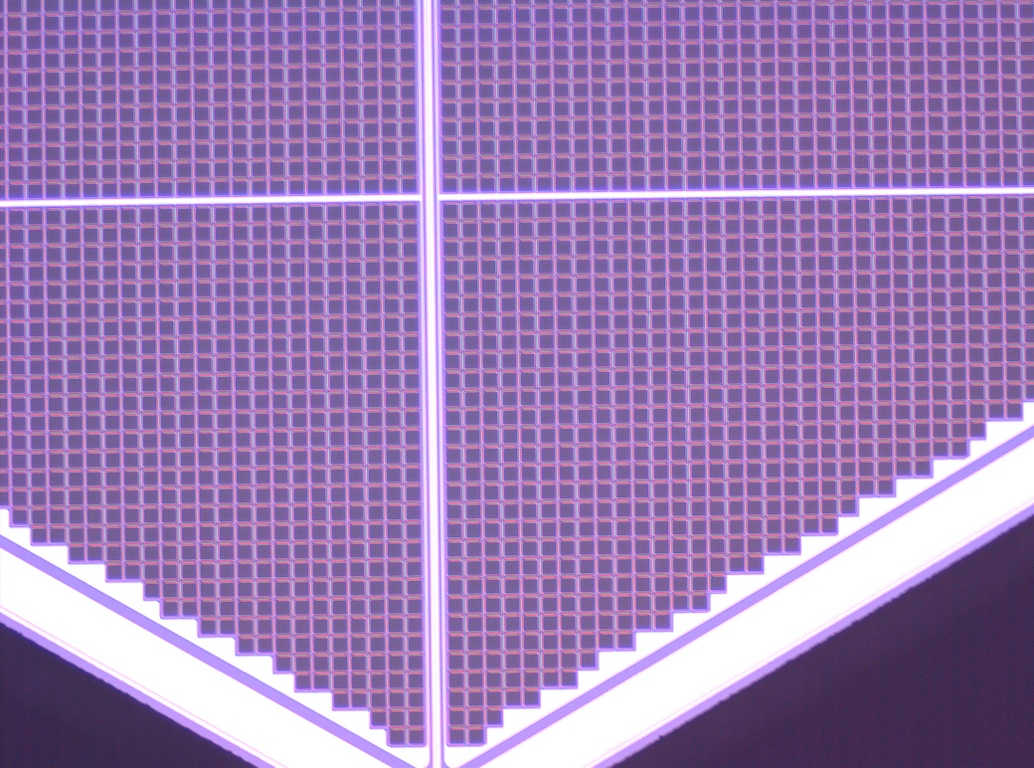}
        \caption{A microscope picture of the sensor showing a region of the microcells matrix. The separation into four channels is visible.}
	\label{sensor cells}
\end{figure}
The \ac{SiPM} is a matrix of 36'840, 50~$\mu$m-size square microcells (see Fig.~\ref{sensor cells}). The main drawback of the large area sensors with respect to smaller sensors is the related high capacitance (3.4~nF) which induces  long signals.  For the hexagonal sensors, signals last of order of 100~ns, a value which does not fit within the \ac{CTA} required 80~ns integration window. To reduce the capacitance, the 36'840 cells are grouped into four channels of 9'210 cells each, with a capacitance of 850~pF each. 

Although the SiPM production technique is well established, it has been important to characterize the hexagonal large surface device thoroughly to ensure that it meets the expected performance. 
\begin{figure}
	\centering
	\includegraphics[width=0.45\textwidth]{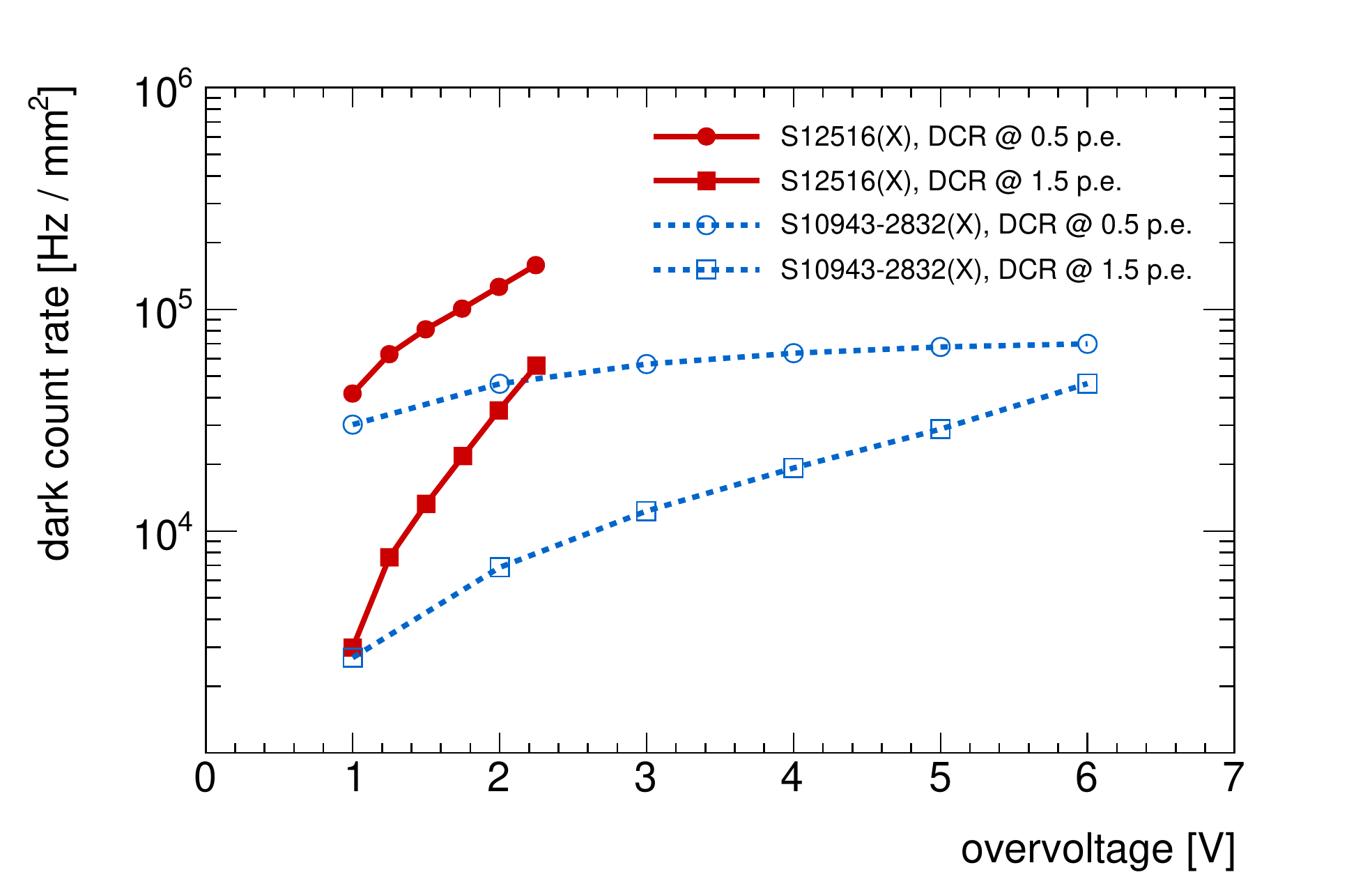}
        \includegraphics[width=0.45\textwidth]{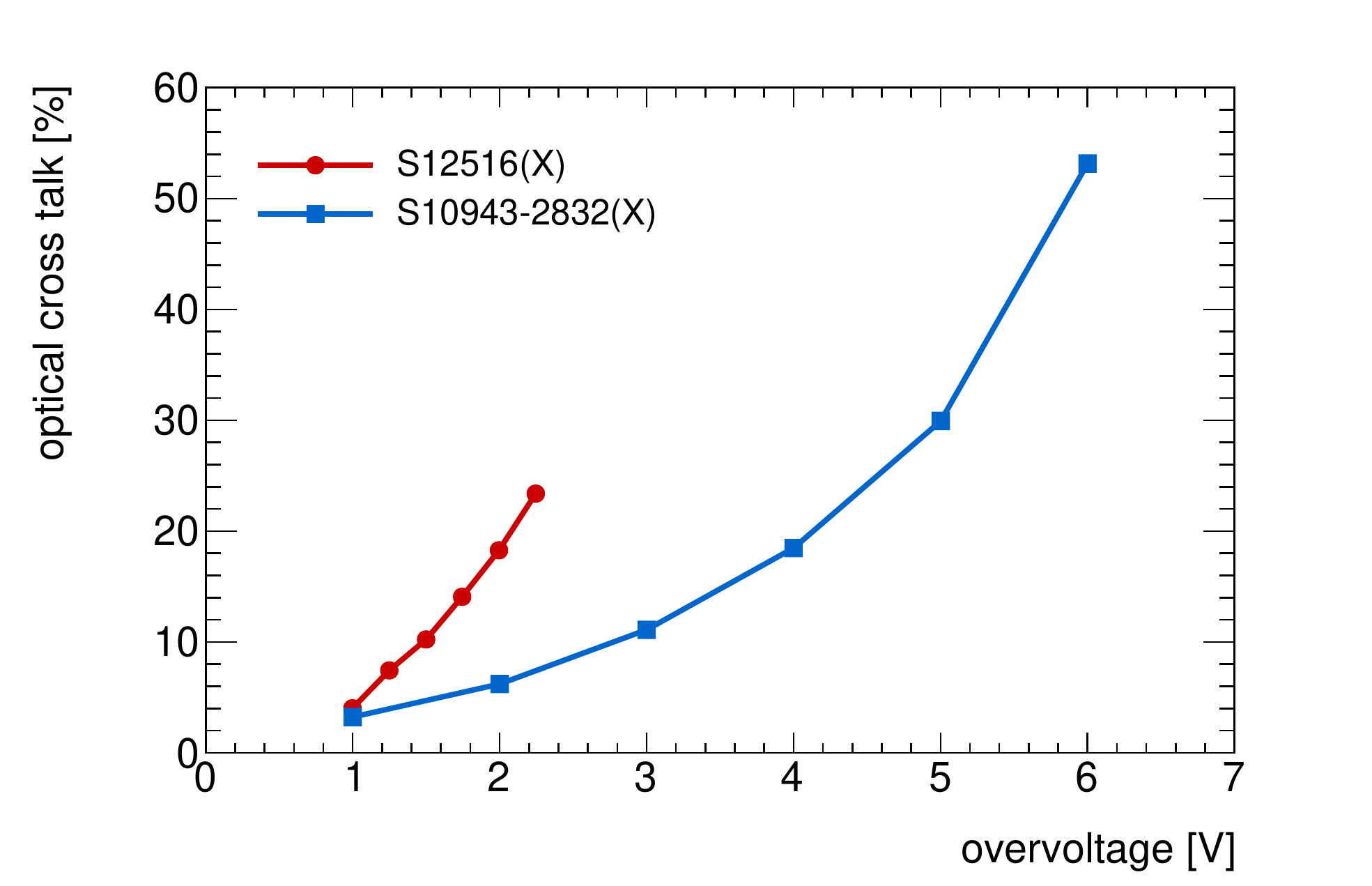}
        \includegraphics[width=0.45\textwidth]{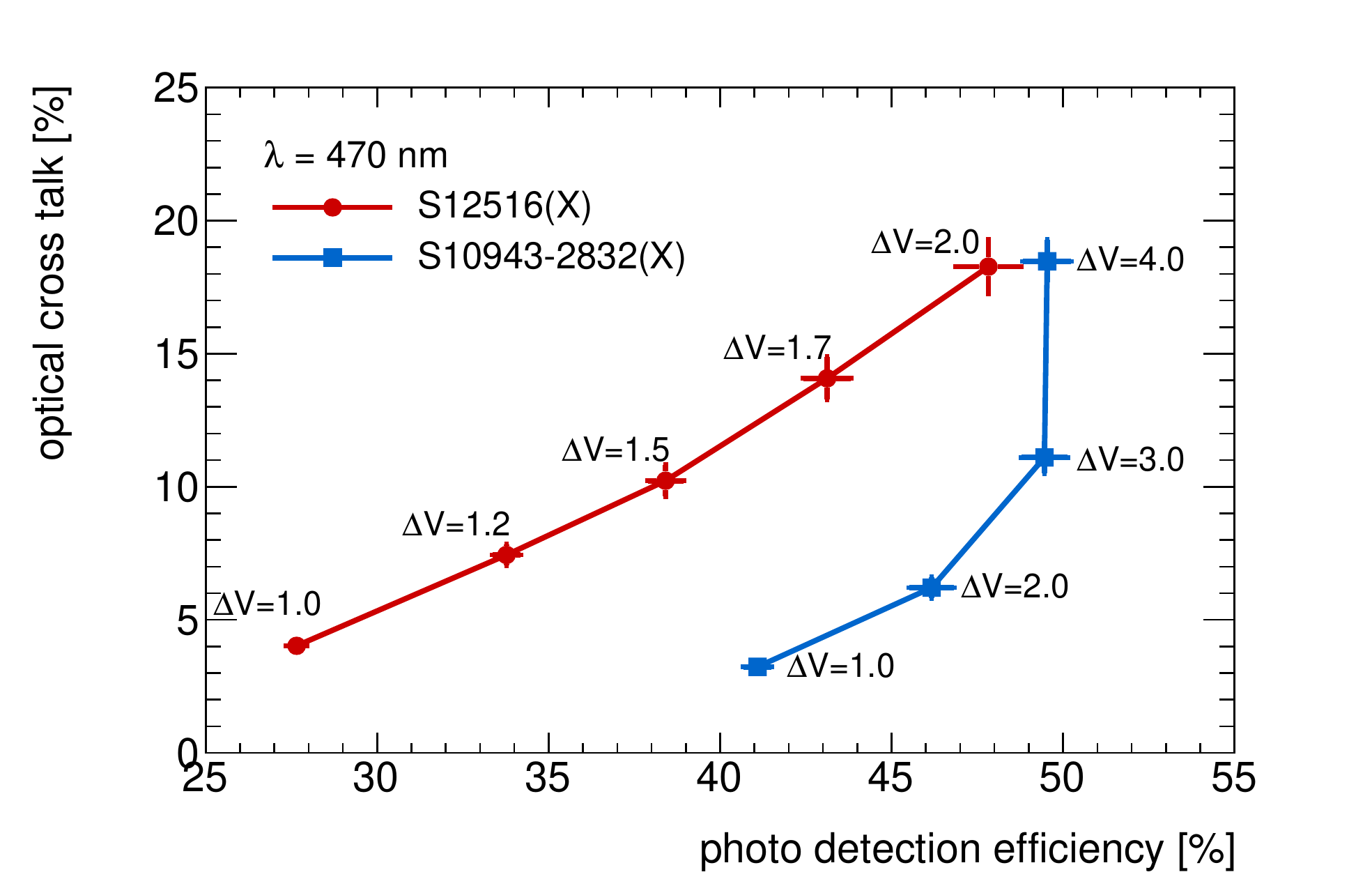}              
        \caption{Top: Dark noise per unit area at two different \ac{p.e.} thresholds for the first batch of sensors received (red curves) and the selected low technology cross-talk sensors (blue curves). The operation over-voltage of the camera will be at around 2.8 V.
	Middle: optical cross-talk versus over-voltage for the two sensor types.
        Bottom: Optical cross-talk versus \ac{PDE} for the two types of sensors. The saturation of the PDE is visible in the S10943-2832-050 sensors data. A summary of all measurements of the fist batch custom sensors is in Ref.~\cite{MPPC}.}
	\label{PDE_XT}
\end{figure}
As an example, the measured \ac{PDE}, dark count rate and cross-talk of sensors are shown in Fig.~\ref{PDE_XT}. At the time of these measurements, a system to monitor the temperature was not available and no cooling was implemented. Later evaluations showed that the sensors had been operated at an average temperature of 40$^\circ$. As a result, the values measured for dark count rate are sensibly higher than the ones that will be presented later on in Sec.~\ref{sec:tests}, for which the sensors were operated at around 20$^\circ$. However, these results still provide a valid comparison between the two sensor types, since the measurement conditions were the same for both. A future publication will discuss extensively these measurements and a previous one discussed the properties of the first version of the sensor \cite{MPPC}.

As for the light concentrators, testing each SiPM in the laboratory was not feasible. Hence, our strategy has been to characterize a sub-sample of the sensor total production to verify the reliability of the Hamamatsu data-sheets.
A dedicated measurement campaign has thus been carried out to measure the basic functional properties and the values of the main parameters on some sensors: I-V curves, optical cross-talk, dark-count rate, gain, breakdown voltage, \ac{PDE} and pulse shape analysis.  

\begin{figure}
	\centering
  \includegraphics[width=0.4\textwidth]{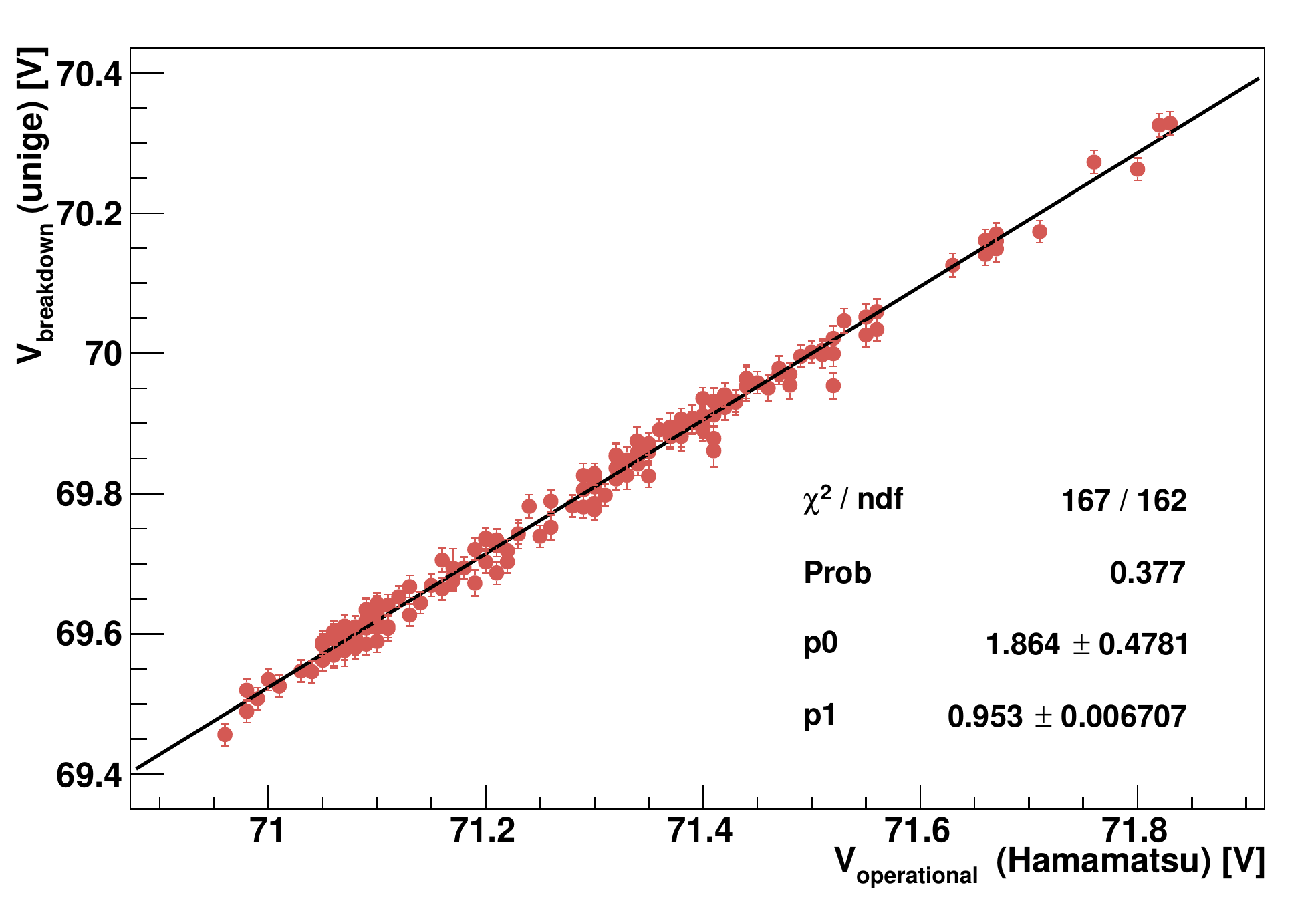}
    \raisebox{-1mm}{\includegraphics[width=0.45\textwidth]{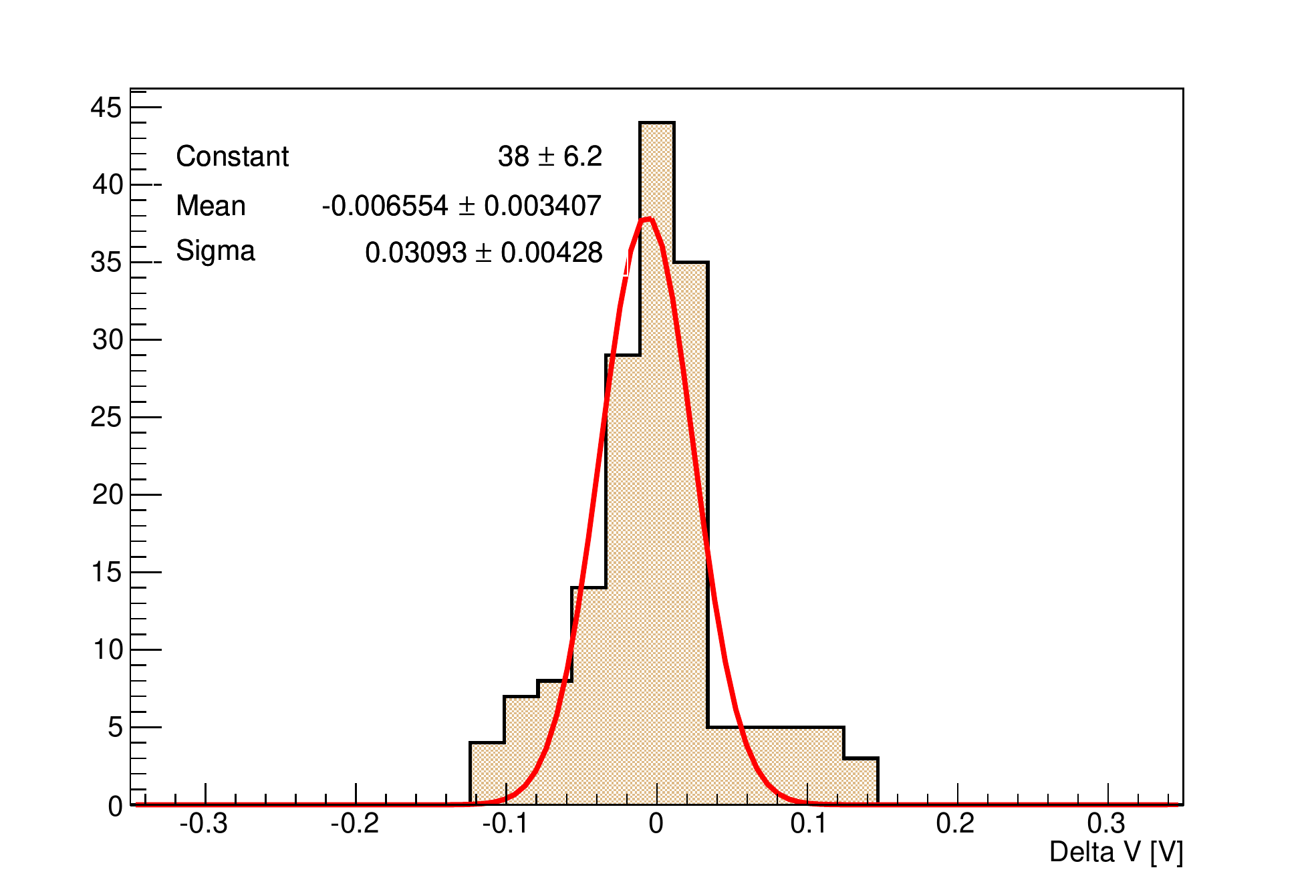}}
    \includegraphics[width=0.4\textwidth]{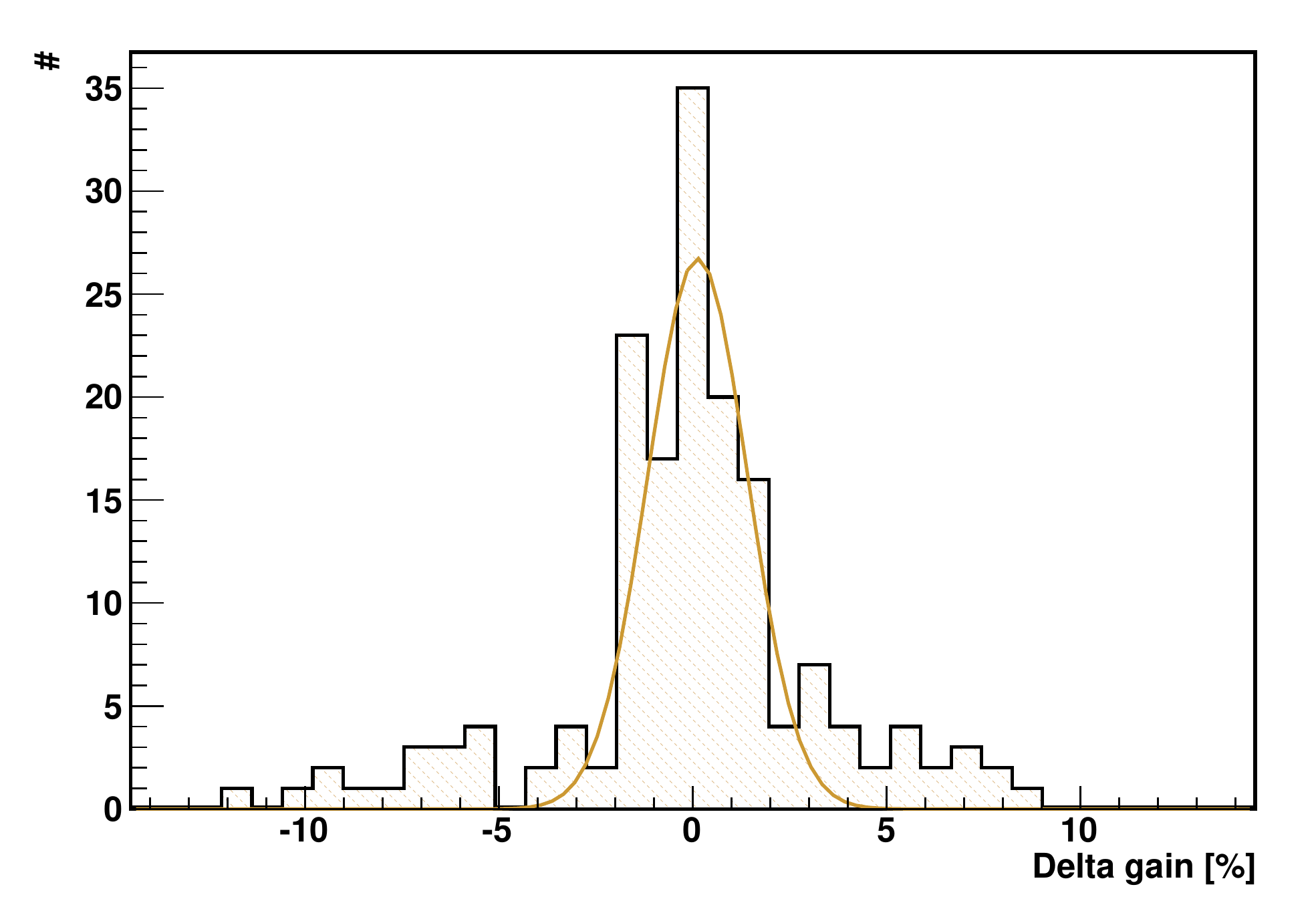}
   \caption{Top: correlation between the operational voltage provided by Hamamatsu and the breakdown voltage measured at UniGE for 42 sensors.
      Middle: $\Delta V$ is the residual of $V_{break}$ measured at $25 ^\circ$C for each of the 4 channels of a sensor with respect to their average value.       The spread varies between 10 mV and 140 mV and a Gaussian fit of the distribution returns a standard deviation of 30 mV.
   Bottom: spread in the pixel's channel gain determined with respect to the average per sensor operated at the $V_{ov}$ provided by Hamamatsu.
    The RMS of the distribution is 3\% and the Gaussian fit returns a $\sigma = 1.2\%$. Outliers of the distribution are about 20 channels beyond the 5\% variation belonging to 13 channels of 42 sensors (4 channels each).}
    \label{fig:gainelisa}
\end{figure}

In particular, from the measurement of the I-V characteristics the breakdown voltage and the quenching resistance can be extracted. 
It has been verified that the information in the Hamamatsu data-sheets on the operational voltage is well correlated with our measurements of the breakdown voltage for 42 sensors, as shown in Fig.~\ref{fig:gainelisa}-top. The conclusion of this campaign  was that the sensors' homogeneity is high, that the values of the operational voltage at a fixed gain (of $7.5 \times 10^5$ for the first type of sensors and $1.6 \times 10^6$ for the second type) provided by Hamamatsu are highly reliable. As a matter of fact, their suggested operational voltage is the best working point in terms of compromise between the \ac{PDE}, the cross-talk and the dark count rate. 
Moreover, the main parameter values of the custom designed sensors correspond to the ones expected by extrapolation from smaller area devices, which indicates that the large area hexagonal sensors are in fact performing as a conventional (smaller area, square) \ac{SiPM}. 

In the Hamamatsu data sheets the value of the operational voltage at fixed gain is reported for each of the four channels of a sensor. 
Nonetheless, the four channels share a common cathode, which implies that for the operation of the camera, an average bias voltage is applied to the four channels, rather than an individual bias per channel. 
Therefore, it has been necessary to check the values of the differences of the break down voltages among the four channels.
The typical breakdown voltage spread between channels in a sensor was less than 300~mV, that is less than 0.5\% if compared to the typical bias voltage at around 57~V. 
Fig.~\ref{fig:gainelisa}-middle shows the residual of $V_{break}$ for each of the four channels of a sensor, that is the difference between the bias voltage that a single channle would require and the average bias voltage that is applied.
This is the main parameter affecting the gain uncertainty (see Fig.~\ref{fig:gainelisa}-bottom) and hence the charge resolution of the camera.
 In Sec.~\ref{section: charge resolution} it will be shown that, indeed, this feature does not have relevant consequences on the single photon sensitivity and charge resolution of the sensors.
Simulations indicate that for a Gaussian distribution of the $V_{break}$ residuals with $\sigma/\mu = 5\%$, the charge resolution is inside the CTA required values (see section~\ref{section: charge resolution}).
Consequently, a specification value for Hamamatsu has been estimated in order to reject sensors with channel spreads $\Delta V > 300$ mV. 

The high number of microcells in the sensor provides a high dynamic range of the collected light. Measurements have demonstrated that, for the foreseen light intensities (up to a few thousand photons per pixel at most, see simulation results in Sec.~\ref{expectedPH}), the deviation from linearity is negligible. What could affect this feature is the presence of the non-imaging light concentrators, which results in a non-homogeneous distribution of the incoming light onto the sensor surface (see Ref.~\cite{CONES}). However, studies on ray-tracing simulations and preliminary measurements have shown that the effect is negligible.

Each sensor has an \ac{NTC} probe\footnote{Negative Temperature Coefficient means that the resistance decreases with increasing temperature.} in the packaging used by the front-end electronics to readout the instantaneous temperature of the device. This information is used by the \ac{PDP} slow control system to stabilize the working point of individual sensors as a function of temperature (see Sec.~\ref{electronics section}).
Employing a climatic chamber, the breakdown voltage as a function of temperature was studied, which allowed us to verify that the relation is linear with slope 54~mV$/^\circ$C as expected.


\subsection{The front-end electronics}
\label{electronics section}
The need to operate many large area \acp{SiPM} within the compact \ac{PDP} structure has posed a few challenges in the design of the front-end electronics. Due to space constraints, this has been implemented in two levels, so that each 12 pixel module is provided with a preamplifier and a slow control board. Both boards have been designed to use low-power and low-cost components. A full description of the front-end is reported in Ref.~\cite{electronics_paper}.

\begin{figure}
	\centering
        \includegraphics[angle=0, width=0.4\textwidth]{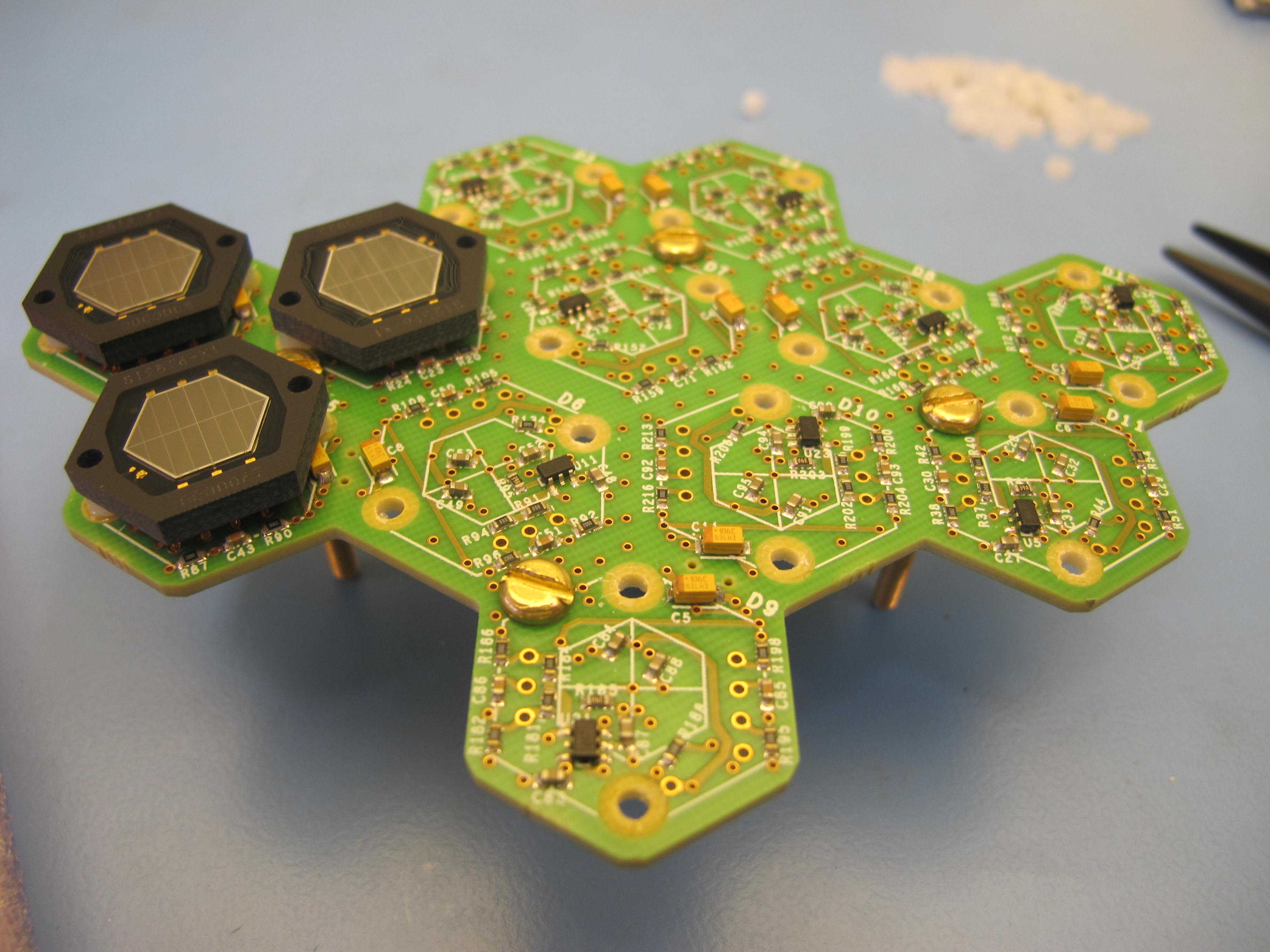}
	\caption{A picture of the preamplifier board at the sensors' side, with three out of twelve sensors mounted.}
	\label{preAmp}
\end{figure}
\begin{figure}
	\centering
	\includegraphics[width=0.5\textwidth]{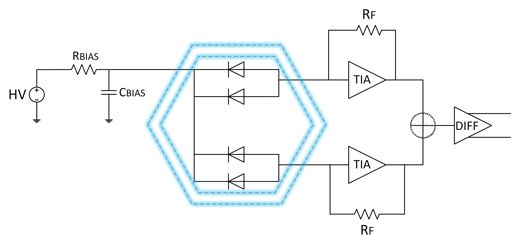}
	\caption{Preamplification topology scheme.}
	\label{front-end}
\end{figure}

The preamplifier board (see Fig.~\ref{preAmp}), holds together the 12 pixels of a module and implements the amplification scheme shown in Fig.~\ref{front-end}. 
Due to space, power and cost constraints, it was not possible to provide each of the four channels of a sensor with a low-noise amplifier. As a solution, the signals from the four channels are summed via two low-noise trans-impedance amplification stages followed by a differential output stage. The values of the parameters of this circuitry have been fine-tuned (through simulations validated by measurements) as a compromise between gain and bandwidth, to achieve well behaved pulses over the full dynamic range.
\begin{figure}
	\centering
	\includegraphics[width=0.5\textwidth]{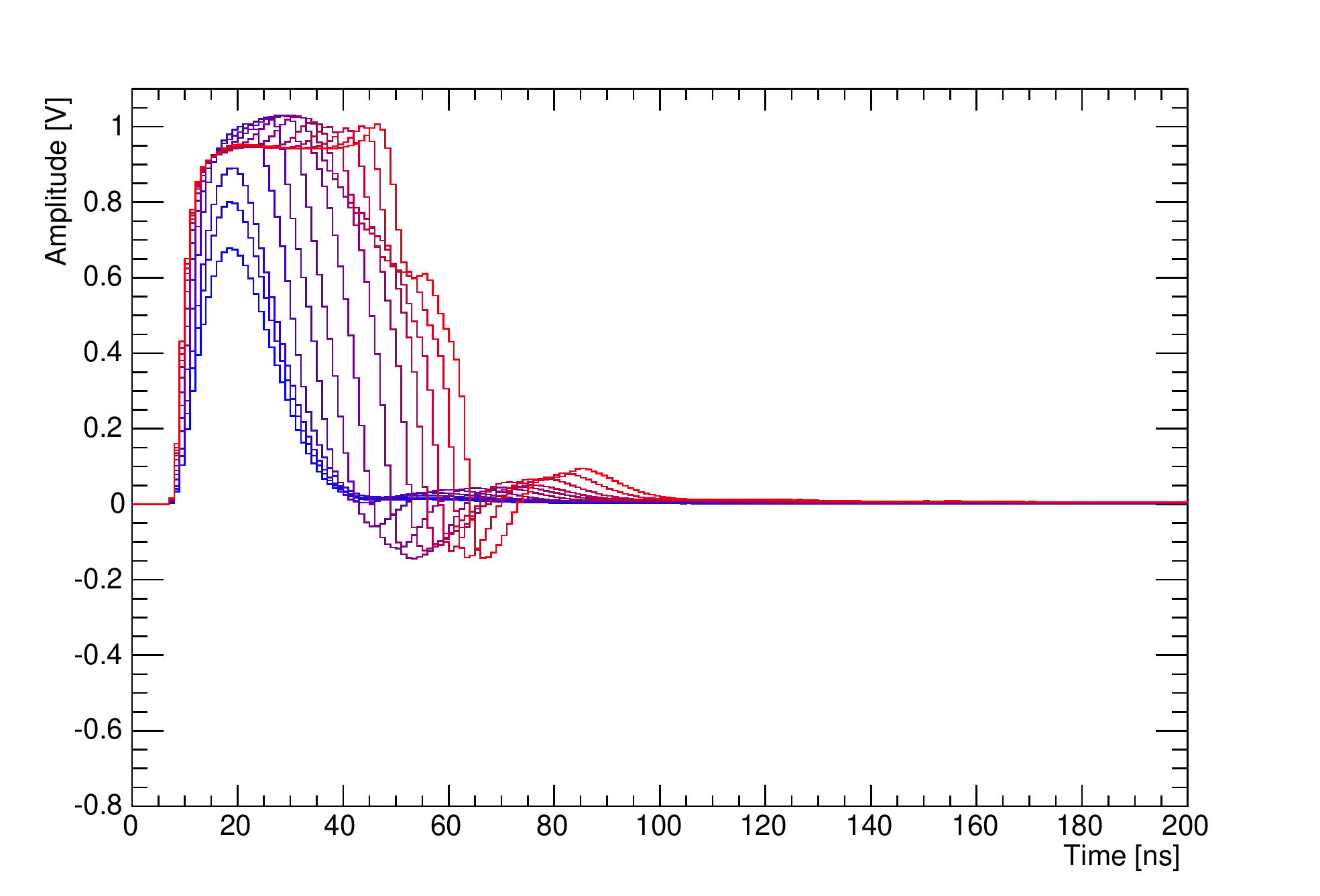}
	\caption{Amplified pulses for increasingly high light levels. The saturation effect is visible and can be corrected for with proper analysis of pulses.}
	\label{pulses}
\end{figure}
The pulse shapes produced by the preamplifier are shown in Fig.~\ref{pulses} for increasingly high light levels. The system provides a linear response up to around 750 photons, after which saturation occurs. Despite this loss of linearity, it will be shown later in Sec.~\ref{section: charge resolution}, that the signals up to few thousands of photons (that is, the range foreseen for the SSTs) can be reconstructed with a resolution that is still well within the CTA requirements.
\newline
A peculiarity of the pre-amplification scheme is the DC coupling of the sensor to the preamplifier, which gives the possibility to measure directly the \ac{NSB} during observation on a per-pixel basis. In fact, the \ac{NSB} is expected to be of the order of 20-30~MHz per pixel in dark nights, reaching up to 600~MHz in half-moon nights, considerably higher than the sensor dark noise rate of about few MHz (see Fig.~\ref{PDE_XT}). As a net effect, the signal baseline position shifts towards higher values as a function of the \ac{NSB} level. Therefore, the latter can be estimated by measuring the position (in addition to the noise) of the baseline thanks to the DC coupling.
The capability of measuring the \ac{NSB} could be used to keep the trigger rate constant and this feature can be implemented in DigiCam thanks to its high flexibility (see Sec.~\ref{sec:DigiCam}).
\begin{figure}
	\centering
        \includegraphics[angle=90, width=0.4\textwidth]{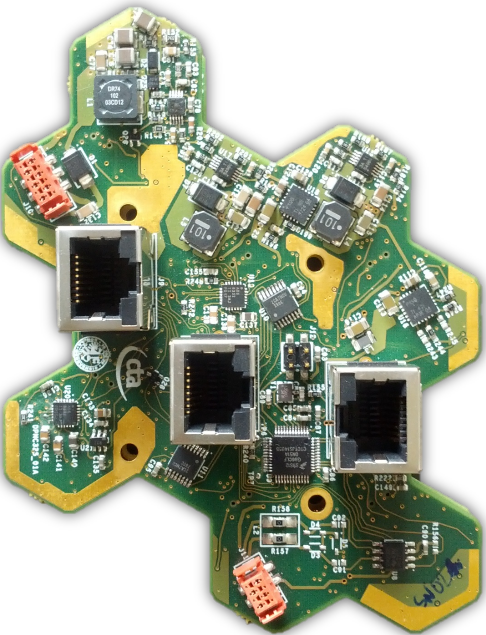}
        \caption{A picture of the \ac{SCB} from the side of the RJ45 connectors.}
	\label{SCB}
\end{figure}
\newline
The \ac{SCB} (see Fig.~\ref{SCB}) has a more complex design than the preamplifier board and features both analog and digital components. Its functions are:
\begin{itemize}
	\item to route the analog signals from the preamplifier board to DigiCam via the three RJ45 connectors,
	\item to read and write the bias voltages of the 12 sensors individually,
	\item to read the 12 \ac{NTC} probes encapsulated in the sensors,
	\item to stabilize the operational point of the sensors,
	\item to allow the user to retrieve the high-voltage and temperature values, as well as the values of the various functional parameters, via a CAN bus interface.
\end{itemize}
\begin{figure}
	\centering
        \includegraphics[width=0.4\textwidth]{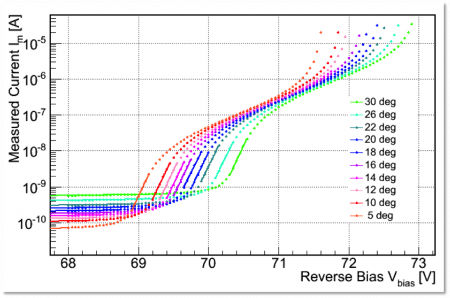}
        \caption{I-V characteristics of a channel of a SiPM at different temperatures.}
	\label{IV vs T}
\end{figure}
Stabilization of the sensor's operational point is a key feature of the camera design. The breakdown voltage variation with temperature has been measured on a few sensors in a climatic chamber by extracting the breakdown voltage from the I-V characteristics at different temperatures. Fig.~\ref{IV vs T} shows an example of I-V curves measured at different temperatures. The results give a coefficient $k$ of the order of 54~mV/$^\circ$C.
Temperature variations produce gain changes and gain non-uniformities across the pixels. A stabilization of the working point is thus necessary, and can be achieved, in principle, in two ways: either by maintaining the temperature constant or by actively adapting the bias voltage according to the temperature variations, in order to keep the over-voltage constant. The implementation of a precise temperature stabilization system was challenging and would have been costly due to the complexity of the camera and the heterogeneity of the different heating sources. Therefore, the choice has been to build a water cooling system (described in Sec.~\ref{sec:cooling}) that keeps the temperature within the specified operation range during observation (between {-15}$^\circ$C and +25$^\circ$C), and to use a dedicated correction loop, implemented in a micro-controller on the \ac{SCB}, to compensate for temperature variations at the level of single pixels. In the compensation loop, the NTC probe of each sensor is read at a frequency of 2~Hz and, according to a pre-calibrated look-up table, the bias voltage of individual sensors is updated at a frequency of 10~Hz to compensate the working point for temperature variations of less than 0.2~$^\circ$C. With such a system, the over-voltage of each sensor is kept stable, as well as the gain and the PDE.
This concept was successfully proven by FACT~\cite{FACT}, with a lower number of temperature sensors (31 distributed homogeneously over the \ac{PDP} and read every 15~s~\cite{FACT2}). A detailed description of the compensation loop of the \ac{SCB} is reported in Ref.~\cite{electronics_paper}.

As a design validation test of the front-end hardware, the electronic cross-talk of a full module (cones + sensors + preamplifier board +  \ac{SCB}) has been measured. A single pixel has been biased and set in front of a calibrated LED source, while all the other pixels have been left unbiased, and the signal induced on these pixels was measured. The results of the test 
shows a very low level of electronic cross-talk. Small induced pulses on pixels sharing the same connector as the illuminated pixel, corresponding to a signal between 1 and 2 \acp{p.e.}, can be observed only when around 3000 \acp{p.e.} are injected in the illuminated pixel. Although the effect is, in fact, negligible, in a future re-design of the front-end boards the electronic cross-talk could be further minimized (if not eliminated) by a more appropriate choice of these connectors.

To qualify the electronic components of the camera, standard industrial techniques have been developed in house, where dedicated electronic boards, test setups, firmwares and softwares have been produced. The design of both the preamplifier board and the  \ac{SCB} has been accompanied by the parallel development of \acp{PCB} designed to perform a full functional test of the two boards at the production factory. Test setups and analysis software have been developed in order to provide a user friendly interface that could be used by the operators at the factory to assess the quality of the production prior to the shipping of the boards. The functional test automatically produces a report and uploads it on the web for an easy real-time monitoring of the progress. In the case of the \ac{SCB}, the functional test also performs a first calibration, that is then completed in the laboratory in Geneva.
\newline
\begin{figure}
	\centering
	\includegraphics[width=0.5\textwidth]{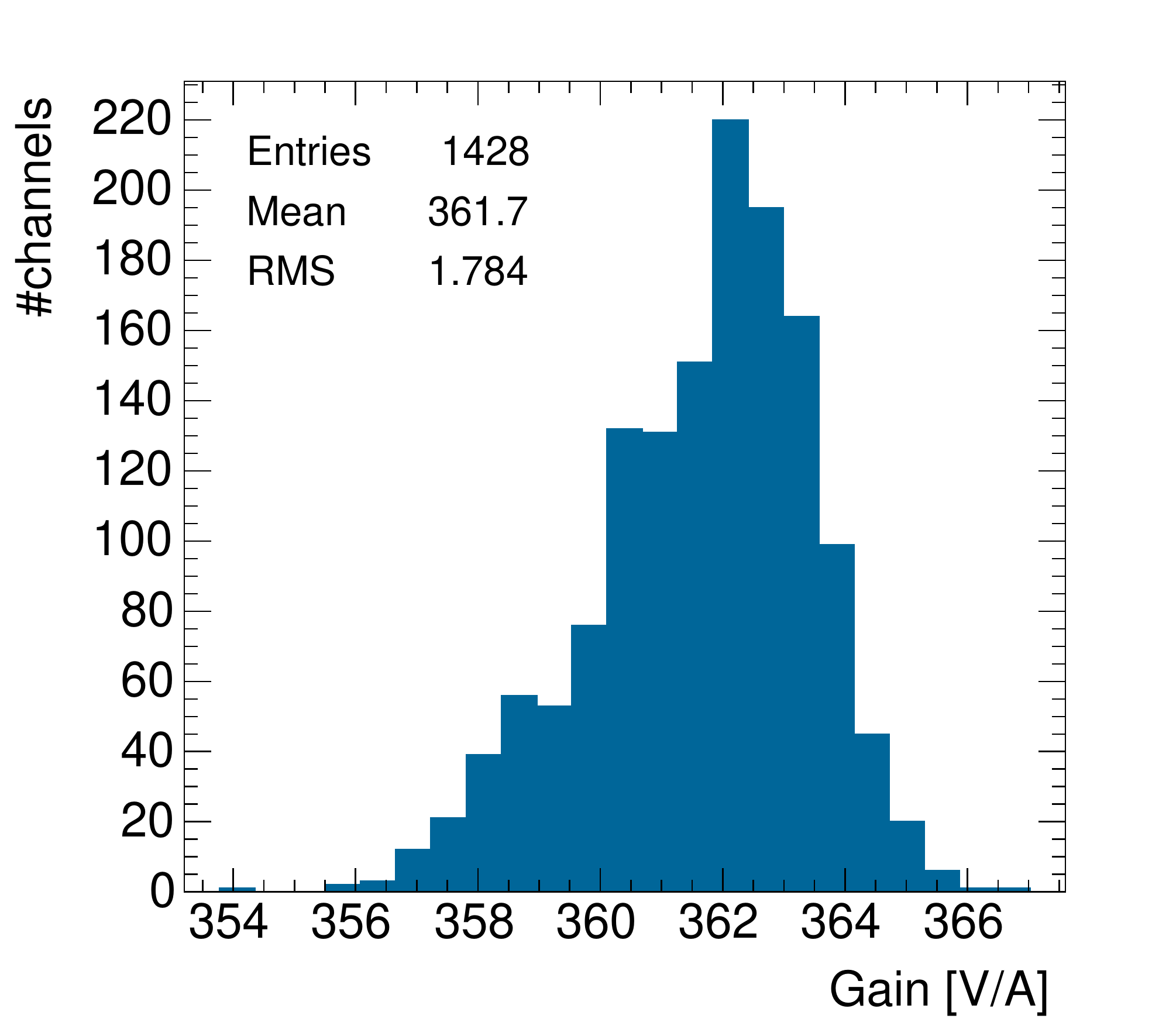}
	\caption{Distribution of the preamplifier gains.}
	\label{gain uniformity preAmp}
\end{figure}
\begin{figure}
	\centering
	\includegraphics[trim = 0mm 3mm 0mm 0mm, clip, width=0.5\textwidth]{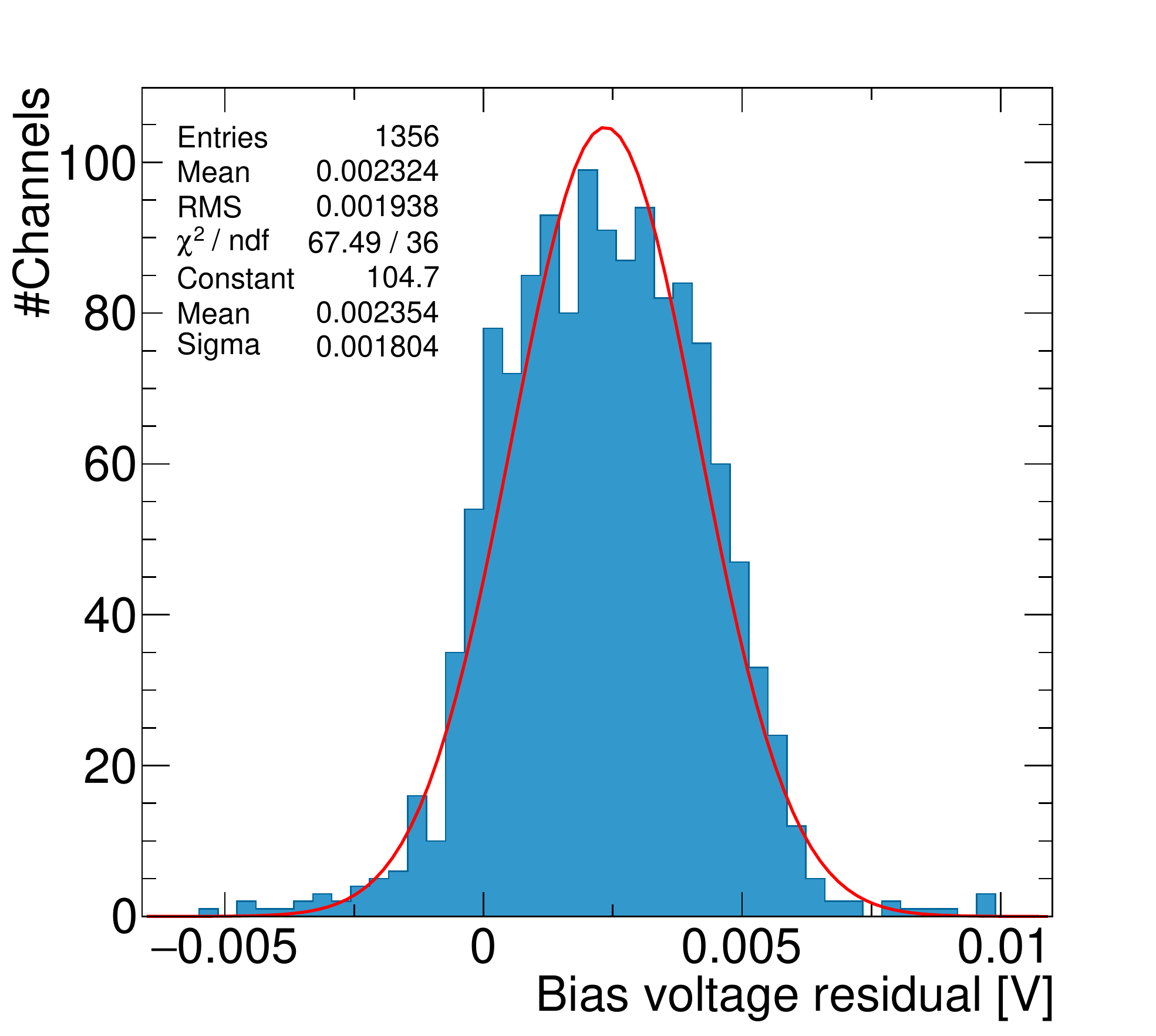}
	\caption{Distribution of the residuals on the applied sensor bias voltage with respect to the one measured by the  \ac{SCB} closed loop.}
	\label{HV calibration error SCB}
\end{figure}
The results of the functional test allow us to establish the overall quality of the production. A typical example is presented in Fig.s~\ref{gain uniformity preAmp} and~\ref{HV calibration error SCB}. Fig.~\ref{gain uniformity preAmp} shows the distribution of the measured gains of the preamplifiers over a full batch of preamplifier boards. The 0.5\% relative dispersion demonstrates the high homogeneity of the production. In Fig.~\ref{HV calibration error SCB}, the distribution of the residual on the applied sensor bias voltage, with respect to the one measured by the slow control closed loop, gives a value of $2.3\pm1.9$~mV, very small when compared to the typical values of the bias voltage, of the order of 57~V. For more details, and for a results on first tests of the compensation loop, see Ref.~\cite{electronics_paper}.


\section{The DigiCam readout and trigger electronics}
\label{sec:DigiCam}
DigiCam is the fully digital readout and trigger system of the camera using the latest \ac{FPGA} for high throughput, high flexibility and dead-time free operation. 
Here we summarize the relevant features of DigiCam, but for a complete overview see Ref.~\cite{ICRC_DigiCam}.
\begin{figure}
	\centering
        \includegraphics[width=0.45\textwidth]{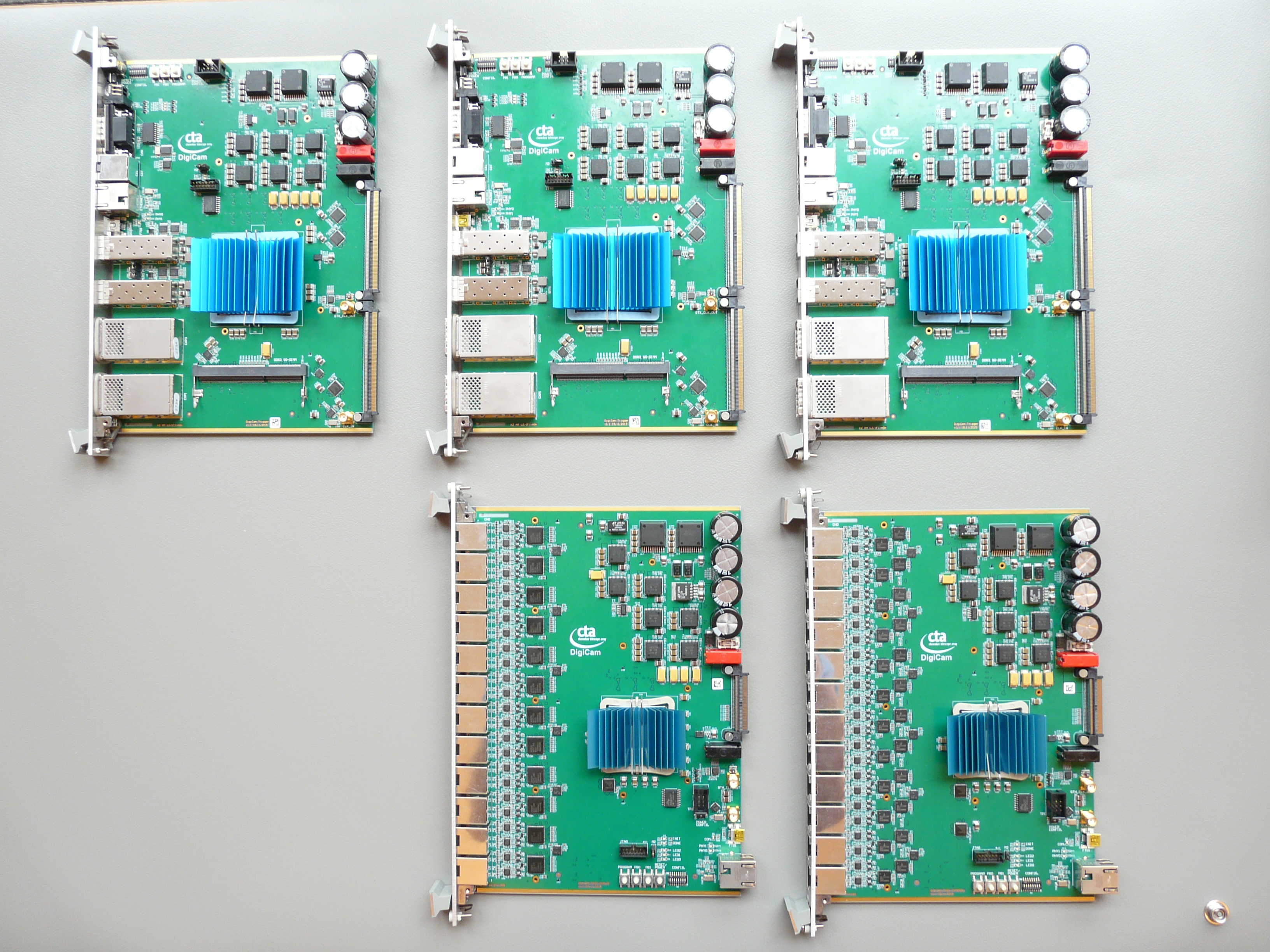}
        \caption{Picture of three DigiCam trigger boards (top) and two digitizer boards (bottom).}
	\label{DigiCam boards}
\end{figure}
\begin{figure}
	\centering
        \includegraphics[width=0.45\textwidth]{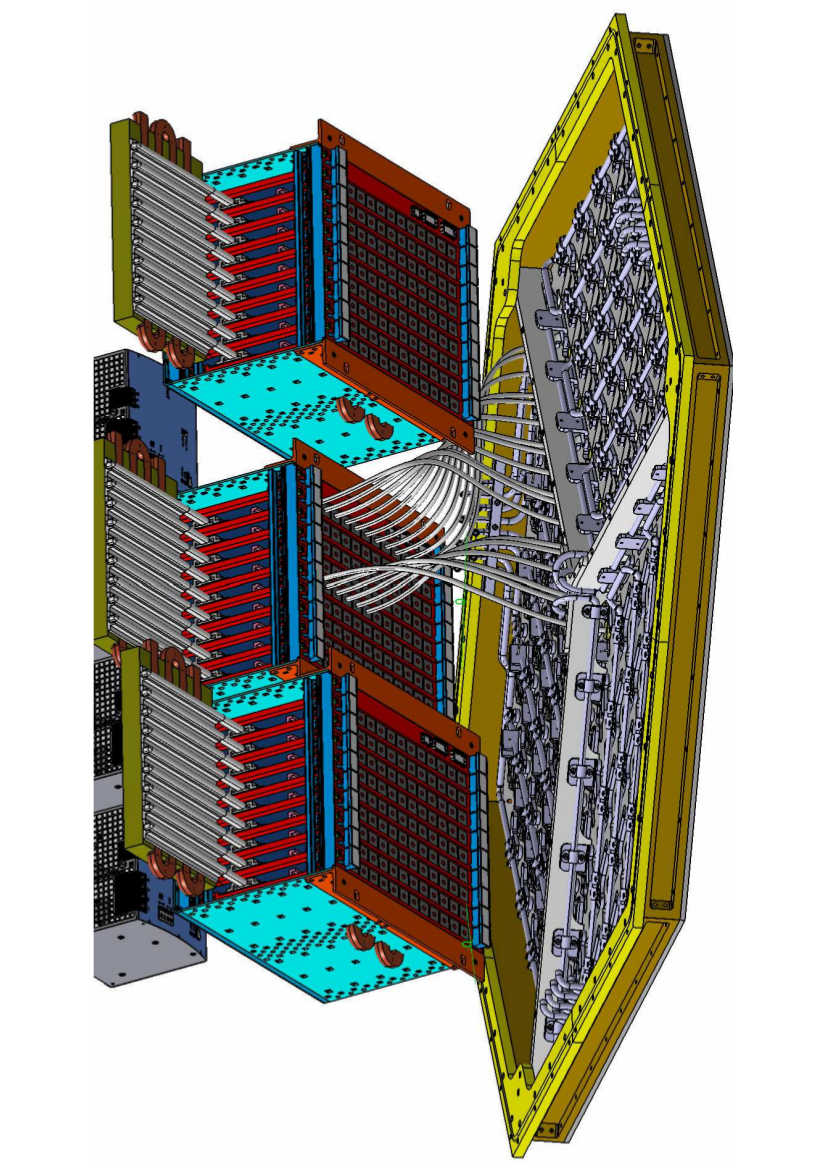}
	\caption{Drawing of the three micro-crates hosting the DigiCam readout and trigger electronics, installed behind the \ac{PDP}.}
	\label{microcrates}
\end{figure}
\newline
The DigiCam hardware consists of 27 digitizer boards and three trigger boards (see Fig.~\ref{DigiCam boards}) arranged in three micro-crates (see Fig.~\ref{microcrates}), each containing 9 digitizer boards and one trigger board. 
\begin{figure}
	\centering
        \includegraphics[width=0.45\textwidth]{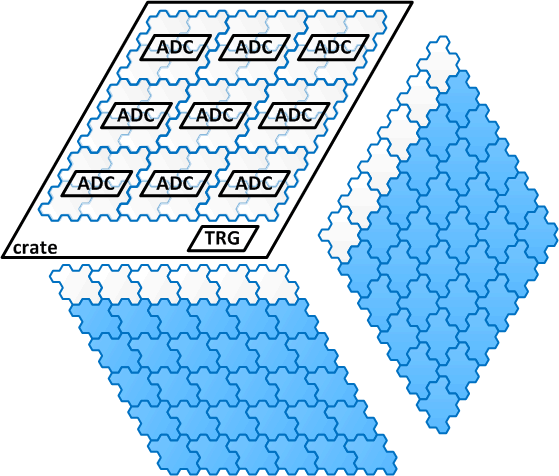}
        \caption{Subdivision of the PDP into logical sectors in DigiCam.}
	\label{sectors}
\end{figure}
From the point of view of the readout, the \ac{PDP} is divided into three logical sectors (432 pixels, 36 modules, see Fig.~\ref{sectors}), each connected to one micro-crate. The three micro-crates are connected with each other through the three trigger boards. Data are exchanged between crates in order for the trigger logic to be able to process images where Cherenkov events have been detected at the boundary between two (or all three) sectors. One of the trigger boards is configured as master, with the function of receiving the signals used for the trigger decision and the corresponding data of the selected images from the slave boards and of sending them to the camera server.

The analog signals from 4 modules (48 pixels) of the \ac{PDP} are transferred to a single digitizer board via standard CAT5 cables, where the signals are digitized at a sampling rate of 250~MHz (4~ns time steps) by 12-bit \acp{FADC}. ADCs from both Analog Devices, Inc, and Intersil have been tested in order to verify possible performance and cost benefits. Preliminary results are given in Sec.~\ref{section: charge resolution}. The 250~MHz sampling rate has been proven to be adequate for a sufficiently precise photo-signal reconstruction already by FlashCam~\cite{ICRC_DigiCam}, which deals with \ac{PMT} signals that are faster than the \ac{SST-1M} camera \ac{SiPM} signals (for which, therefore, the sampling is more accurate). 

The digitized samples are serialized and sent in packets through high speed multi-Gbit serial digital  GTX/GTH interfaces to the Xilinx XC7VX415T FPGA, where they are pre-processed and stored in the local ring buffers. 

The data from the 9 FADC boards of a micro-crate are copied and sent to the corresponding trigger board, where they are stored into 4GB external DDR3 memories. 
Without accounting for the entire readout chain, i.e. only at the trigger board level, with a trigger rate of 600~Hz (resp. 2~kHz), the events can be stored 154~s (resp. 46~s) before being readout. This calculation assumes an event size of 43.2~kB. 

In order to reduce the size of the data received and processed by the trigger card, the digitized signals are first grouped in sets of three adjacent pixels (called triplets) and re-binned at 8~bits.

The trigger board features a Xilinx XC7VX485T FPGA where a highly parallelized trigger algorithm is implemented. The algorithm is applied within the PDP sector managed by the micro-crate, plus the neighboring pixels from the adjacent sectors, whose information is shared thanks to the intercommunication links between the three trigger boards via the backplane of the microcrate. The trigger decisions are taken based on the recognition of specified geometrical patterns among triplets over threshold in the lower resolution copy of the image. A high flexibility is ensured in the implementation of different trigger algorithms for the recognition of multiple pattern shapes (e.g. circles and ellipses for gamma-ray events and rings for muon events) without significantly increasing the level of complexity. If an event is selected, the corresponding full resolution data stored in the digitizer boards are sent to the central acquisition system of the telescope by the master trigger card via a 10Gbps ethernet link.

As for the front-end electronics, testing hardware and protocols have been developed also for DigiCam, which are used to check the internal communication and the proper functioning of the \acp{FADC}, this latter by injecting test pulses.



\section{The cooling system}
\label{sec:cooling}
The camera will need about 2~kW of cooling power, of which about 500~W will be needed by the \ac{PDP} (0.38~W per channel) and about 1200~W by DigiCam, the rest being dissipated by auxiliary systems within the camera structure, such as the power supplies. The challenge in the design of the cooling system has been the necessity of efficiently extracting the dissipated heat from such a compact camera while complying with the IP65 insulation requirement. Such a demand rules out the possibility of using air cooling, and a water-based cooling system has been adopted as a solution, to extract the heat from both the \ac{PDP} and DigiCam.


\subsection{\ac{PDP} cooling}
\begin{figure}
	\centering
        \includegraphics[width=0.5\textwidth]{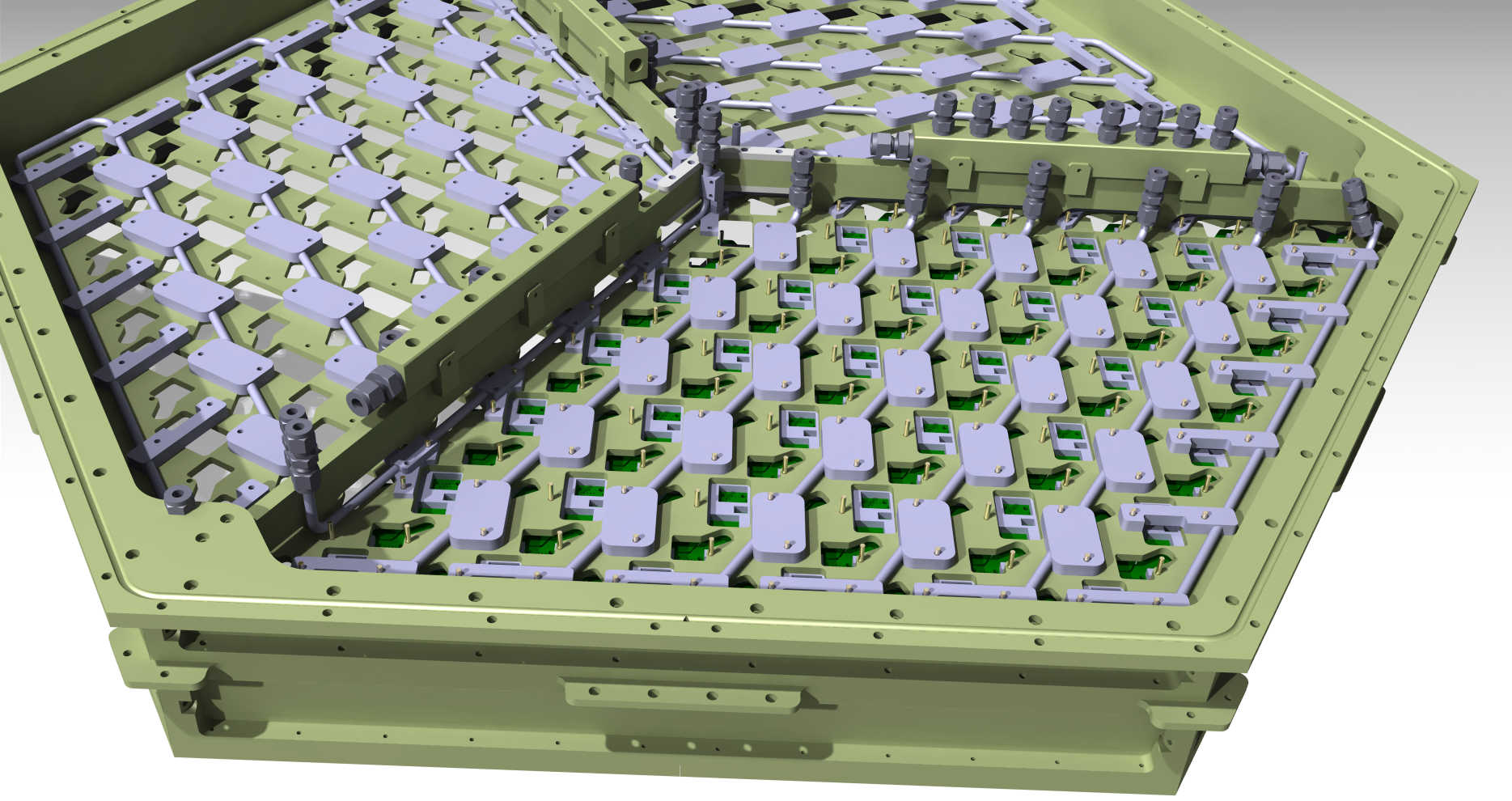}
                \includegraphics[width=0.5\textwidth]{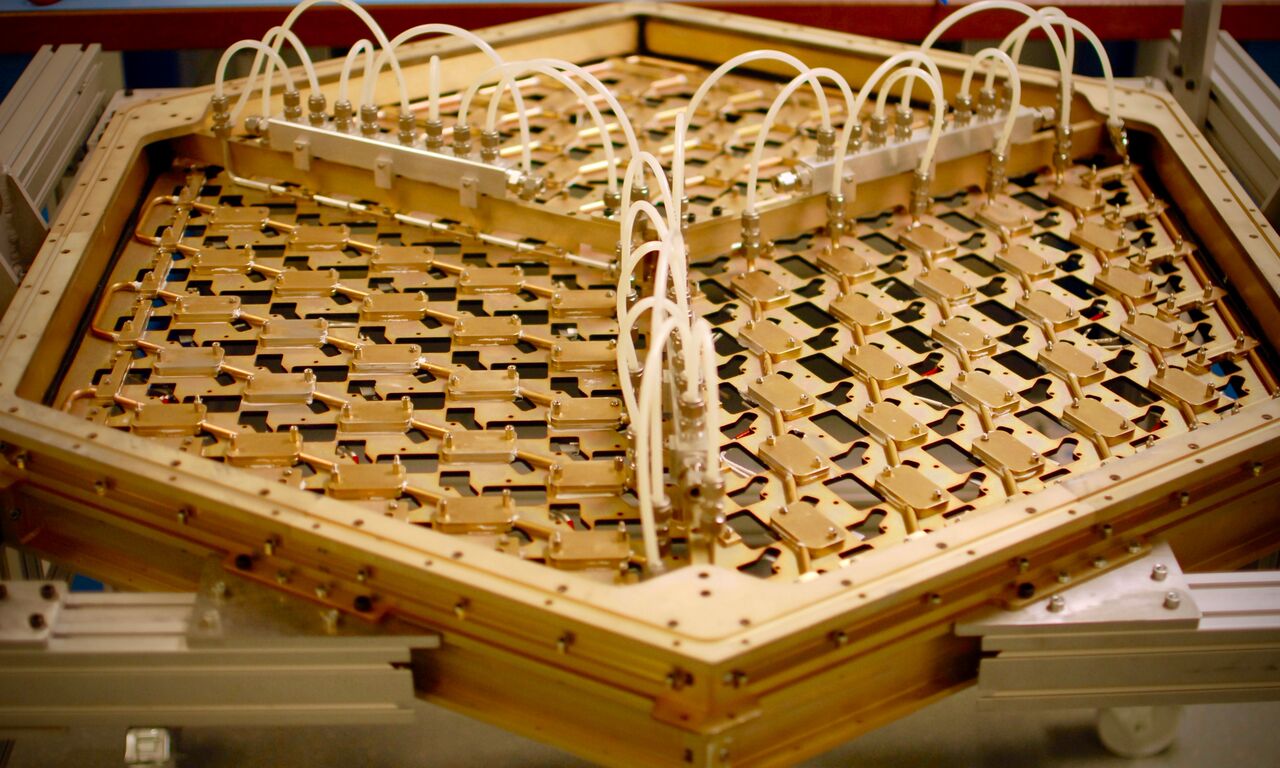}
	\caption{Top: A CAD drawing of the connection of the cooling pipes to the PDP backplate via aluminium blocks. Bottom: a photograph of the backplate and pipes.}
	\label{cooling}
\end{figure}
The \ac{PDP} is cooled by a constant flow of cold water mixed with glycol, that keeps the temperature at around 15-20~$^\circ$C. Fluctuations on the temperature of individual sensors, that translate into fluctuations of their operational point, are managed by the compensation loop of the slow control system as described in Sec.~\ref{electronics section}. The water is cooled at around 7~$^\circ$C by a chilling unit installed outside the camera on the telescope tower head. The liquid flows through aluminium pipes that are connected to the backplate of the \ac{PDP} via aluminium blocks (see Fig.~\ref{cooling}). The backplate itself thus acts as a cold plate for the whole \ac{PDP}. The contact between the backplate and the front-end electronics boards (the preamplifier and the \ac{SCB}) is realized via the four mounting screws of each module, that act as cold fingers.

To homogenize the heat distribution over the surface of the two electronics boards, both \acsp{PCB} have been realized with a thicker copper layer (72~$\upmu$m instead of the conventional 18~$\upmu$m). Furthermore, a thermally conductive material (TFLEX 5200 from LAIRD technologies) is inserted between the two boards and between the full module and the backplate.
\begin{figure}
	\centering
        \includegraphics[trim = 10mm 0mm 0mm 20mm, clip, width=0.5\textwidth]{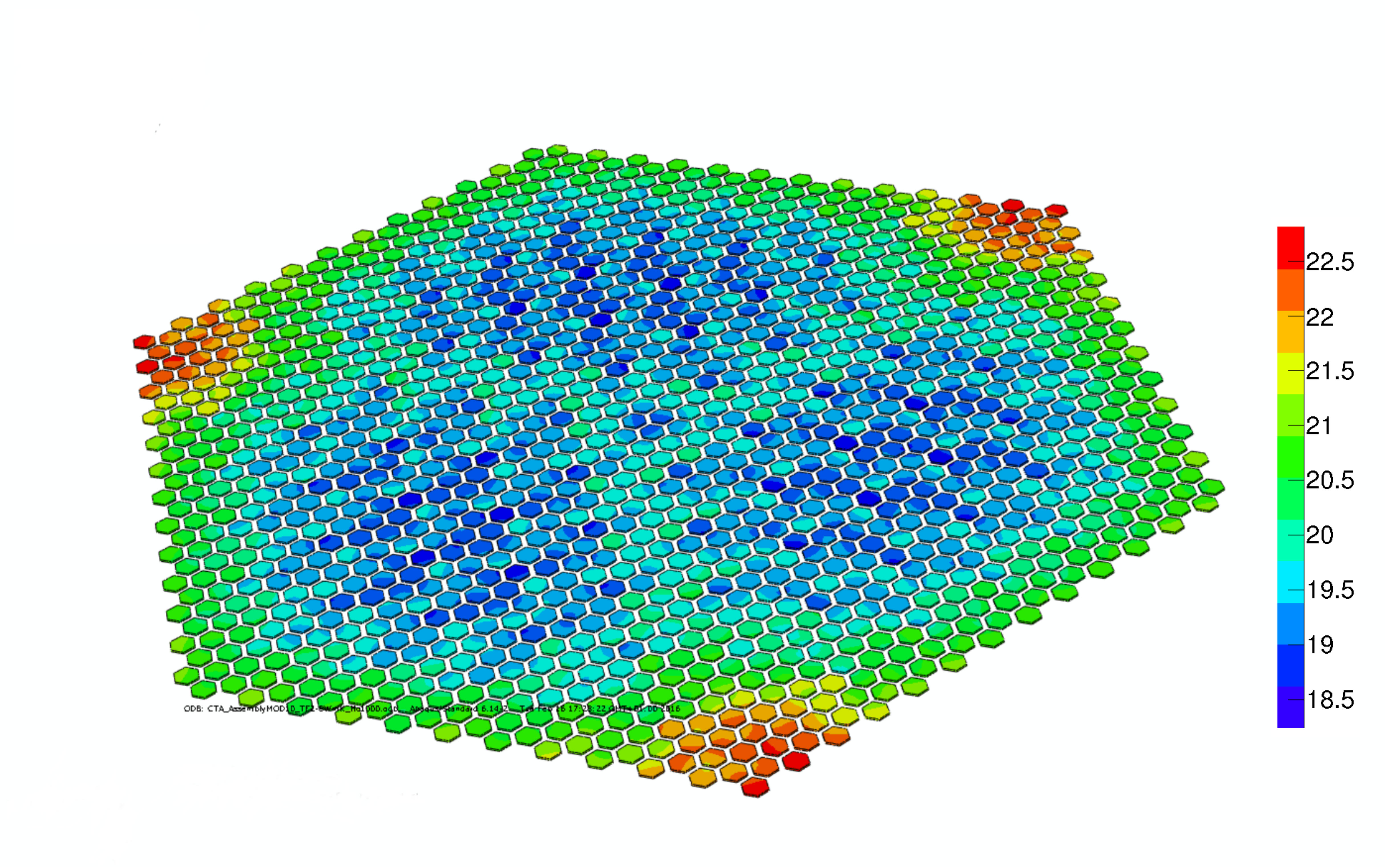}
        \caption{\ac{FEA} calculation of the \ac{PDP} temperature when the cooling system is operating with water at 7~$^\circ$C. The color scale is in $^\circ$C.}
	\label{cooling result}
\end{figure}
Fig.~\ref{cooling result} shows an \ac{FEA} calculation of the temperature distribution over the 1296 pixels during operation of the cooling system with water at 7~$^\circ$C. 
\begin{figure}
	\centering
        \includegraphics[width=0.45\textwidth]{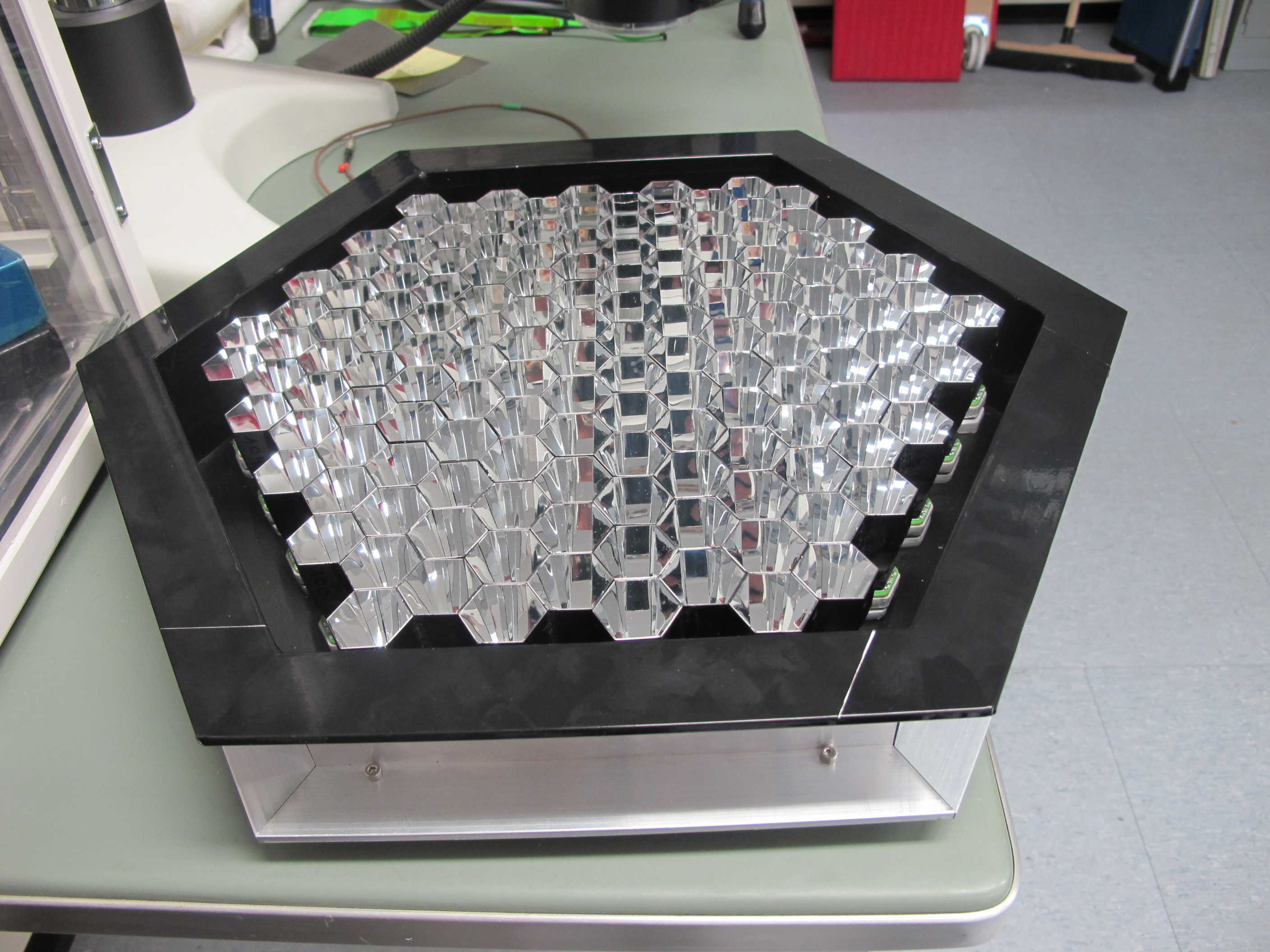}
        \caption{Mockup of the \ac{PDP}, used for testing the \ac{PDP} cooling system.}
	\label{mockup}
\end{figure}
\begin{figure}
	\centering
        \includegraphics[width=0.4\textwidth]{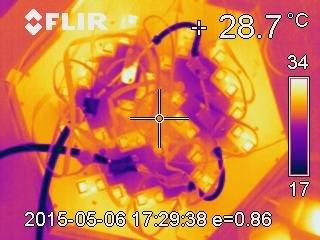}
        \caption{IR image of the mockup (view from the backplane) during the test of the cooling system.}
	\label{mockup IR}
\end{figure}
The concept has been validated on a mock-up of the \ac{PDP} with 12 of the 108 modules installed on a size-reduced \ac{PDP} mechanical structure (see Fig.s~\ref{mockup} and~\ref{mockup IR}). 
\begin{figure*}
	\centering
	\includegraphics[width=\textwidth]{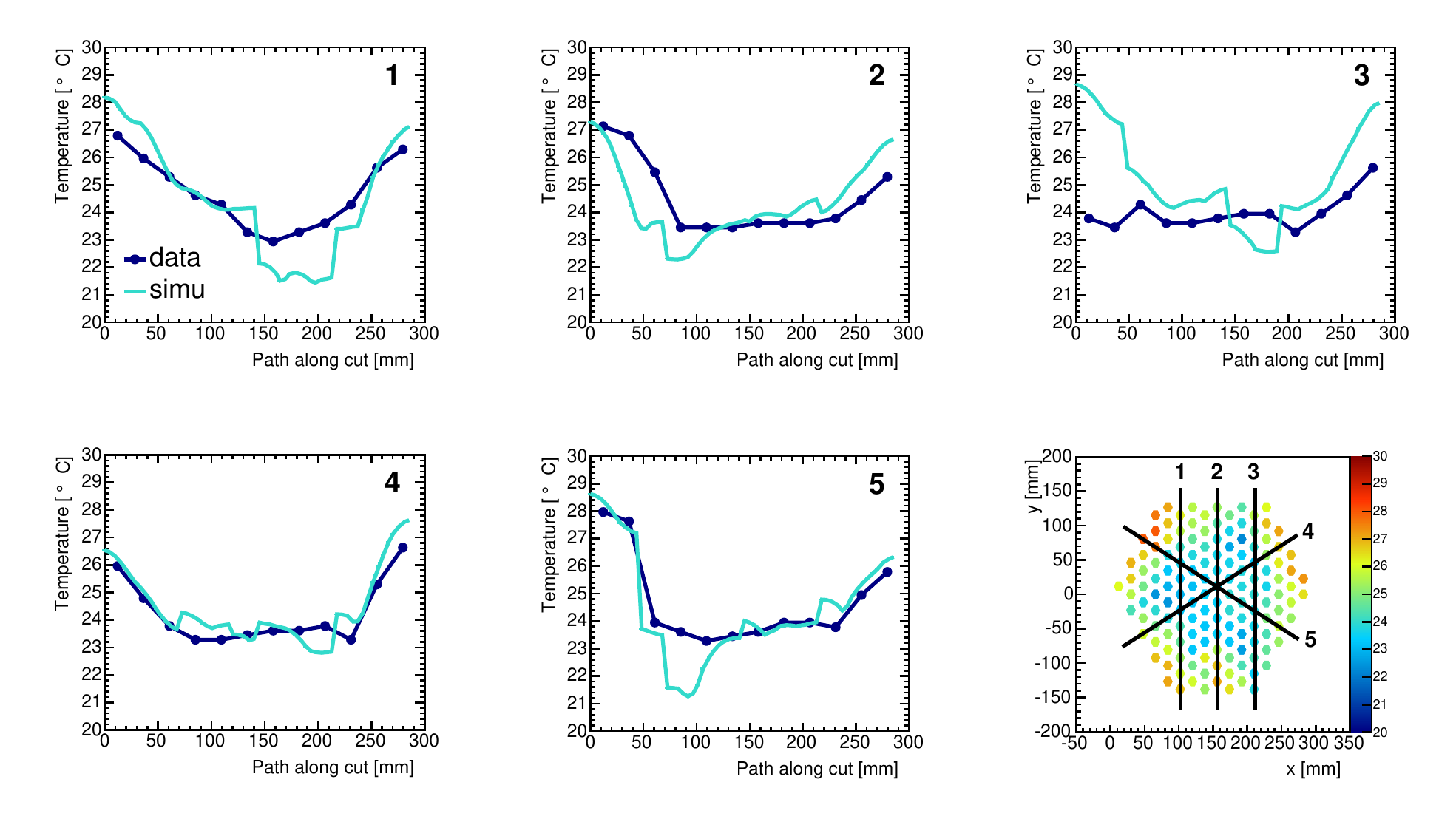}
	\caption{Results of the \ac{PDP} cooling tests on a 1:10 mockup of the \ac{PDP}. Comparison between data and the \ac{FEA} calculation are shown for groups of pixels belonging to different sections (1,2,3,4,5) along the surface as shown in the bottom right plot. The agreement between data and simulation is good. The discrepancy that is visible in the first 50~mm along direction 3 (top right plot) is due to the fact that in the actual setup the cooling pipe was locally touching the backplane, which is not accounted for in the \ac{FEA}.}
	\label{cooling test results}
\end{figure*}
The results of the test are presented in Fig.~\ref{cooling test results}.

\begin{figure}
	\centering
        \includegraphics[width=0.45\textwidth]{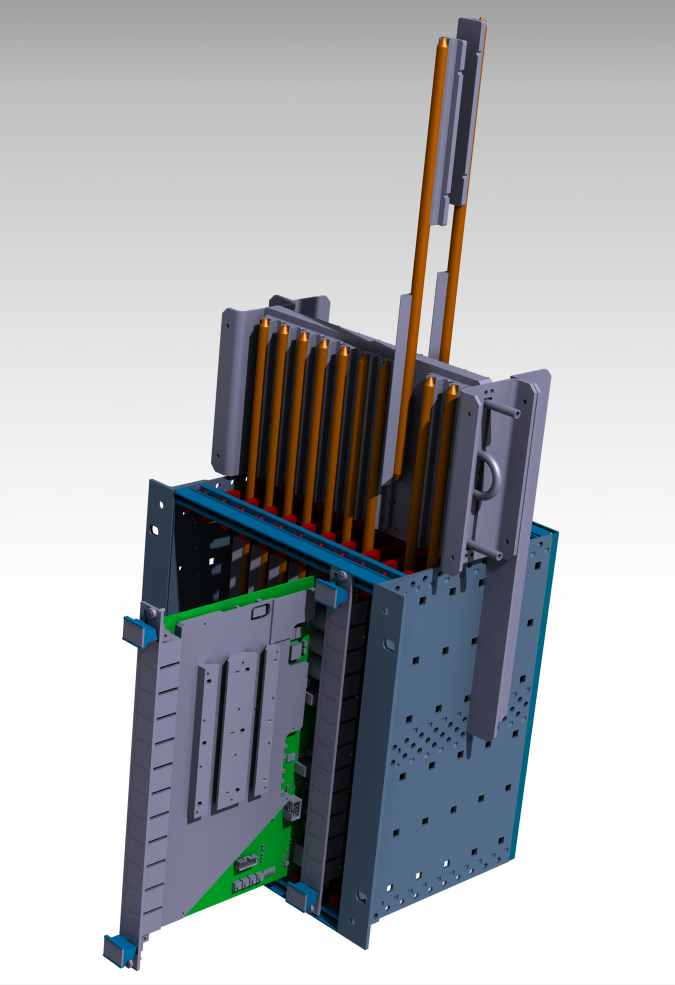}
        \includegraphics[width=0.45\textwidth]{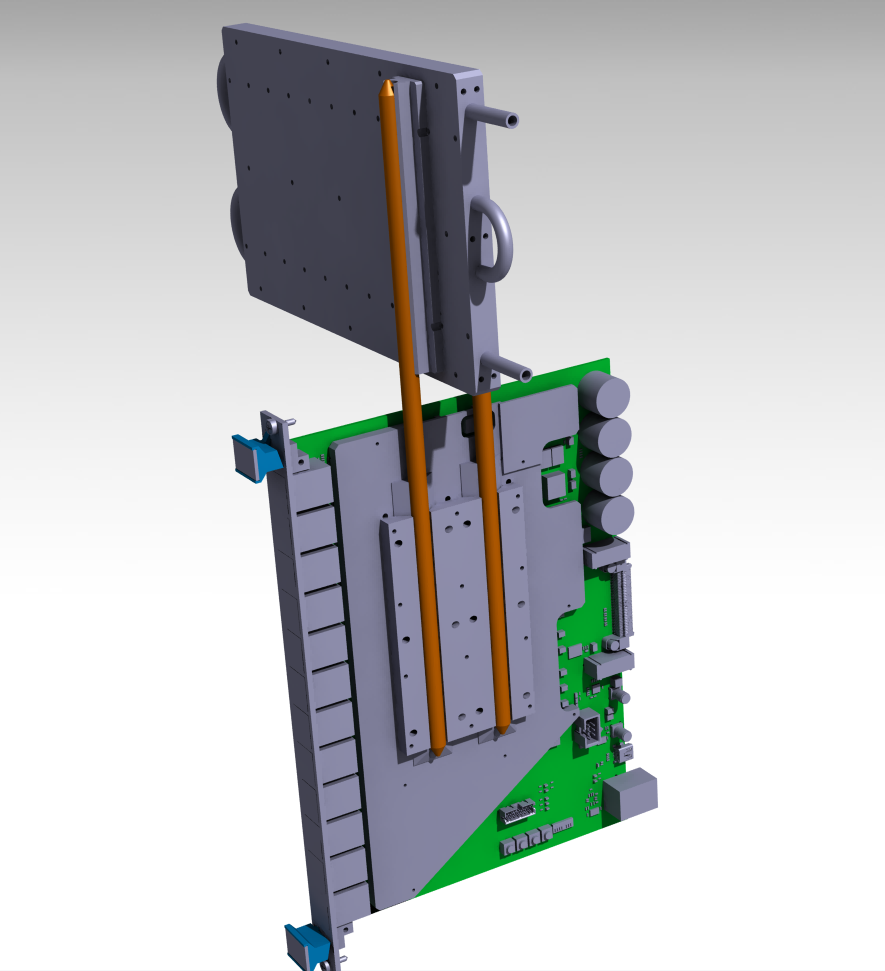}
        \caption{Drawings of the cooling system for one of the DigiCam micro-crates. In the top figure, the heat exchanger sitting above the micro-crate, is visible, while the bottom figure shows in detail the connection between the heat pipes and one of the digitizer boards.}
	\label{DigiCam cooling}
\end{figure}

\subsection{DigiCam cooling}
Due to the compact design of the micro-crates, the DigiCam boards cannot be cooled with  standard water pipes, so they are cooled with heat pipes. 
The mechanics of the DigiCam cooling system is shown in Fig.~\ref{DigiCam cooling}. Metal blocks act as heat exchangers by connecting the heat pipes to the water cooling pipes. Two heat pipes, each capable of absorbing 25~W, are connected to each digitizer and trigger board, in contact with the \acp{FADC} of the formers and the \acp{FPGA} of both.

The efficiency of the heat pipes is influenced by gravity, since the return of the coolant liquid is usually produced via capillarity or gravity itself. In the mechanical design of the camera structure, the DigiCam micro-crates are installed with an inclination of 45$^\circ$ (see Fig.~\ref{microcrates}). This configuration ensures that the heat pipes will work properly irrespective of the inclination of the telescope.

\section{The camera housekeeping system}
\label{sec:safety}
The camera has been designed to be long term stable and reliable during its lifetime on site. Day-night temperature gradients as well as any possible weather condition must be carefully accounted for to avoid permanent damages. For this reason, the camera is provided with a housekeeping system that continuously monitors its conditions, in particular during non-operation in daytime, and reacts accordingly when potentially dangerous conditions are recognized.

While the IP65 compliant design will provide major protection against water and dust, the chance of condensation inside the sealed structure is still high, especially outside of operation time, when the camera is turned off and information on temperature from the SiPM \ac{NTC} probes and from DigiCam  is not available. To avoid damages due to water condensation or moist, other temperature, pressure and humidity sensors are installed inside the camera and are continuously (also in daytime) readout by a dedicated housekeeping board. If a condensation danger is detected, the housekeeping board sends a signal to the safety PLC, which activates a heating unit installed inside the camera. To avoid over-pressure conditions, the camera chassis implements an IP65 Gore-Tex\textsuperscript{\textregistered} membrane, that allows for the air exchange with the environment but prevents water to flow inside. Another solution using a compact desiccant air dryer is under study. 


\section{Camera test setups}
\label{sec:tests}
An aspect that has been taken care of during the design of the camera, is the development of dedicated test setups and test routines for the validation of each component of the camera prior to its final installation, both for individual elements (cones, sensors, electronics boards, etc., as presented in the previous sections), for assembled parts (e.g.\ modules, as it is shown in the following section), and for the assessment of the homogeneity and reproducibility of the production. When needed, the same tests are used to characterize the object (e.g.\ the measurement of sensor properties during the module optical test, see Sec.~\ref{sec:opticalTest}) or even to calibrate it (such as in the test of the slow control board~\cite{electronics_paper}). 
\subsection{Optical test of full modules}

\begin{figure}
	\centering
	\includegraphics[width=0.45\textwidth]{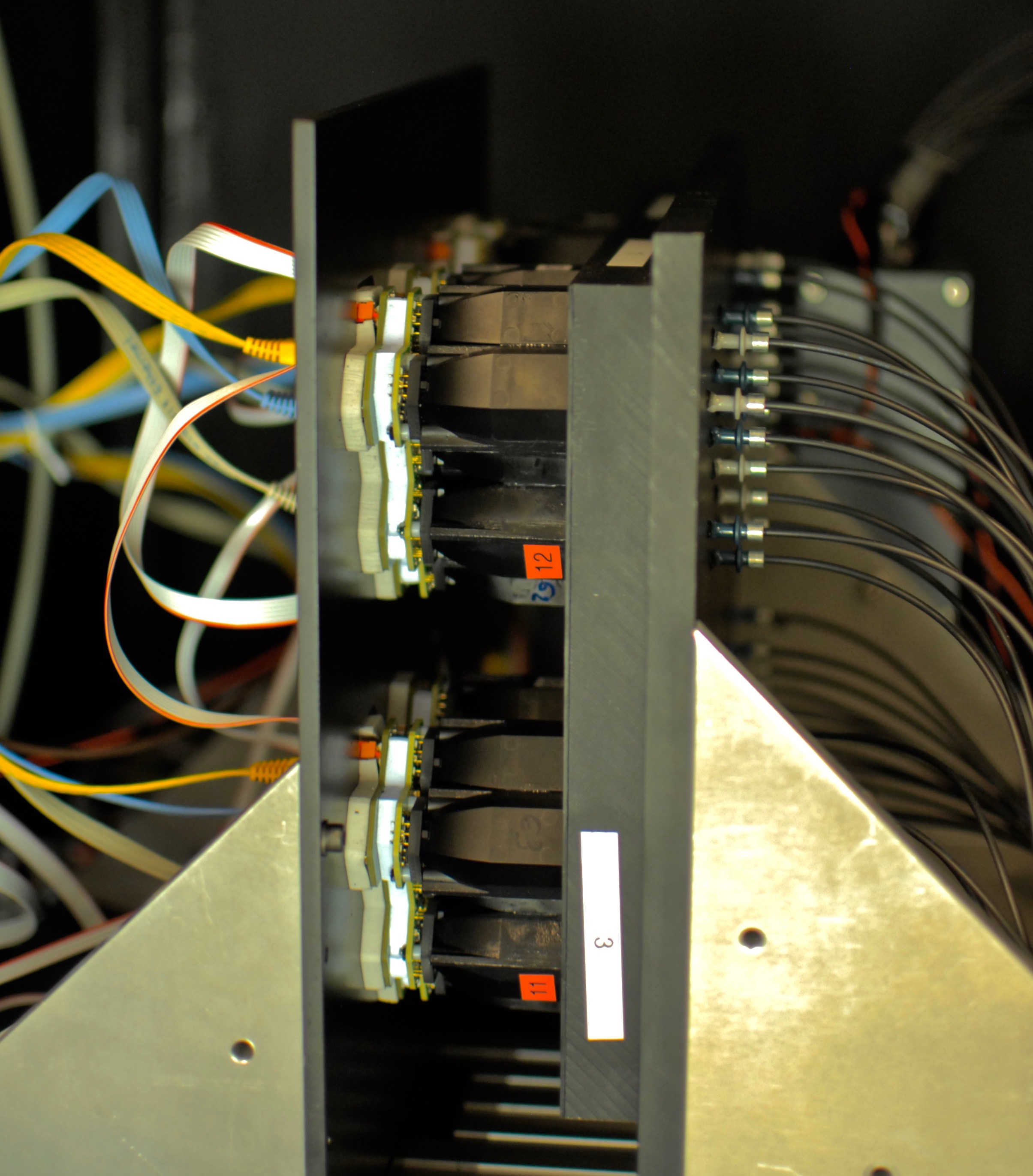}
	\includegraphics[width=0.45\textwidth]{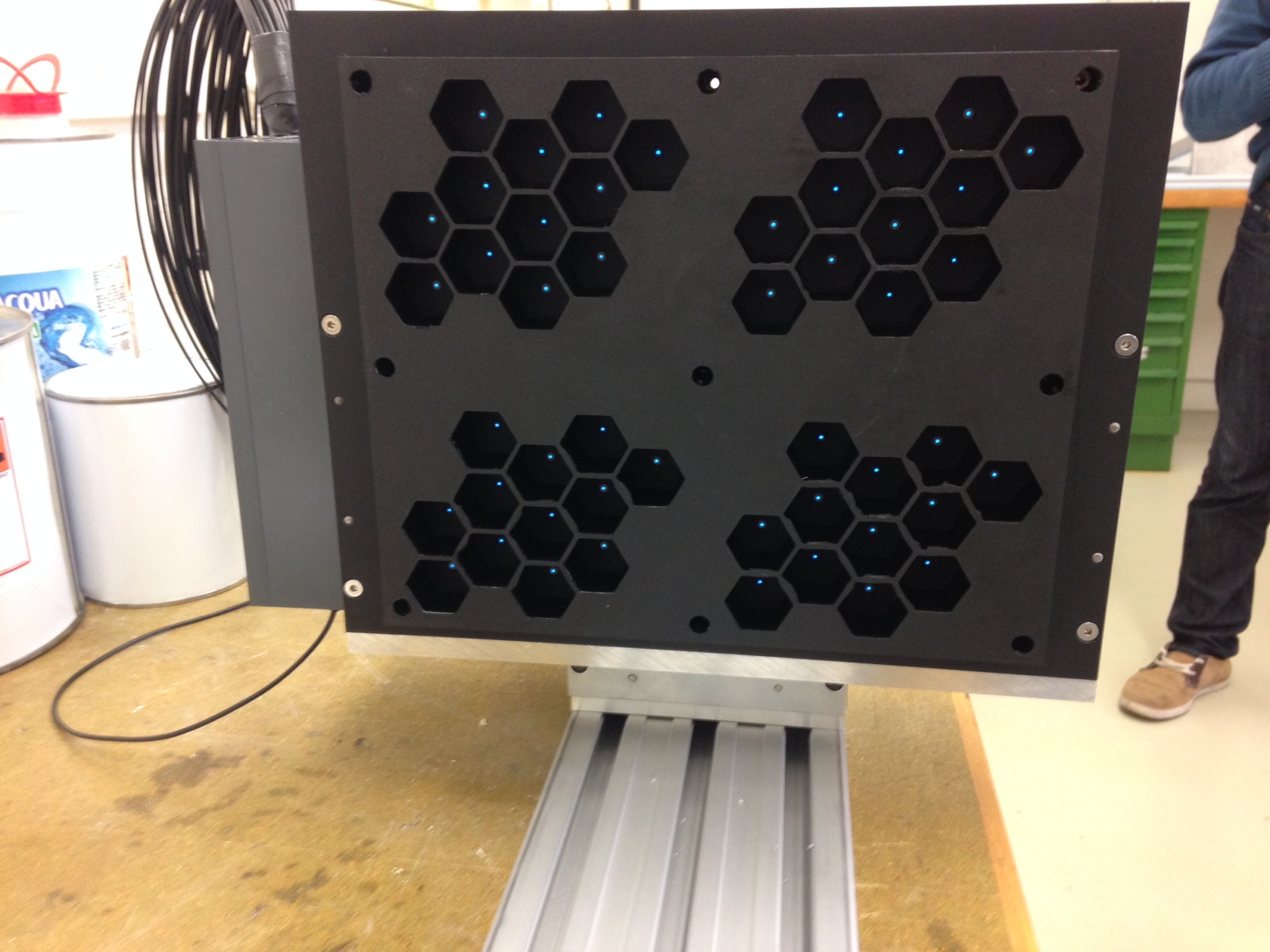}
	\caption{Top: a photo of the optical test setup with modules installed. Bottom: the front panel of the setup, where the 48 illuminated optical fibers are visible.}
	\label{optical module}
\end{figure}

Following its assembly and prior to its final installation on the \ac{PDP}, each 12-pixel module undergoes an optical test using the setup shown in Fig.~\ref{optical module}. A pulsed 470~nm LED source is connected to 48 optical fibers whose outputs are aligned with the center of the 48 pixels of four modules fixed on a support structure. The setup is enclosed in a light tight box. A replica of the PDP cooling (see Sec.~\ref{sec:cooling}) system is used to cool the modules via the metal plate of the support structure. Using an external chiller, the system stabilizes the temperature of the modules while the control loop is running during testing.

The setup is used to qualify the overall functioning of the modules, but also to characterize each pixel in terms of basic performance parameters. For this purpose, four types of data are taken. Dark runs are used to extract the dark count rate and the cross-talk; low light level runs are used to reconstruct the Multiple PhotoElectron (MPE) spectrum (see Sec.~\ref{sec:SPE}), from which parameters such as the gain can be extracted; high light level runs yielding signals below saturation are used to monitor the signal amplitude, rise time and fall time, and to study the baseline position and noise; very high light level runs produce pulses above saturation, useful to check the saturation behavior of the channel. These data also allow us to monitor the proper functioning of the entire readout chain (the modules are readout by DigiCam FADC boards in their final version or using a demonstrator board), and can also provide preliminary calibration data.

\begin{figure*}
	\centering
	\includegraphics[width=0.45\textwidth]{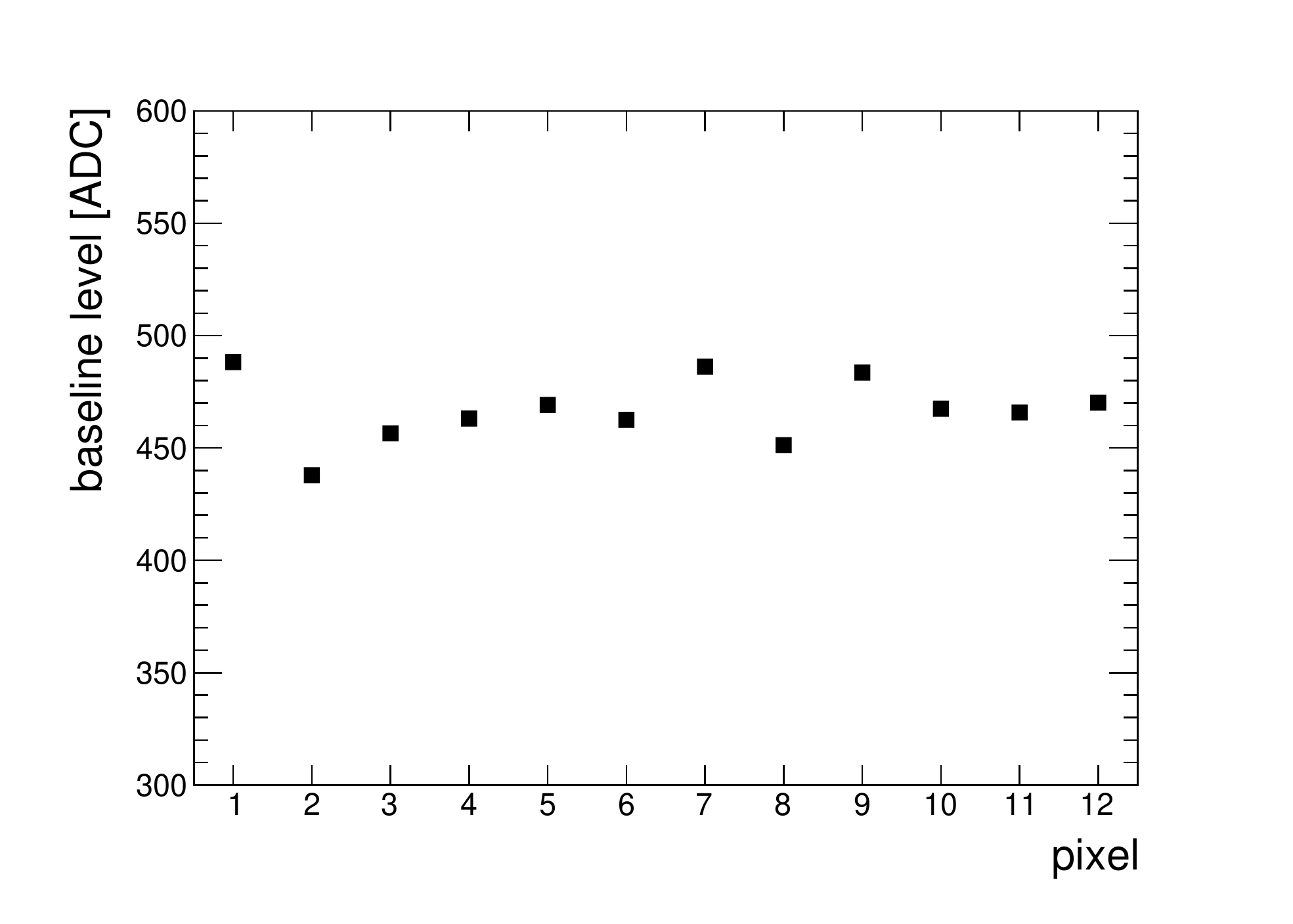}
	\includegraphics[width=0.45\textwidth]{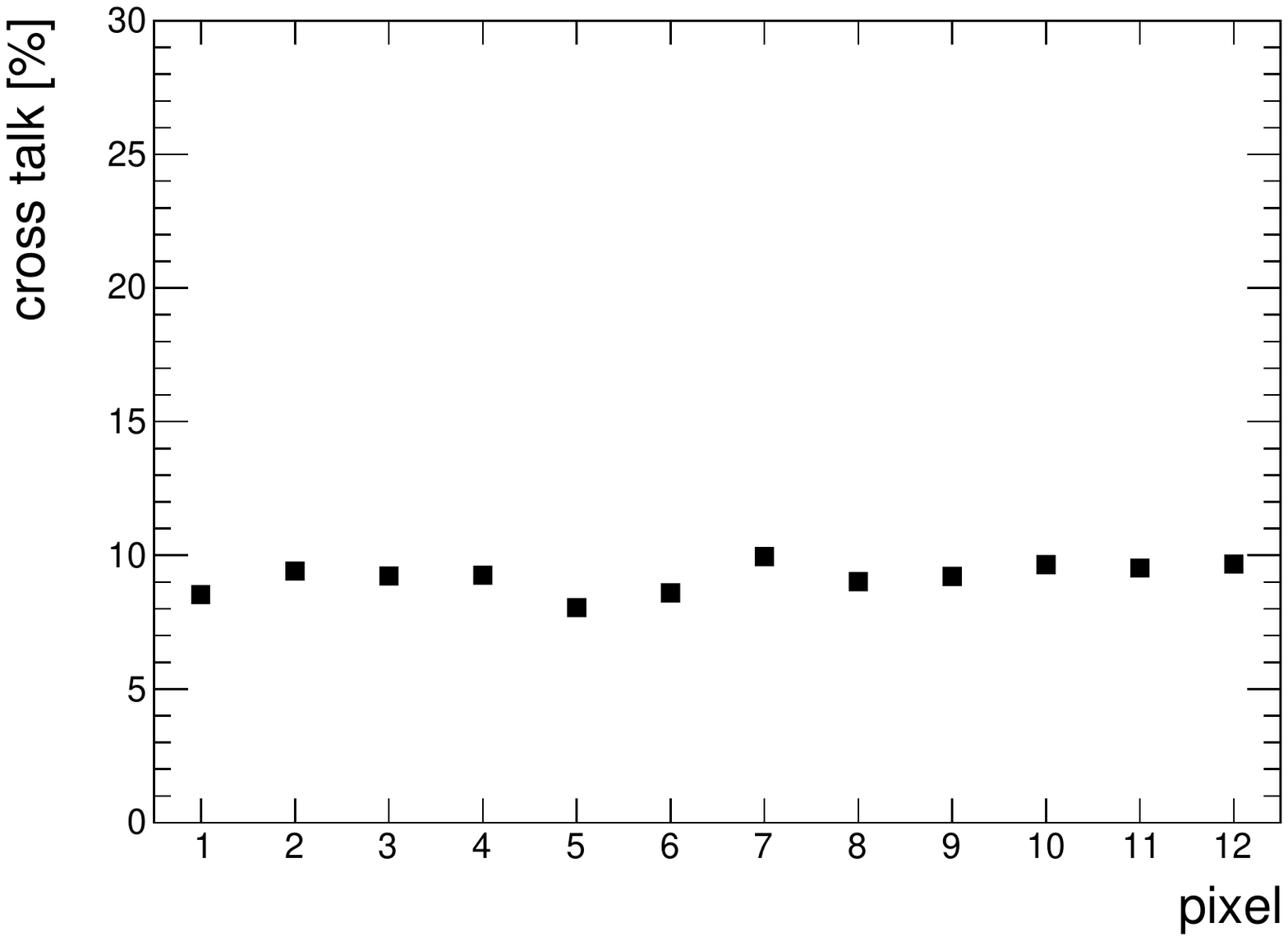}
	\caption{Some typical results from the optical test of a module: baseline level and cross talk are measured for each of the 12 pixels. The spread observed on the baseline level is related to the non-equalization of the DigiCam ADC offsets.}
	\label{module optical test}
\end{figure*}

\begin{figure*}
	\centering
	\includegraphics[width=0.45\textwidth]{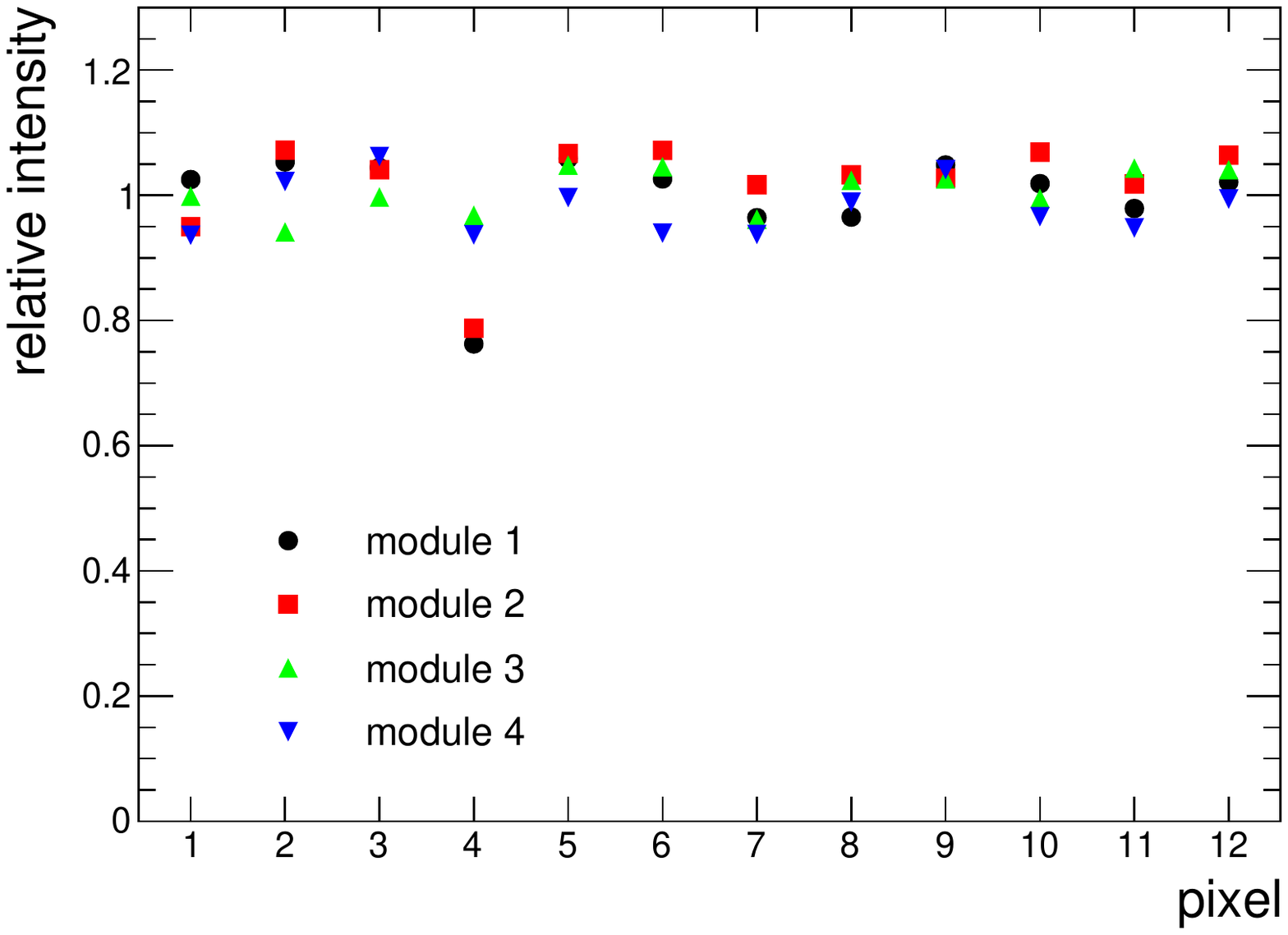}
	\includegraphics[width=0.45\textwidth]{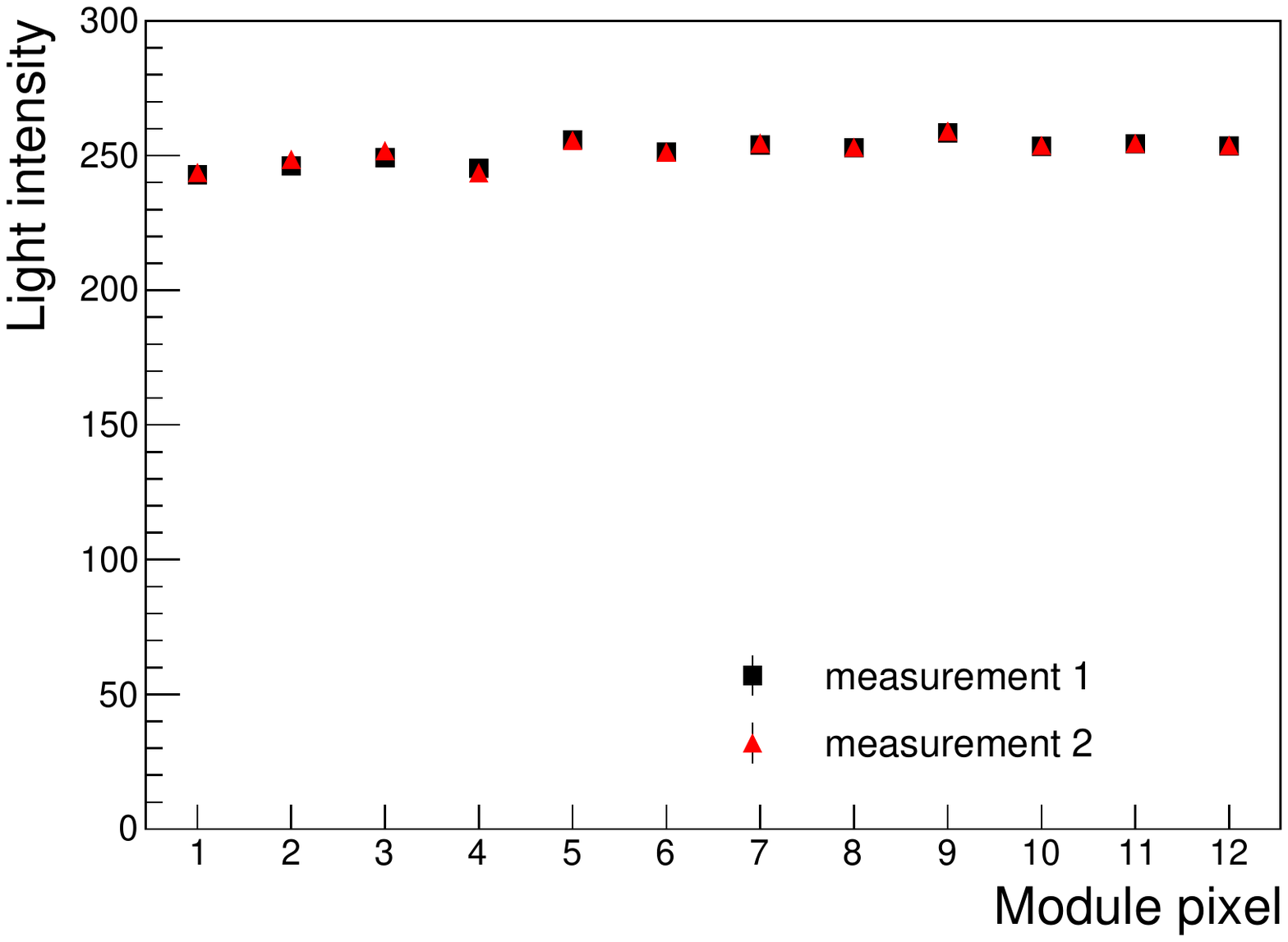}
	\caption{Left: measurement of the light yield of the 48 optical fibers of the module optical test setup. Right: two measurements of the signal yield from the 12 channels of the DigiCam demo-board, mapped onto the module pixels.}
	\label{module optical test calibration}
\end{figure*}

A few examples of the typical results from the module optical test are shown in Fig.~\ref{module optical test}. 
The data are taken using a LabVIEW interface to control the hardware units (including the power supplies, the LED pulse generator, the CAN bus communication with the \acl{SCB} and the ethernet connection to DigiCam for the readout). A C++ program analyses the data and produces automatically a report that the user can scroll to quickly check the proper functioning of the module or, conversely, to spot possible problems. In the analysis, the data are corrected for the relative light yield of the optical fibers, that has been calibrated using a single \ac{SiPM} coupled to a light guide to measure the light intensity of individual fibers (Fig.~\ref{module optical test calibration}, left). A correction,
derived from the calibration of the individual \acp{FADC} of the DigiCam, is also applied. The correction is measured by injecting the same analog pulse to each \ac{FADC} channel and by comparing the amplitudes of the corresponding digitized signals (Fig.~\ref{module optical test calibration}, right).
Both the LabVIEW interface for data taking and the analysis software are designed to be run with minimal intervention of the user.

\label{sec:opticalTest}

\subsection{Cabling test setup}\label{camera test setup section}
\begin{figure}
	\centering
	\includegraphics[width=0.45\textwidth]{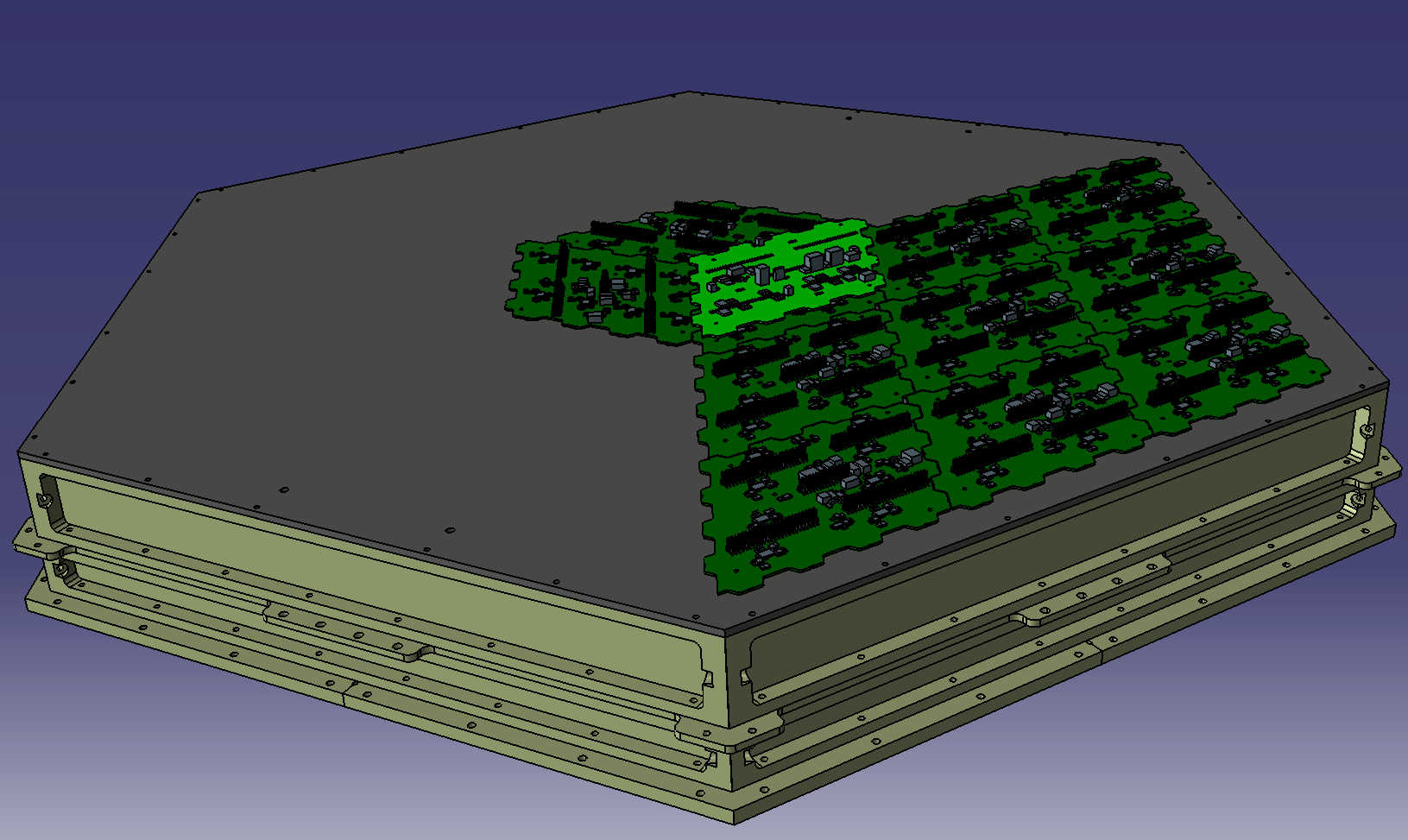}
	\caption{Schematics of the cabling test setup.}
	\label{camera test setup}
\end{figure}

A test setup has been developed to check the proper cabling of the PDP to DigiCam prior to the camera installation on the final telescope structure.
The setup is shown in Fig.~\ref{camera test setup}. A mechanical structure covers one third plus the central region of the \acl{PDP} and hosts a matrix of 420~nm LED sources located on its surface that illuminate each pixel individually (the setup will be rotated in steps of 120$^\circ$ to cover the full PDP). By illuminating each pixel at a time, it is possible to check the proper routing of the signal in order to spot possible errors in the connection between the PDP and DigiCam. Although this system was originally conceived to solely test the cabling, it will also be used for calibration and flat fielding (see Sec.~\ref{sec:calibration}). For this reason, the LED carrier boards have been designed with two LEDs pointing to each pixel, one pulsed and one in continuous light mode. The former simulates pulses of Cherenkov light, the latter emulates the \ac{NSB}. By switching on and off each LED individually, and by adjusting their light level in groups of three, it will be possible to reproduce most of the foreseen calibration conditions. Moreover, light patterns can be programmed in order to test the trigger logic.


\section{Performance validation}
\label{sec:validation}

Preliminary measurements prior to the final camera assembly have been carried out to validate the performance with respect to the goals and requirements set by CTA. The main performance parameters to be checked are the sensitivity to single photons and the charge resolution. The former is crucial for a \ac{SiPM} camera, because single photon spectra and multiple photon spectra are regularly used to extract calibration parameters, such as the gain of the sensors, the dark count rate and cross talk; the latter affects the energy and the angular resolution that are of primary importance for the CTA physics goals. 
Such measurements have been crucial also to compare the different FADCs provided by Analog Devices and Intersil mounted on the prototype DigiCam digitizer boards. Moreover, for a given FADC type, different gain settings could be evaluated and optimized.

In the analysis of the data that is carried out to extract the camera performance parameters, a few systematic effects have been taken into account, among which the effect of cross talk and dark counts in the reconstruction of the signals. To estimate such effects, a toy Monte Carlo to simulate the signals produced by the \acp{SiPM} has been developed, as described in the following section.

\subsection{The toy Monte Carlo}
\label{sec:toyMC}
In the toy Monte Carlo, single pulses produced by detected photons are generated using waveform templates taken from measurements, and taking into account Poisson statistics, cross talk, electronics noise, dark counts and NSB. The input values for cross talk and electronic noise levels and dark count rate were derived from measurements.
Hence charge spectra are built.
\begin{figure}
	\centering
	\includegraphics[width=0.45\textwidth]{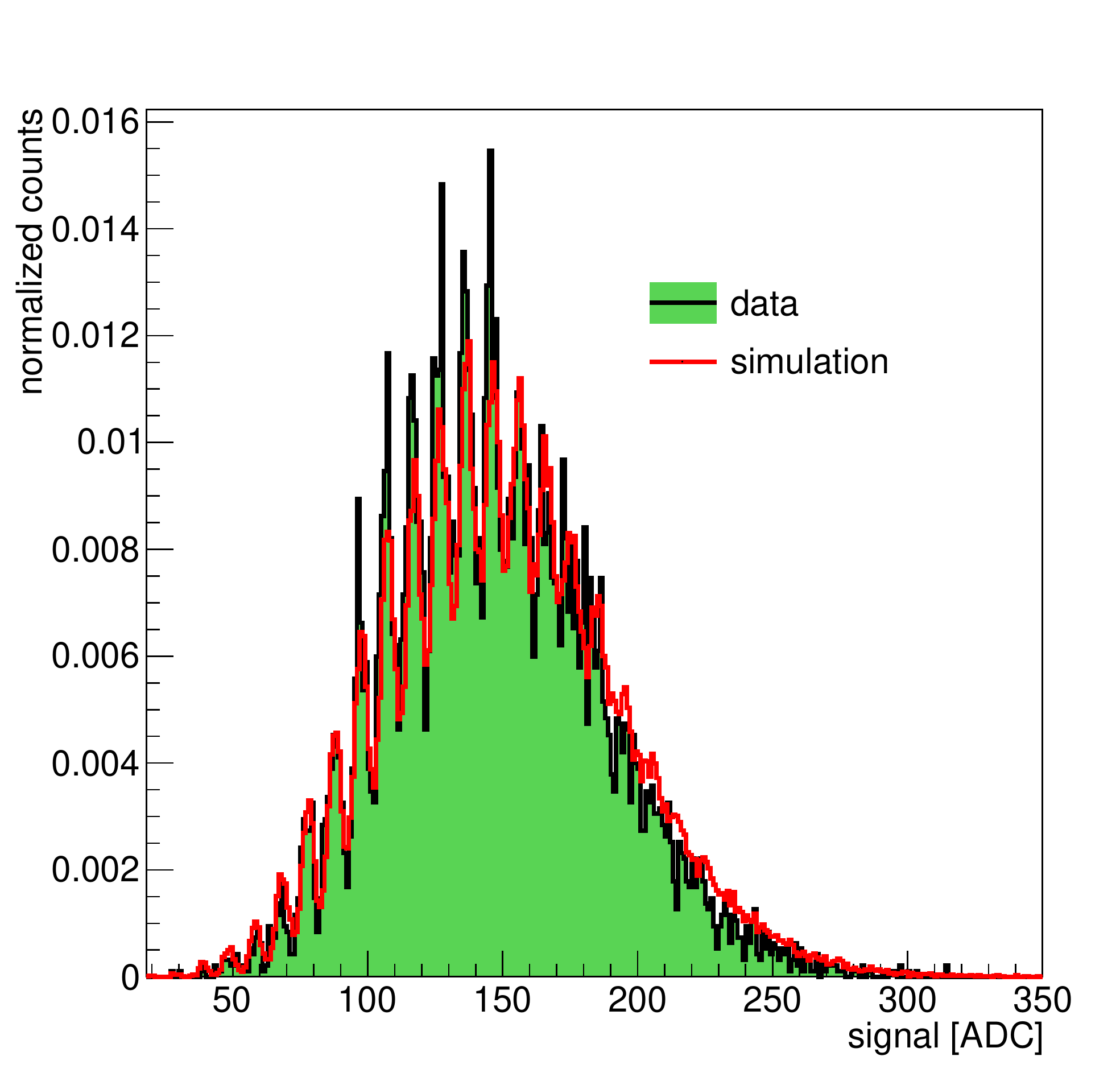}
	\caption{Comparison between a MPE spectrum measured directly from data and the corresponding one generated through the toy Monte Carlo using the measured parameters (gain, mean number of \acp{p.e.}, electronic noise, etc.).  }
	\label{toyMC vs data}
\end{figure}
As Fig.~\ref{toyMC vs data} shows, the toy Monte Carlo well reproduces the typical shape of the \ac{MPE}.
\newline
\begin{figure}
	\centering
	\includegraphics[width=0.45\textwidth]{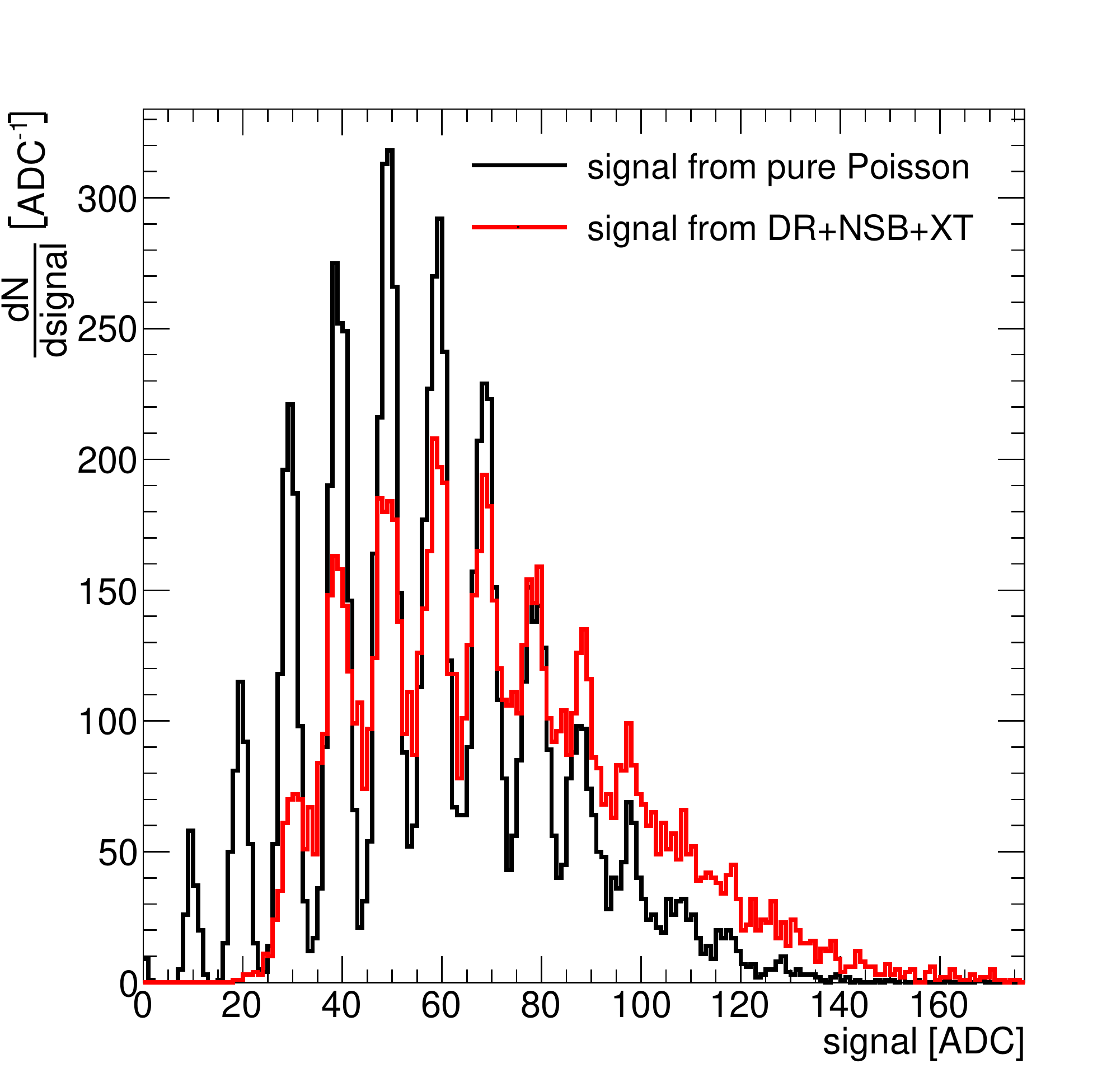}
	\includegraphics[width=0.45\textwidth]{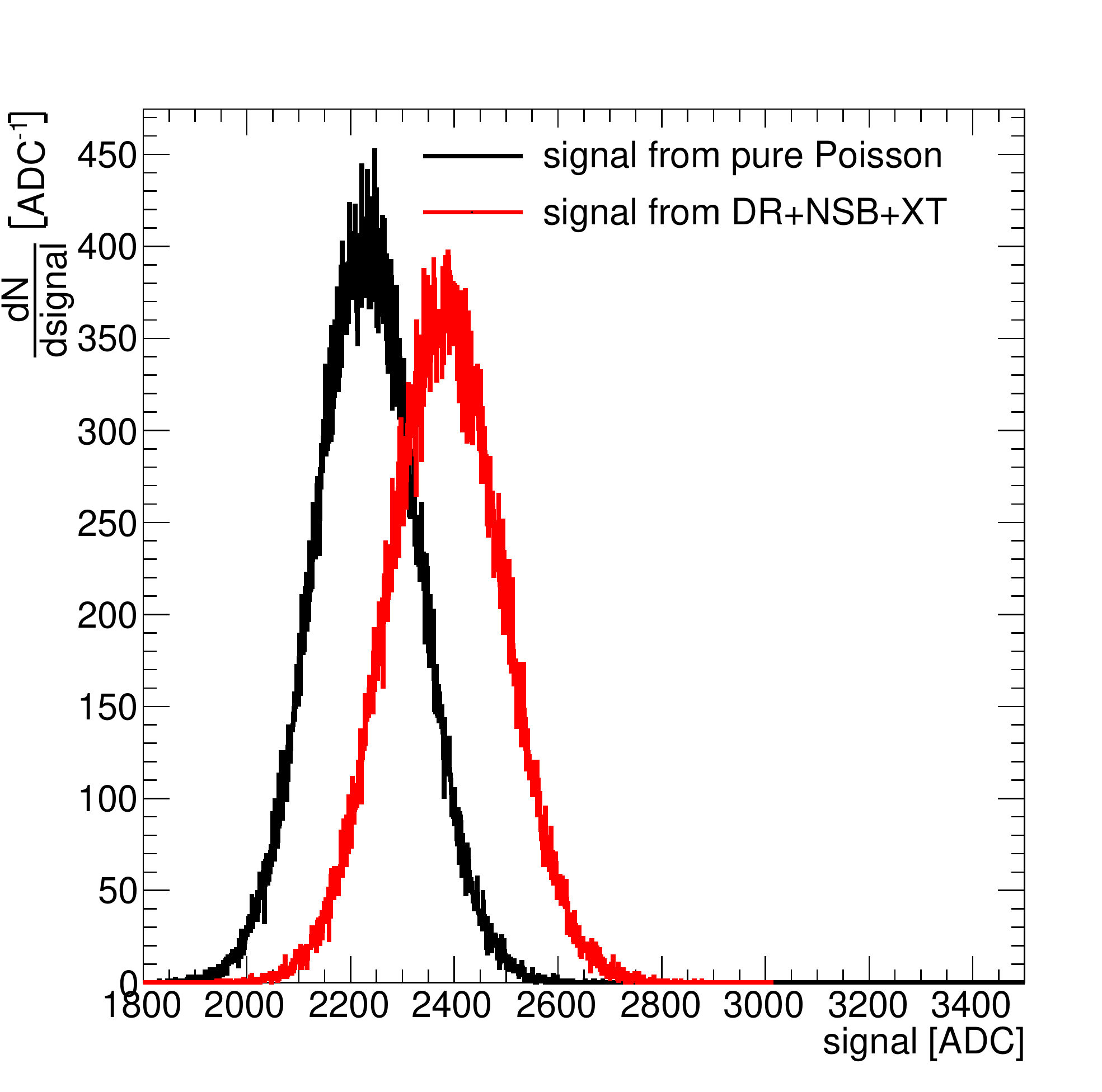}
	\caption{Comparison between distributions of pulse heights calculated with the toy Monte Carlo in the pure Poisson regime, in black, and after including dark count rate, NSB and cross talk (XT), in red, for the case of low light levels (top plot 6~p.e.) and high light levels (bottom plot 519~p.e.).}
	\label{toyMC distribution shift}
\end{figure}
Simulated datasets can be used to study how cross talk and dark count rate influence the shape of the charge distributions. An example is shown in Fig.~\ref{toyMC distribution shift}, where a pure Poisson distribution gets distorted and its mean value shifts towards a higher level. The main contribution to this effect arises from cross talk, i.e. a cross talk level of 10\% (as it was in the case of this simulation) results in a shift of the Poisson mean of at least the same amount. This systematic shift has to be taken into account when the actual signal has to be extracted from fits to the distributions of pulse amplitude or area.
\begin{figure}
	\centering
	\includegraphics[width=0.45\textwidth]{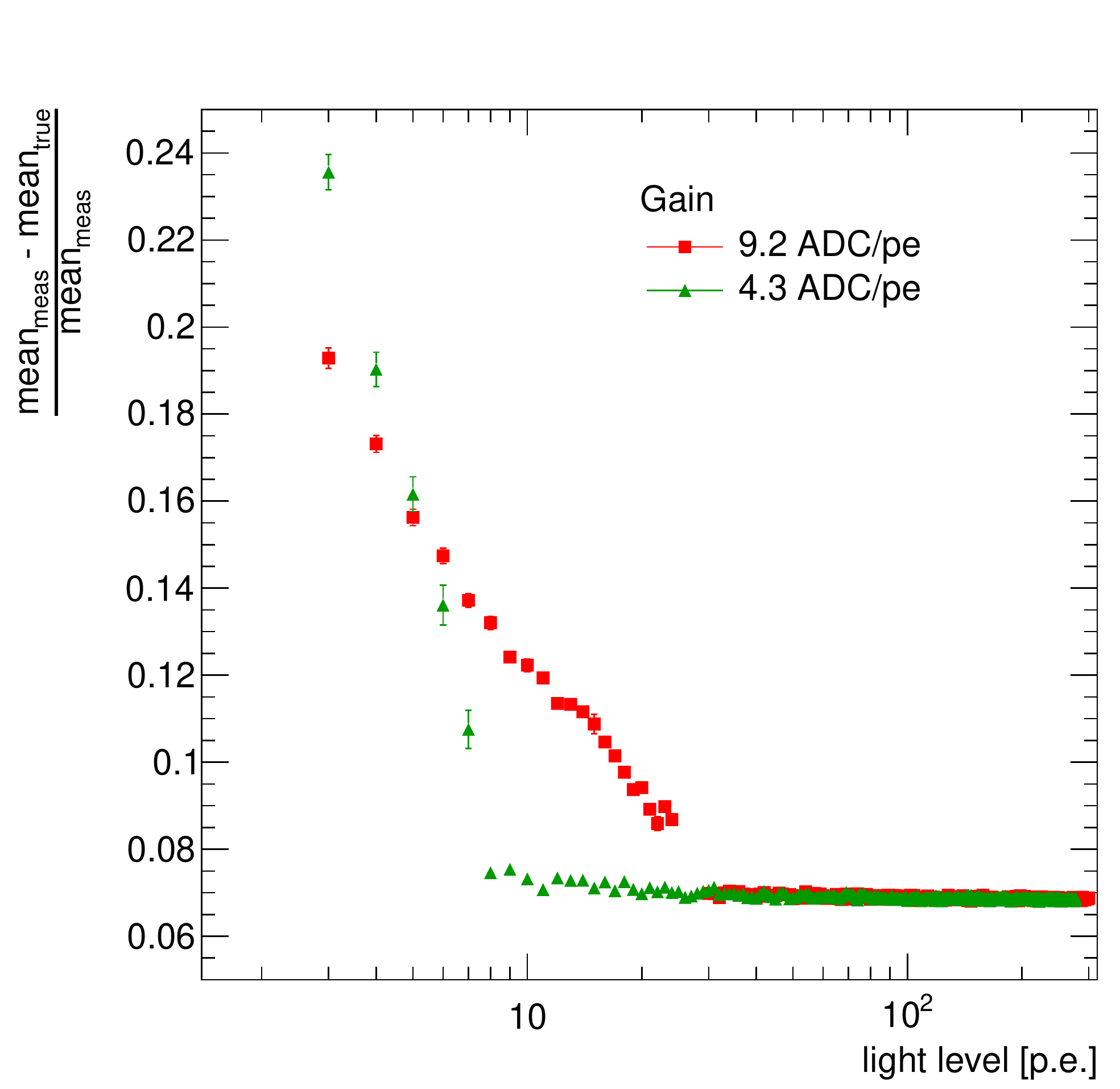}
	\caption{Relative deviation of the measured distribution mean from the true mean, for two different gains, as simulated with the toy Monte Carlo.}
	\label{toyMC mean shift}
\end{figure}
Fig.~\ref{toyMC mean shift}, for example, shows the effect on the determination of the mean of the amplitude distributions in a range up to 300~\acp{p.e.} with a cross talk of 6.4\% and a 2.79~MHz dark count rate\footnote{These values are the typical ones determined from actual measurements on sensors.}, for two different gain settings (9.2~ADC/\ac{p.e.} and 4.3~ADC/\ac{p.e.}). The two different regimes that are visible (a left-most inclined one and a right-most flat one) arise from the two types of distributions that are fitted: multiple photon spectra for lower light levels and Gaussians for higher light levels. While these latter are fitted via a Gaussian function, the former are fitted via the model presented in Sec.~\ref{sec:SPE}. Similar plots are produced to study as well the deviation of the measured gain from the true gain (see Fig. \ref{toyMC gain shift}).


\subsection{Sensitivity to single photons}
\label{sec:SPE}
\begin{figure}
	\centering
	\includegraphics[width=0.45\textwidth]{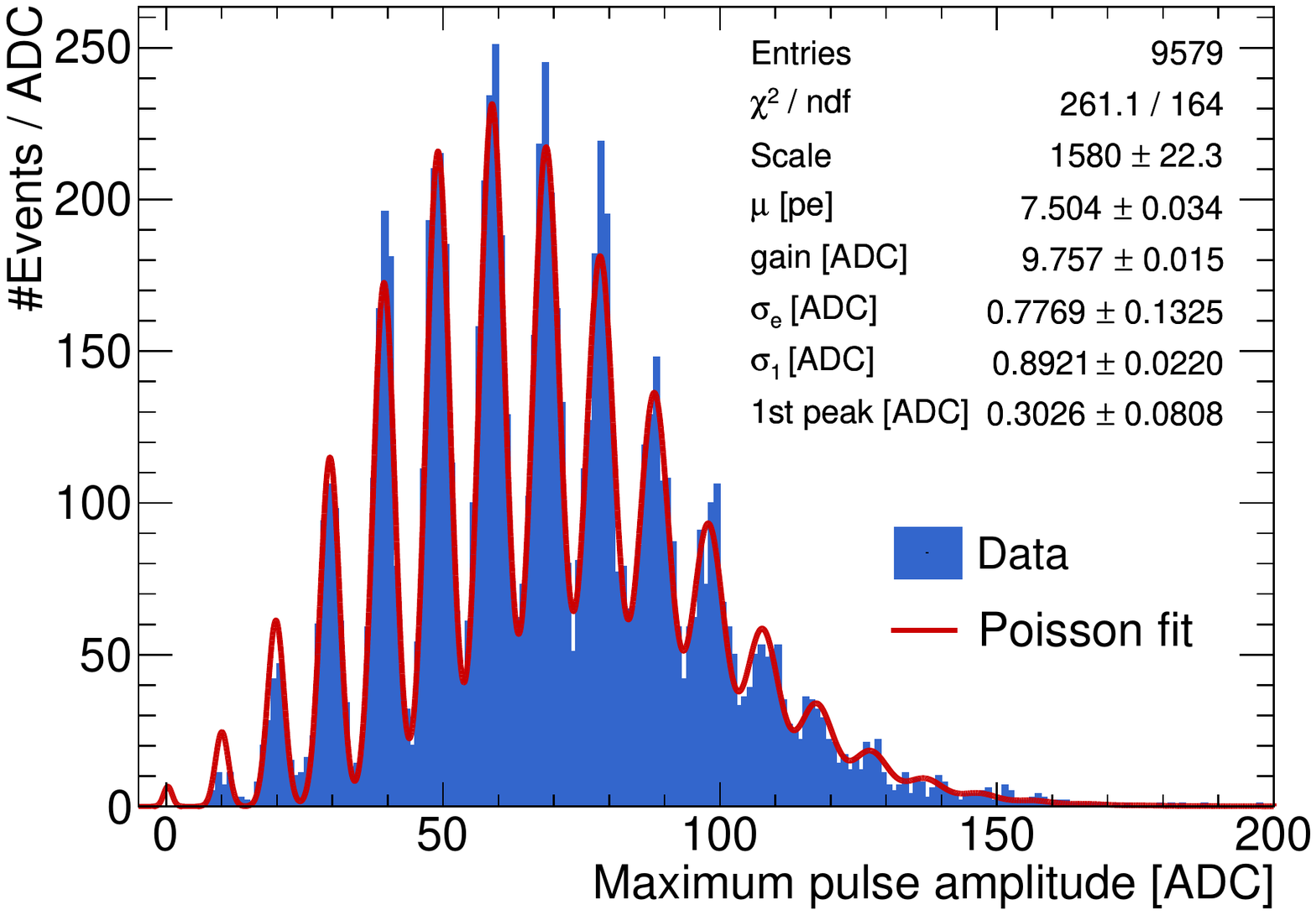}
	\caption{\ac{MPE} spectrum of a \ac{SiPM} obtained pulsing at 1 kHz a 400 nm LED with readout window of 80 ns. The device sees an average of $\sim$7.5 photons (the mean value of the Poisson function is $7.504\pm0.034$). The distance between the photo-peaks gives the gain of the detector, that is $9.757\pm0.015$ ADC counts/\ac{p.e.}.}
	\label{finger plot}
\end{figure}
The sensitivity of a SiPM to single photons can be assessed through the quality of the \ac{MPE} spectrum. An example of a \ac{MPE} spectrum acquired with a sensor mounted on a module and readout with DigiCam is shown in Fig.~\ref{finger plot}. Despite the large capacitance of the \ac{SiPM} (which affects the noise performance) and despite the common cathode configuration of the four channels of the sensor (which causes each of the four channels to be biased at the same average voltage, instead of applying a dedicated bias voltage per channel) the individual photo-peaks are well separated. The performance of such a large area sensor in the detection of single photons is thus comparable to that of conventional \acp{SiPM}.

\ac{MPE} spectra are important in the camera calibration strategy, since they are used to extract the gain of individual sensors together with the overall optical efficiency (sensor+light guides), to be used in the gain flat-fielding of the camera. \ac{MPE} spectra are also acquired during the optical test of each module for individual pixels (see Sec.~\ref{sec:opticalTest}). 

To extract the gain from the \ac{MPE} spectrum we use two methods: a direct fit of the spectrum or the analysis of its \ac{FFT}. In the former case, the \ac{MPE} spectrum can be described, to first approximation, by a function of the form
\begin{equation}\label{finger function}
	f(x) = A \sum_{n=1}^{N} P(n|\mu) \left[ \frac{1}{\sqrt{2\pi}\sigma_n} e^{-\left( \frac{x - n \cdot g}{\sqrt{2}\sigma_n} \right)^2} \right].
\end{equation}
In this formula, $A$ is a normalization constant, $P(n|\mu)$ is the integer Poisson distribution with mean $\mu$ modulating a set of Gaussian distributions for each photo-peak $n$, each centered in $n\cdot g$ where $g$ is the gain, i.e. the conversion factor between ADC counts and number of \acp{p.e.}. The width of the $n$-th photo-peak is given by
\begin{equation}
	\sigma_n = \sqrt{\sigma_e^2 + n \sigma_1^2},
\end{equation}
where $\sigma_e$ is the electronic noise and $\sigma_1$ is the intrinsic noise associated with the detection of a single photon. The fit to the data shown in Fig.~\ref{finger plot}, is done according to this model.

\begin{figure}
	\centering
	\includegraphics[width=0.45\textwidth]{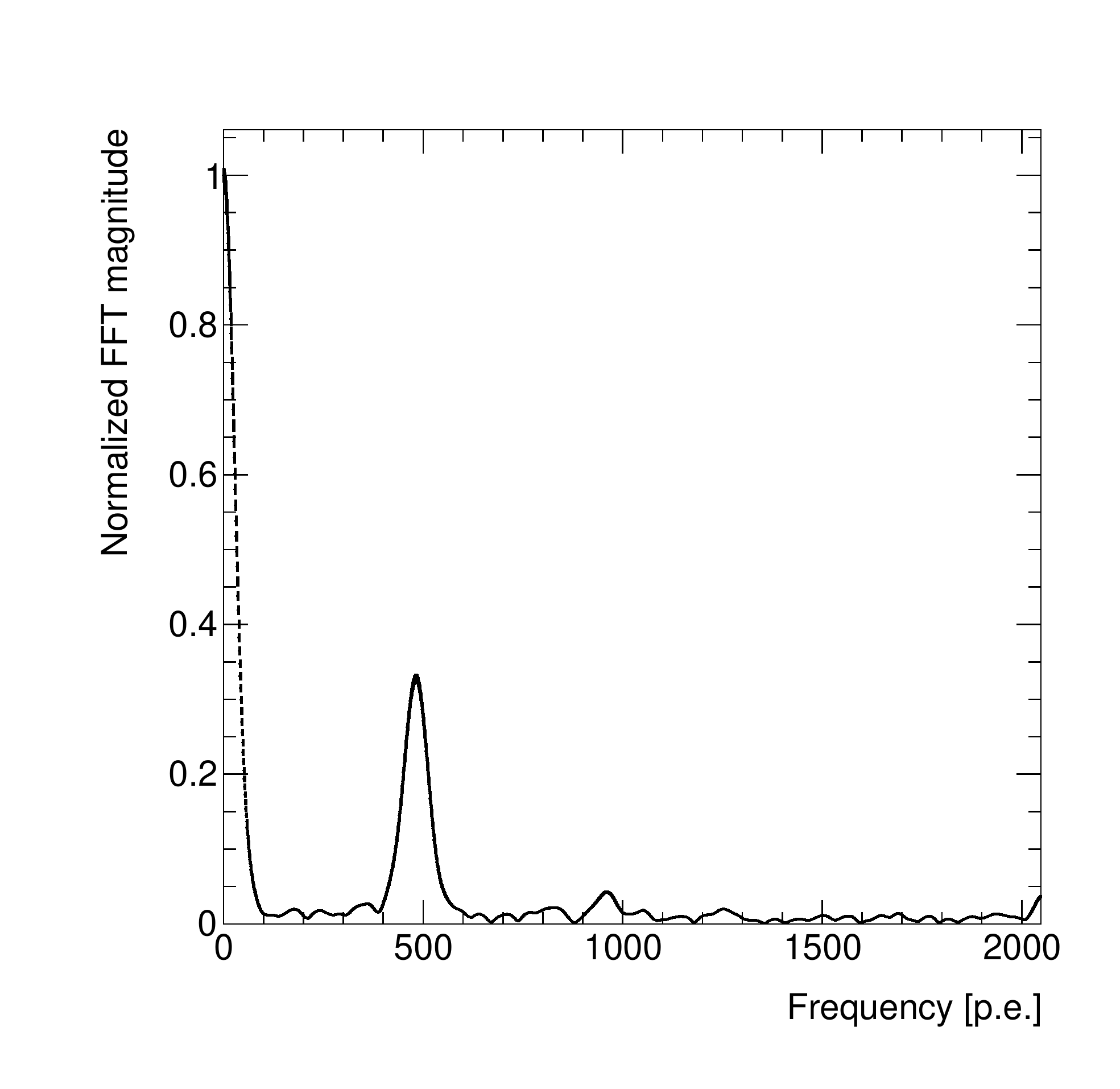}
	\caption{\ac{FFT} of a \ac{MPE} spectrum. The range on the horizontal axis is half the total range (the \ac{FFT} is symmetrical around the center of the range).}
	\label{FFT}
\end{figure}
In the \ac{FFT} method the Fast Fourier transform of the \ac{MPE} spectrum is calculated, as shown in Fig.~\ref{FFT} for a \ac{MPE} spectrum acquired from one sensor readout by DigiCam. The main peak at around 500~\ac{p.e.} corresponds to the main frequency of the single photon peaks, and the gain can be extracted as
\begin{equation}
	g = \frac{\text{ADC range}}{\text{peak position}}.
\end{equation}

A study of the accuracy of either methods (fit and \ac{FFT}) has been carried out in the framework of the toy Monte Carlo. As was shown earlier (see Sec.~\ref{sec:toyMC}), the pure Poisson signal distributions are distorted by cross talk and dark count rate. In the fit method, one could improve the model by adding these effects in some parametrized form, as was already shown in Ref.~\cite{1748-0221-9-10-P10012}. However, this adds parameters and, in general, complexity to the fit. A study on simulated data has been carried out to characterize the quality of the pure Poisson fit when used to estimate the gain (for the estimation of the light level from the same fit, the result was already shown in Fig.~\ref{toyMC mean shift}). 
\begin{figure}
	\centering
	\includegraphics[width=0.45\textwidth]{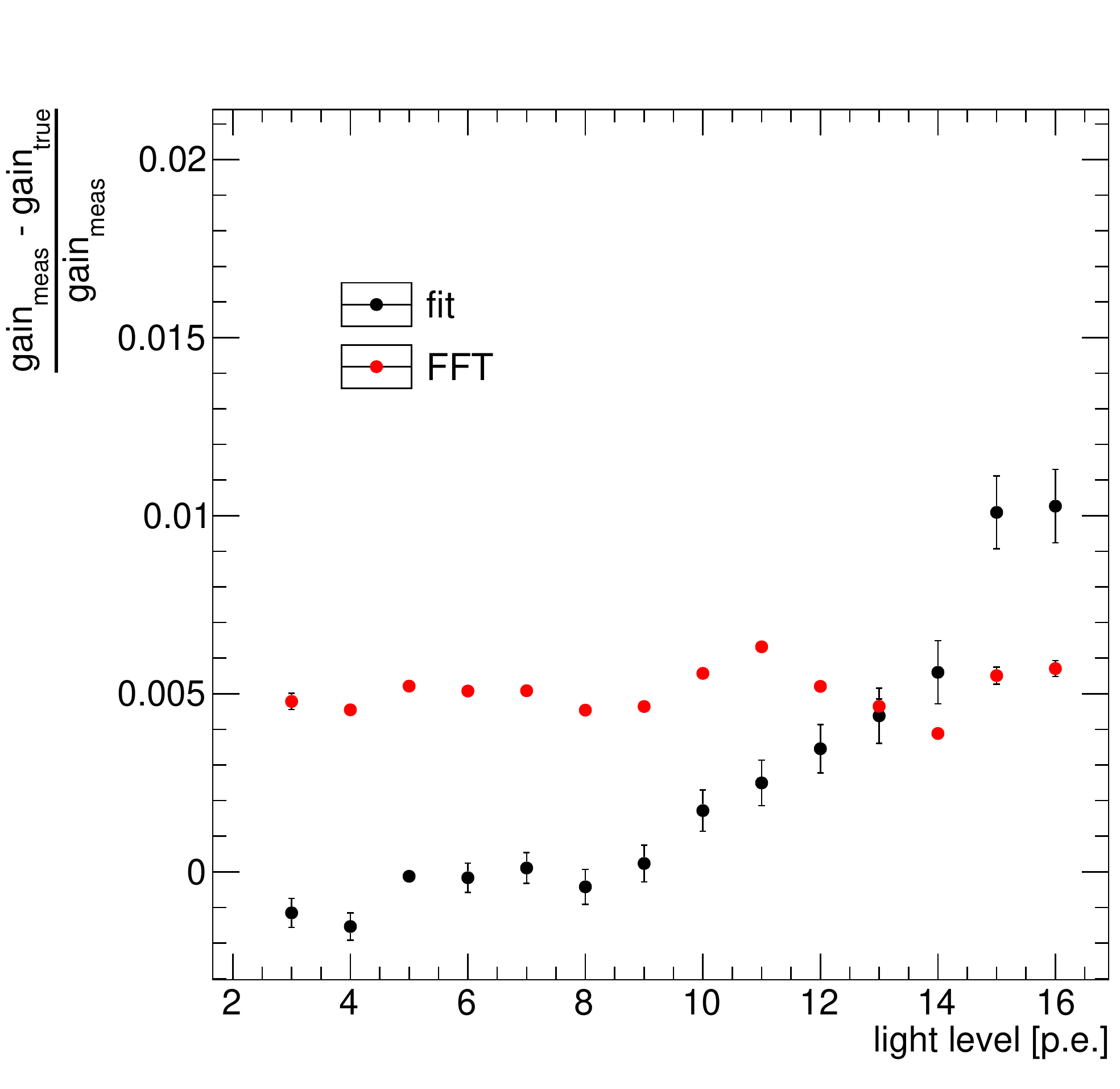}
	\caption{Relative deviation of the measured gain from the true gain of 9.2 ADC/p.e. in the fit method and the FFT method.}
	\label{toyMC gain shift}
\end{figure}
This result is shown in Fig.~\ref{toyMC gain shift} (data simulated with gain of 10~ADC/\ac{p.e.}, cross talk 10\% and dark count rate 5~MHz), and is compared to the performance of the FFT method applied on the same set. The fit method yields a more accurate estimation for low light levels (up to around 13 \acp{p.e.}), and looses accuracy with increasing light. The FFT method is generally less accurate, but the effect does not depend on the light level. Overall, either methods give an uncertainty which is systematically below 1\%. This can be used to set a systematic uncertainty on the extracted gain values. Otherwise, the data points in this plot can be employed as correction coefficients to retrieve a more accurate value of the gain in either methods, as will be done in Sec.~\ref{section: charge resolution}.


\subsection{Charge resolution}
\label{section: charge resolution}
The charge measured by a single pixel in the camera is proportional to the amount of Cherenkov light that has reached the sensor. The charge resolution is determined by the statistical fluctuations of the charge on top of which sensor intrinsic resolution and sensor and electronic noise can contribute significantly. \ac{CTA} provides specific requirements and goals for the fractional charge resolution $\sigma_Q/Q$ in the range between 0 and 2000 \acp{p.e.} (see Fig.~\ref{charge resolution}).

The charge resolution of the camera has been measured on a few sensors using a dedicated LED driver board. In line with the cabling test setup concept described in Sec.~\ref{camera test setup section}, this board hosts two LED sources, one in AC mode to simulate the Cherenkov light pulses of particles, and one in DC mode\footnote{The same LEDs (470~nm) were used for AC and DC mode, as for the DC, the goal was to emulate a defined photoelectron rate and not reproduce the wavelength spectrum.} to simulate the \acl{NSB} after having been calibrated. The data are taken using a fully assembled module readout by a standalone DigiCam digitizer board. The module is mounted on the temperature-controlled support structure of the optical test setup (Sec.~\ref{sec:opticalTest}).

The charge resolution is extracted from the data by analyzing the distributions of pulse amplitude or area (both after baseline subtraction) at different light levels of the AC and DC LEDs. At each level of the DC LED, i.e. at each emulated NSB level (no NSB, 40~MHz - corresponding to dark nights -, 80~MHz and 660~MHz - corresponding to half moon conditions with the moon at 6$^\circ$ off-axis -), the datasets consist of a collection of signals from detected light pulses at increasingly higher levels, from few photons up to few thousand photons. 

\subsubsection{Source calibration}
\begin{figure}
	\centering
	\includegraphics[width=0.45\textwidth]{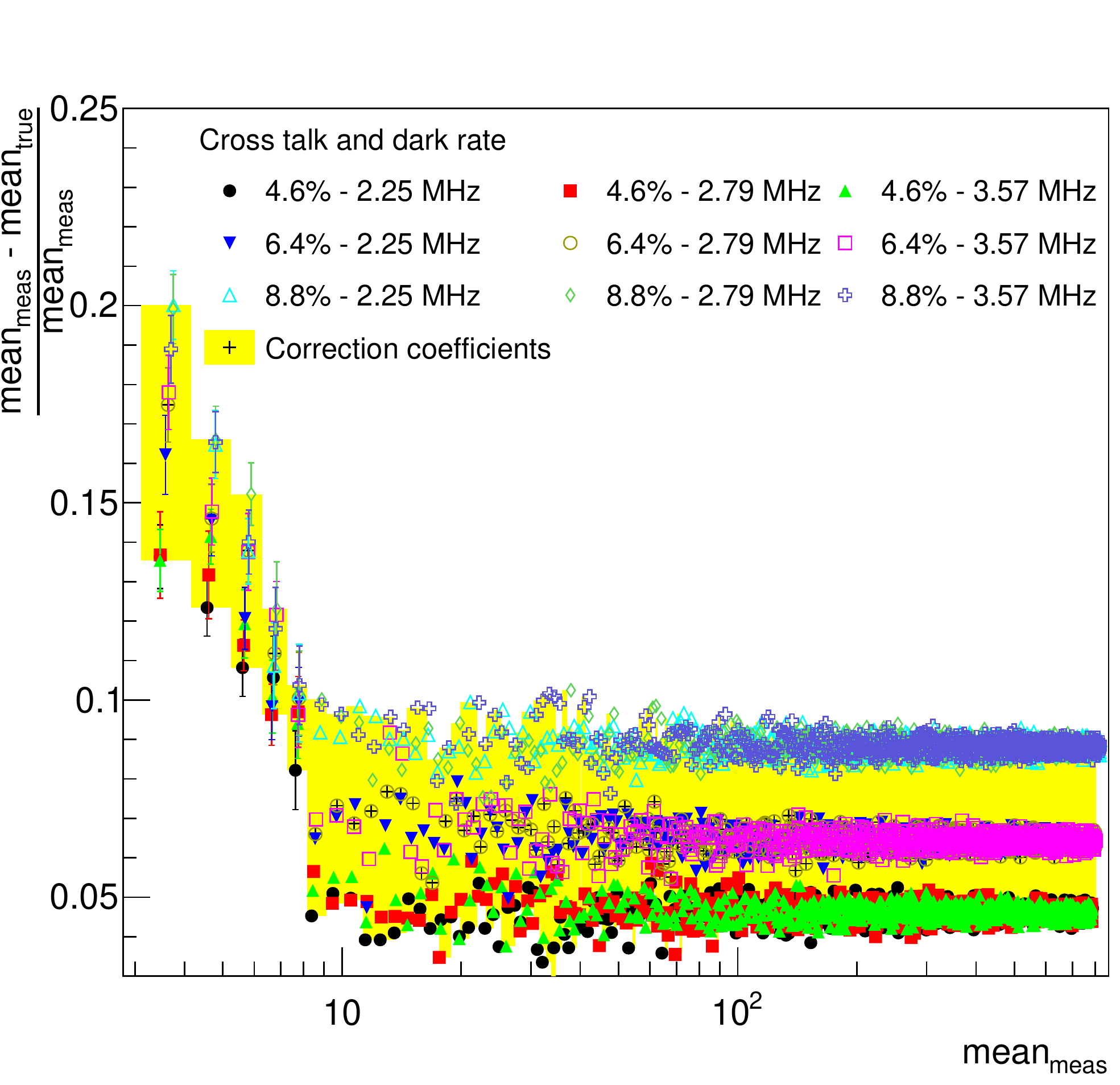}
	\caption{Simulation of the light level correction coefficients for different levels of cross talk and dark count rate according to the measured uncertainty of these quantities. The two regimes (an inclined one on the leftmost part and a flat one on the rightmost part) correspond to the two different fit regimes: \ac{MPE} spectra and Gaussians.}
	\label{correction coefficients}
\end{figure}

While the DC LED was calibrated with a pin diode, the calibration of the AC LED source is derived from the data themselves. For the low intensity data sets, the \ac{MPE} spectra were used to extract the gain with the methods described in Sec.~\ref{sec:SPE}.
\begin{figure}
	\centering
	\includegraphics[trim = 0mm 5mm 0mm 10mm, clip, width=0.45\textwidth]{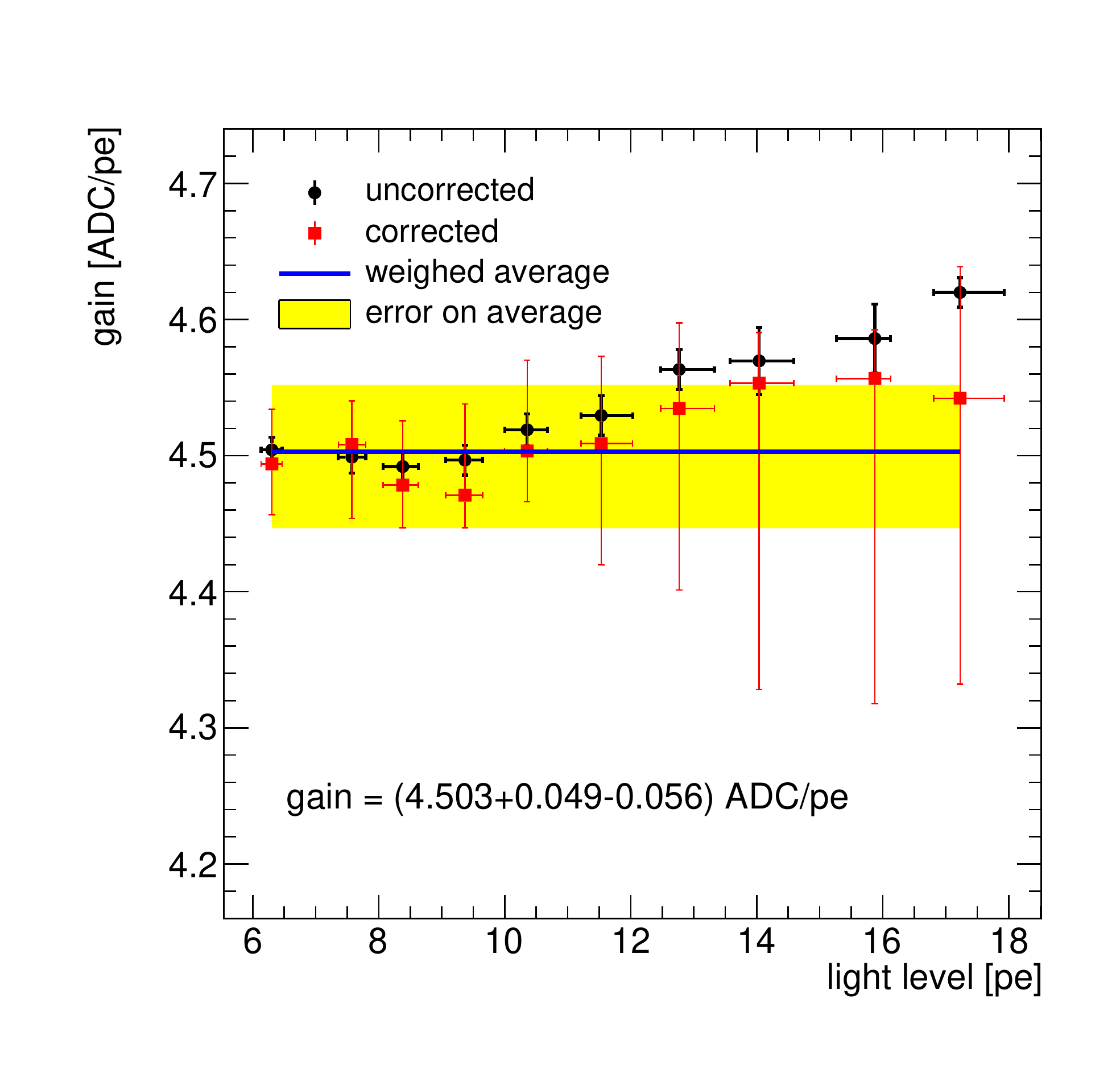}
	\includegraphics[trim = 0mm 5mm 0mm 10mm, clip, width=0.45\textwidth]{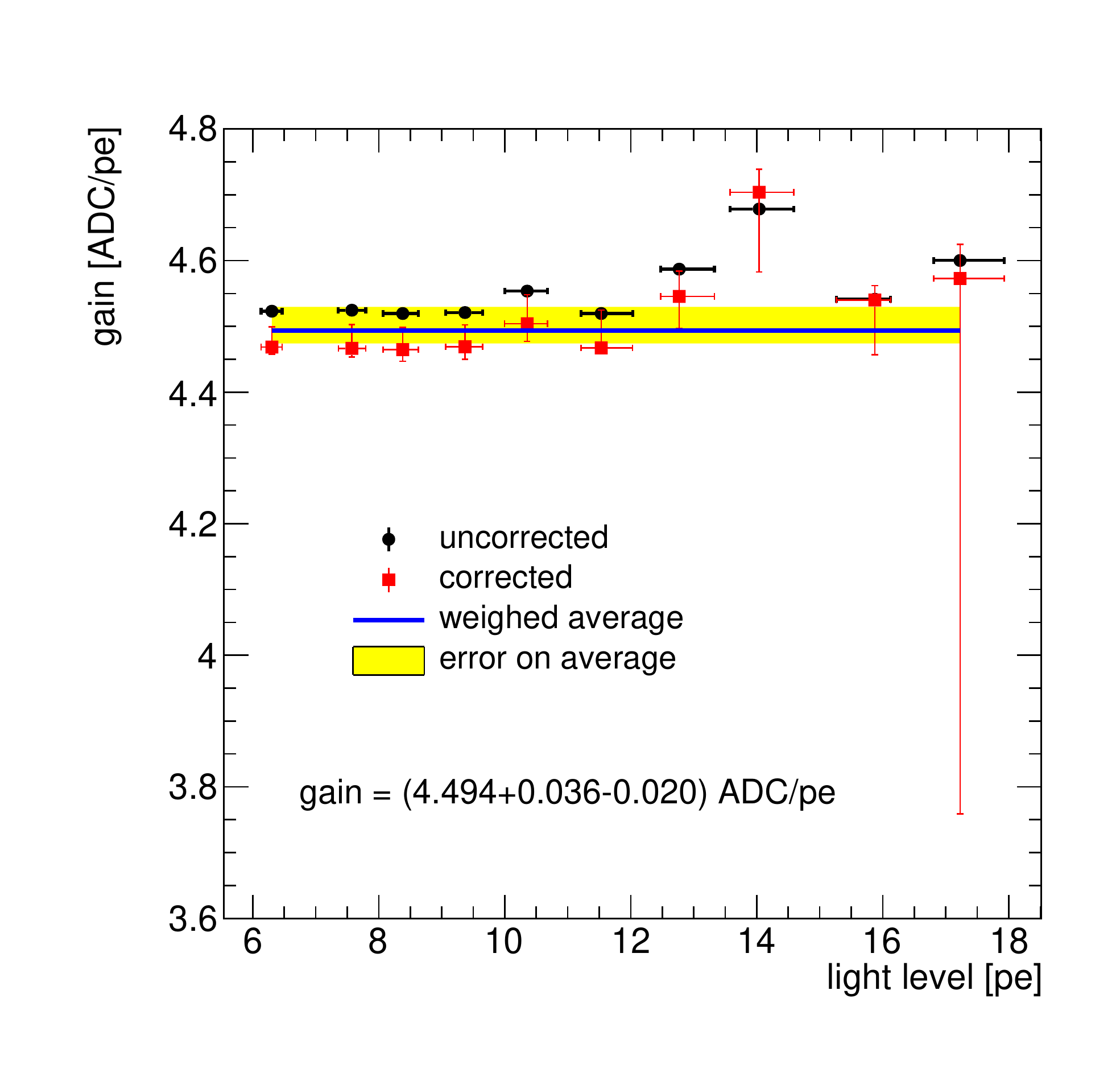}
	\caption{Measurement of the gain from the data taken with the DigiCam digitizer board hosting the FADCs from Intersil configured with gain around 5~ADC/\ac{p.e.}. The gain has been measured from the fit method (top) and the FFT method (bottom). In both cases, the raw values at different light levels are adjusted by the corresponding correction coefficients calculated with the toy Monte Carlo.
}
	\label{gain correction}
\end{figure}
Correction coefficients calculated via the toy Monte Carlo have been used to improve the accuracy of the measured gain as shown, as an example, in Fig.~\ref{gain correction}. The Monte Carlo uses, as input, the values of the parameters extracted from the data (cross talk, dark count rate, electronic noise). The uncertainty on the cross talk and the dark count rate was used to determine the systematic uncertainty on the correction coefficients. 
\newline
The gain value was used to determine the light level from the Gaussian distributions of signal amplitudes below saturation. For the \ac{MPE} spectra, the light levels were retrieved directly from the fits to the distributions as discussed in Sec.~\ref{sec:SPE}. The light levels obtained in either cases (from \ac{MPE} spectra and from Gaussians) were corrected for systematic effects (mostly cross talk) using the correction coefficients calculated from the toy Monte Carlo. As with the case of the correction coefficients for the gain, also in this case the systematic uncertainty on the correction coefficients was estimated from the experimental errors on cross talk and dark count rate. The result of the systematic study to calculate the light level correction coefficients and their uncertainties is shown in Fig.~\ref{correction coefficients}. 
At this stage, a LED calibration curve was built for light levels below the saturation of the detected signals. The extrapolation of the LED calibration curve above saturation was done via a $4^{th}$ degree polynomial\footnote{A more thorough calibration of the AC LED was previously carried out using a dedicated front-end board implementing a pre-amplification stage with a sufficiently high dynamic range to avoid saturation. These measurements showed that the calibration curve is well described by a $4^{th}$ degree polynomial.}. An example of a complete calibration curve is shown in Fig.~\ref{LED calibration}.

\subsubsection{Charge resolution}
\begin{figure}
	\centering
	\includegraphics[trim = 0mm 0mm 0mm 0mm, clip, width=0.45\textwidth]{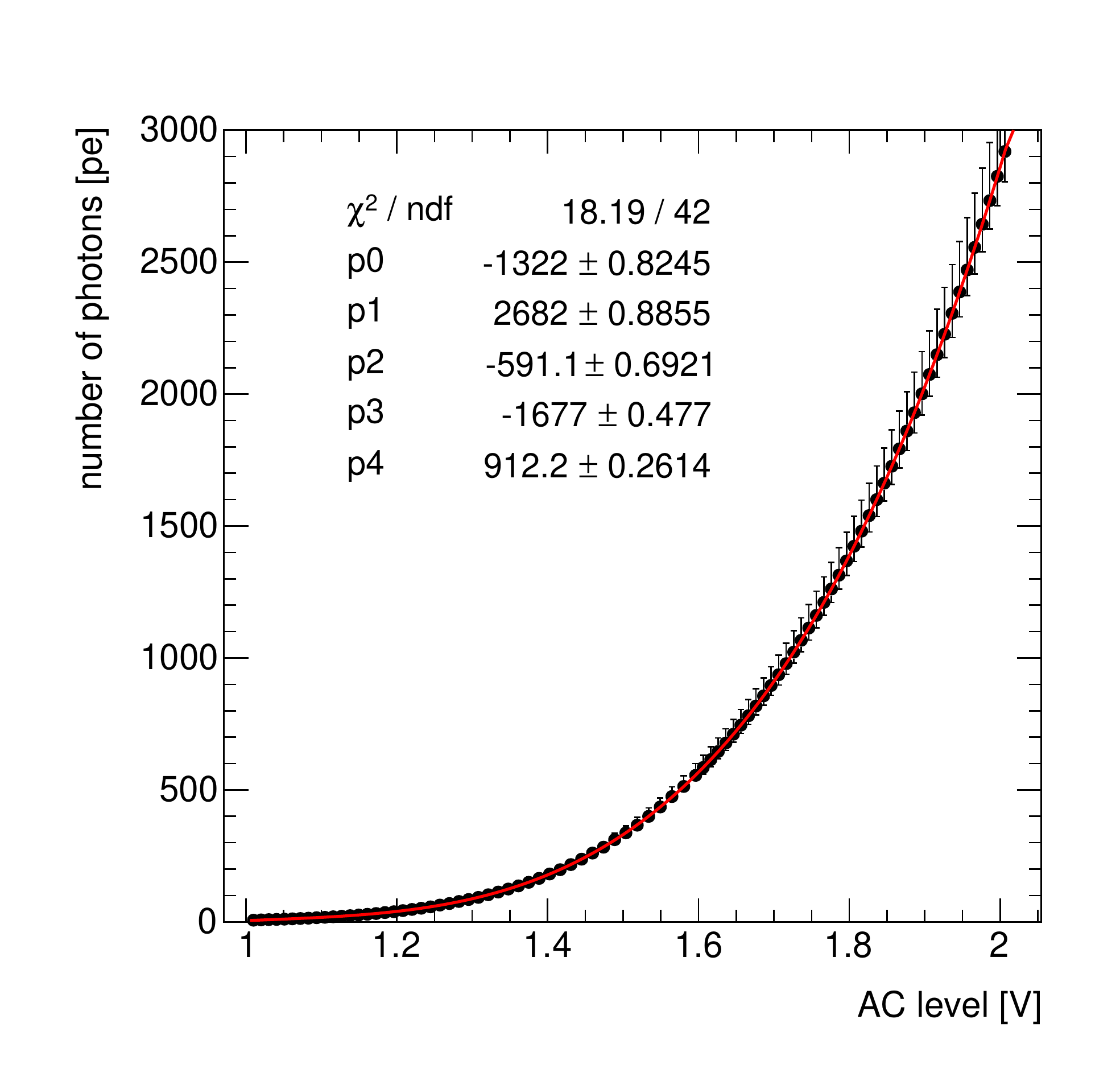}
	\caption{Example of a calibration curve of the AC LED source. Similar curves are produced for each dataset at each level of the DC LED (i.e. at each emulated NSB level).}
	\label{LED calibration}
\end{figure}
\begin{figure}
	\centering
	\includegraphics[trim = 0mm 0mm 0mm 0mm, clip, width=0.45\textwidth]{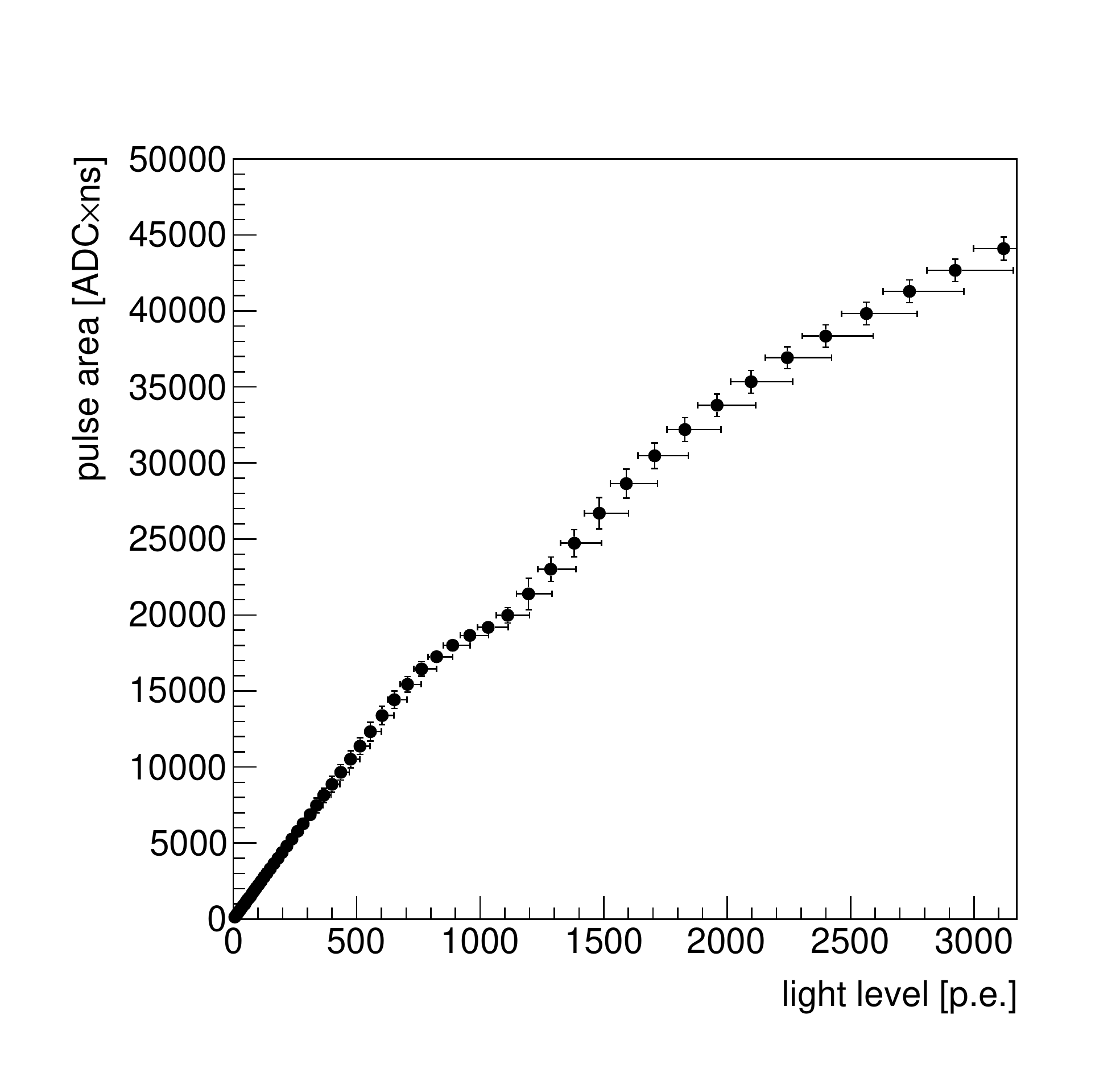}
	\caption{Pulse area as a function of the light level. These data were taken using the low gain configuration (around 5~ADC/\ac{p.e.}) of the DigiCam digitizer board. At around 750 \acp{p.e.}, the effects of the saturation of the pre-amplifier is visible.}
	\label{area calibration}
\end{figure}

At each level of the AC LED, i.e. at each light level calculated according to the calibration curve (as the one in Fig.~\ref{LED calibration}), the charge resolution is determined as
\begin{equation}\label{CR definition}
	\text{CR} = \frac{\sigma}{\mu},
\end{equation}
where $\mu$ and $\sigma$ are the mean value and standard deviation of the charge distribution. Before applying eq.~\ref{CR definition}, the mean value $\mu$ is corrected by the coefficients calculated from the toy Monte Carlo.
\newline
When the distribution is Gaussian (in either non-saturated or saturated regimes), the two quantities are derived directly from a Gaussian fit. In the case of \acp{MPE}, the distributions are fitted by eq.~\ref{finger function}. The $\mu$ parameter is the one determined from the fit, while $\sigma$ is calculated as
\begin{equation}
	\sigma = \sqrt{(\sigma_{68CL})^2 + \left(\frac{\Delta \mu}{2\sqrt{\mu}}\right)^2},
\end{equation}
where $\sigma_{68CL}$ corresponds to the 68\% confidence level around $\mu$, and the second term comes from the propagation of the fit error on $\mu$ for a Poisson-like variance $\sqrt{\mu}$.
\newline
The $\mu$ and $\sigma$ from Gaussian distributions of pulse amplitudes for signals below saturation are used directly to calculate the charge resolution. 
\begin{figure}
	\centering
	\includegraphics[trim = 0mm 0mm 0mm 0mm, clip, width=0.45\textwidth]{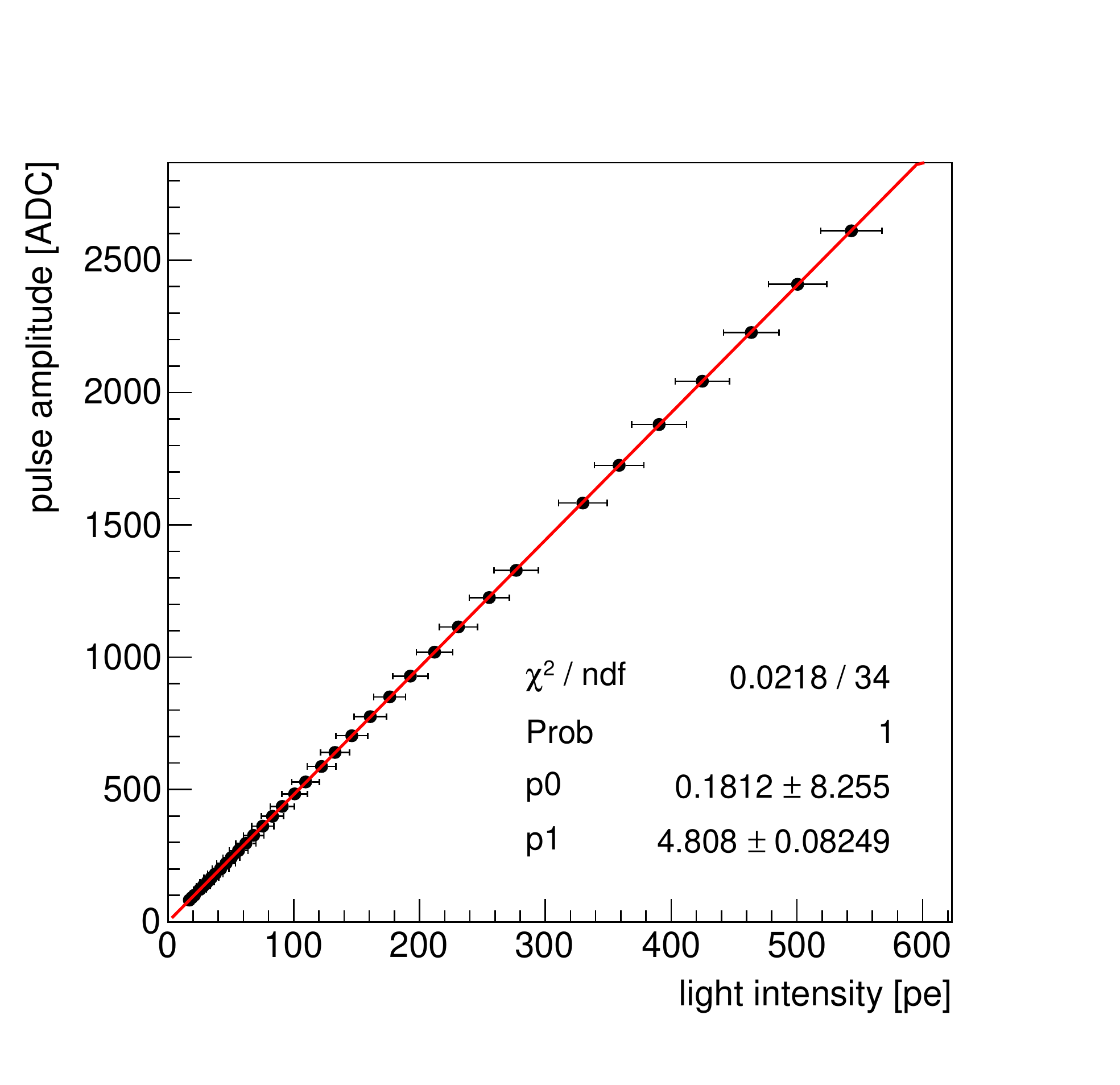}
	\caption{Pulse amplitude versus light intensity for Gaussian charge distributions below saturation (DigiCam low gain data).}
	\label{ampl lin}
\end{figure}
In such a case, the pulse amplitude and the light level are related to each other by a simple conversion factor (see figure~\ref{ampl lin}), and eq.~\ref{CR definition} can be applied on the raw values of the corrected $\mu$ and $\sigma$ in units of ADC counts.
\newline
When the light intensity is high enough for saturation to occur (be it either in the preamplifier or in DigiCam, depending on the gain settings of the FADC used), charge distributions of pulse area rather than pulse amplitude are used (see also~\cite{electronics_paper}). In this case, the relation between charge and light level is not scalar, and the raw $\sigma$ and $\mu$ values retrieved from the area of the Gaussian fit cannot be used directly to calculate the charge resolution according to eq.~\ref{CR definition}, but must be first converted into units of \acp{p.e.}. For this purpose, area~vs~light level curves are built from the data at each NSB level, using the AC LED calibration to estimate the light level at each pulsed light setting, to be correlated to the mean area value of the corresponding pulses. An example of such curve is shown in Fig.~\ref{area calibration}.
\newline
\begin{figure}
	\centering
	\includegraphics[trim = 0mm 6mm 0mm 0mm, clip, width=0.5\textwidth]{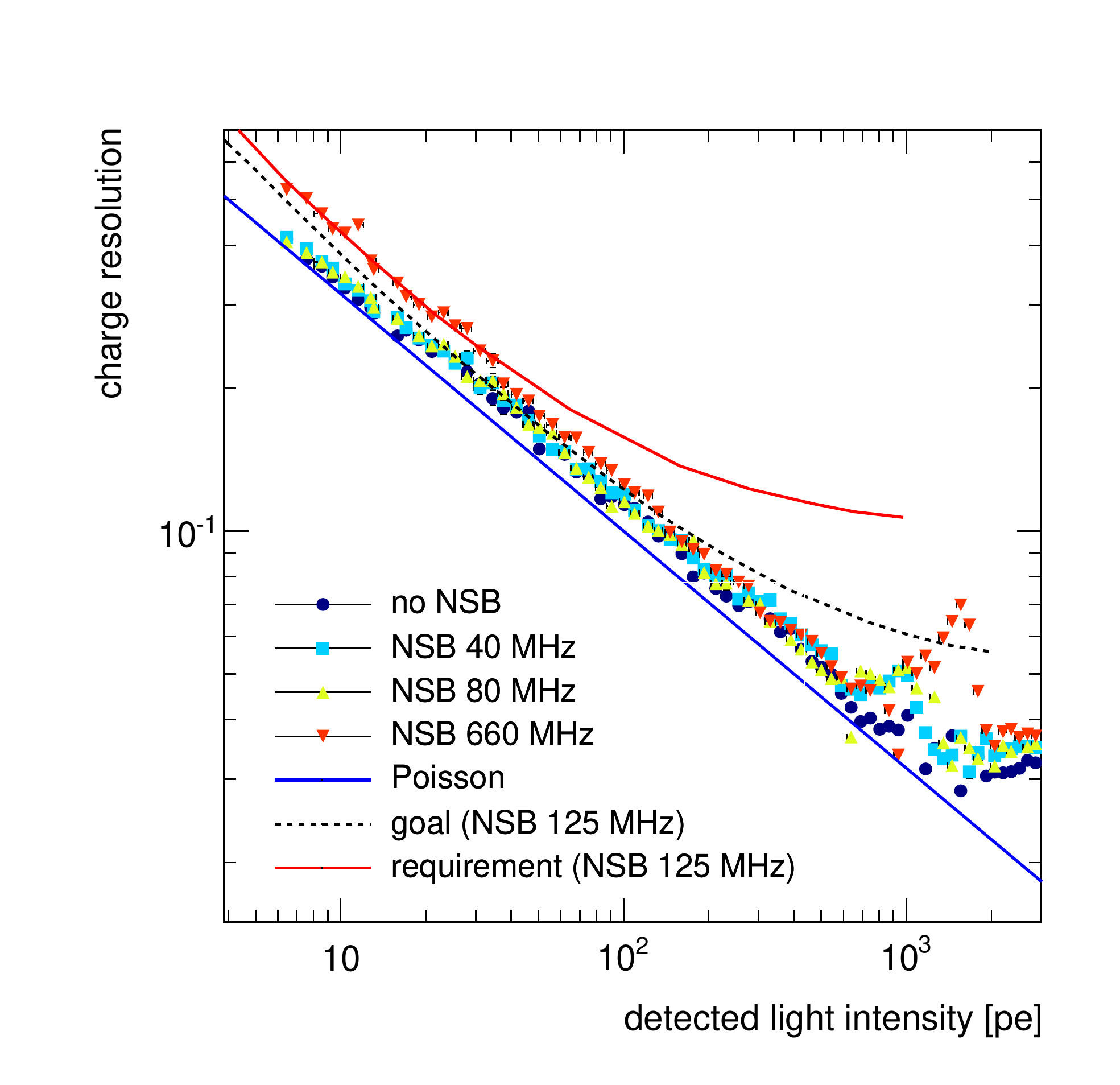}
	\includegraphics[trim = 0mm 6mm 0mm 0mm, clip, width=0.5\textwidth]{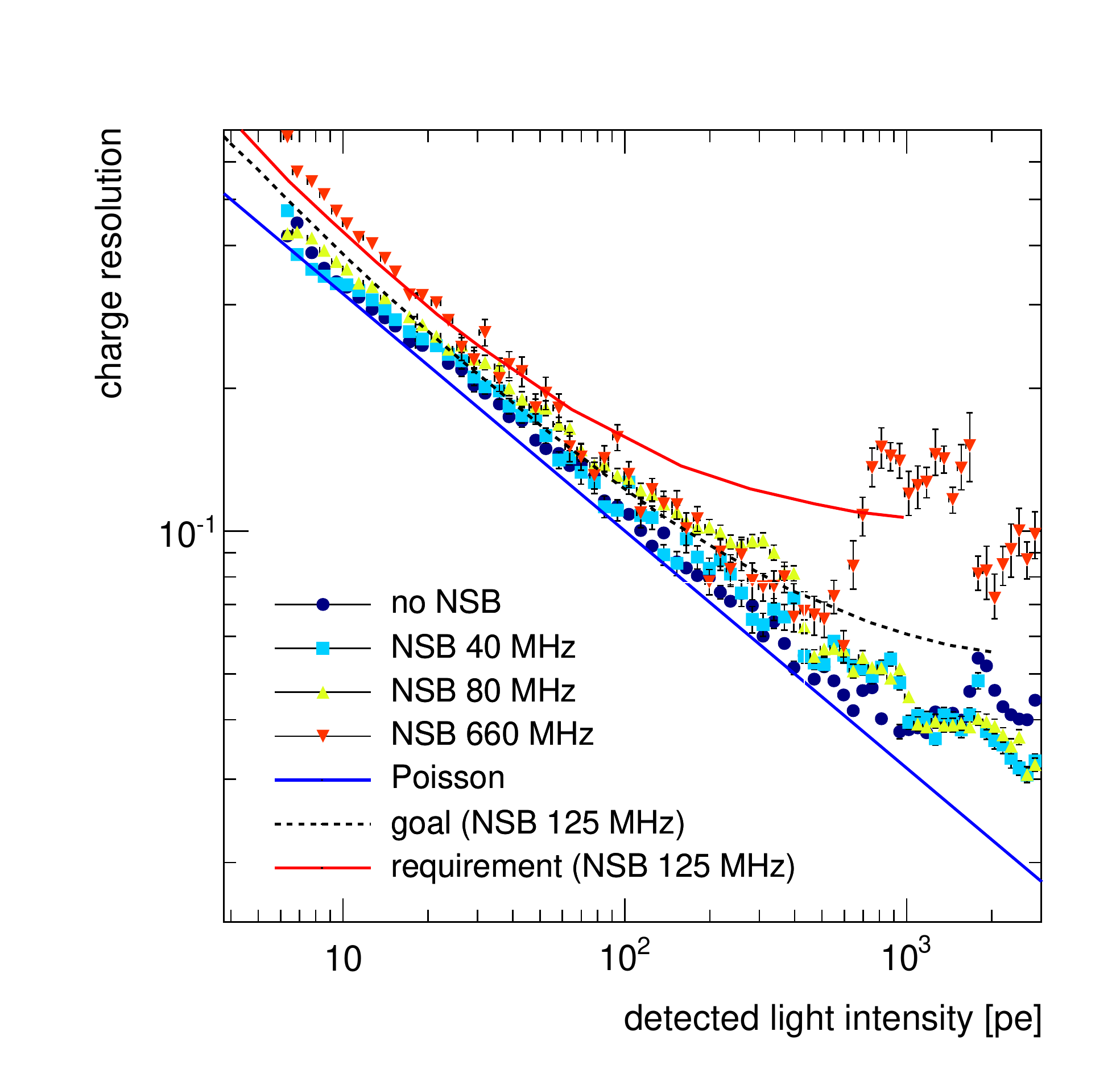}
	\caption{Charge resolution at different emulated \acl{NSB} levels measured on a sensor on a fully assembled module readout by DigiCam in two gain configurations: top, low gain (around 5~ADC/\ac{p.e.}) and, bottom, high gain (around 10~ADC/\ac{p.e.}).}
	\label{charge resolution}
\end{figure}
Fig.~\ref{charge resolution} shows the charge resolution measured at different emulated NSB levels for two gain configurations of DigiCam: low gain (around 5~ADC/\ac{p.e.}) and high gain (around 10~ADC/\ac{p.e.}). The two cases are different in terms of saturation conditions: in the former the preamplifier saturates before DigiCam at around 750 \acp{p.e.}, which means that the full waveforms are always digitized; in the latter DigiCam saturates before the preamplifier at around 400 \acp{p.e.}, and as a consequence the pulses are truncated from this light level on. Notice that in this analysis the possible effect of the LED source fluctuations is not subtracted.
\newline
The results show that, apart from the case at 660~MHz NSB (half moon), all the points fall below the CTA goal curve. In particular there is no sharp transition in correspondence of the saturation points ($\sim$750 \acp{p.e.} for low gain, $\sim$400 \acp{p.e.} for high gain), meaning that, despite the overall non-linearity of the camera, the signals can be reconstructed with equal precision in either non-saturated and saturated regimes. At 660~MHz both gain settings loose performance at low light levels (as it is, in fact, expected for the operation of the telescope in half moon nights), but still keeping below the requirement curve. The 660~MHz high gain data points, however, show, at around 1000~\acp{p.e.}, an increase in resolution above requirement. This effect is to be attributed to the truncation effect in combination with the waveform distortion undergone by signals that exceed the dynamic range of the preamplifier (for more details, see~\cite{electronics_paper}). These results thus show that a low-gain configuration, with no truncation from the digitizers, is to be preferred. The final version of the DigiCam prototype has been produced with a gain configuration where the preamplifier saturates before the FADCs and where the full dynamic range of the FADCs is exploited. As far as performance differences between FADCs from Analog Devices and FADCs from Intersil are concerned, the two turned out to be equivalent. The choice of either company elements will be driven by the cost benefits.

\section{Camera calibration studies}
\label{sec:calibration}

To ensure a homogeneous performance of the camera, the pixels will be calibrated regularly. The determination of the relevant parameters that are necessary to equalize the response of each pixel over the full PDP (also referred to as flat-fielding) will be performed at different timescales and in different measurement conditions, depending on the parameter type (see~\cite{ICRC_Elisa}). While some parameters can be monitored on an event-by-event basis (for instance the measurement of the baseline), others (e.g. dark count rate and cross talk) can be measured with lower frequency. Some of the parameters can be extracted directly from the physics data, others will require special calibration runs, for example dark runs taken with the camera lid closed or data taken by illuminating the camera with a dedicated flasher unit installed on the telescope structure.
The studies performed in the laboratory during the prototyping phase will define the calibration strategy that will be adopted on site. We describe here some relevant aspects of the calibration of the camera.

\begin{figure}
	\centering
	\includegraphics[width=0.45\textwidth]{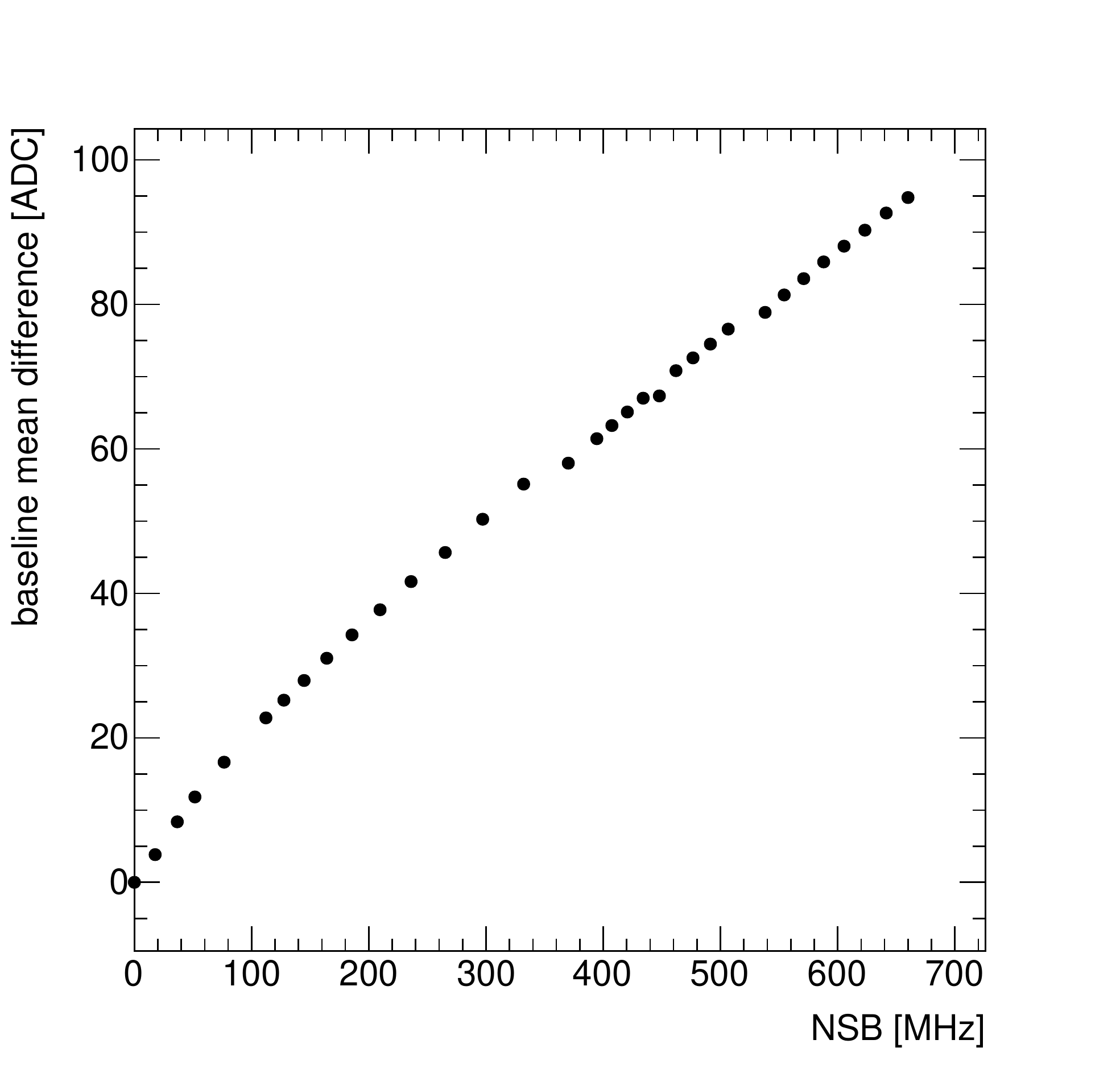}
	\caption{Difference between the baseline position (mean) at a given NSB level and the baseline position in the dark.}
	\label{baseline level vs NSB}
\end{figure}

\begin{figure}
	\centering
	\includegraphics[width=0.45\textwidth]{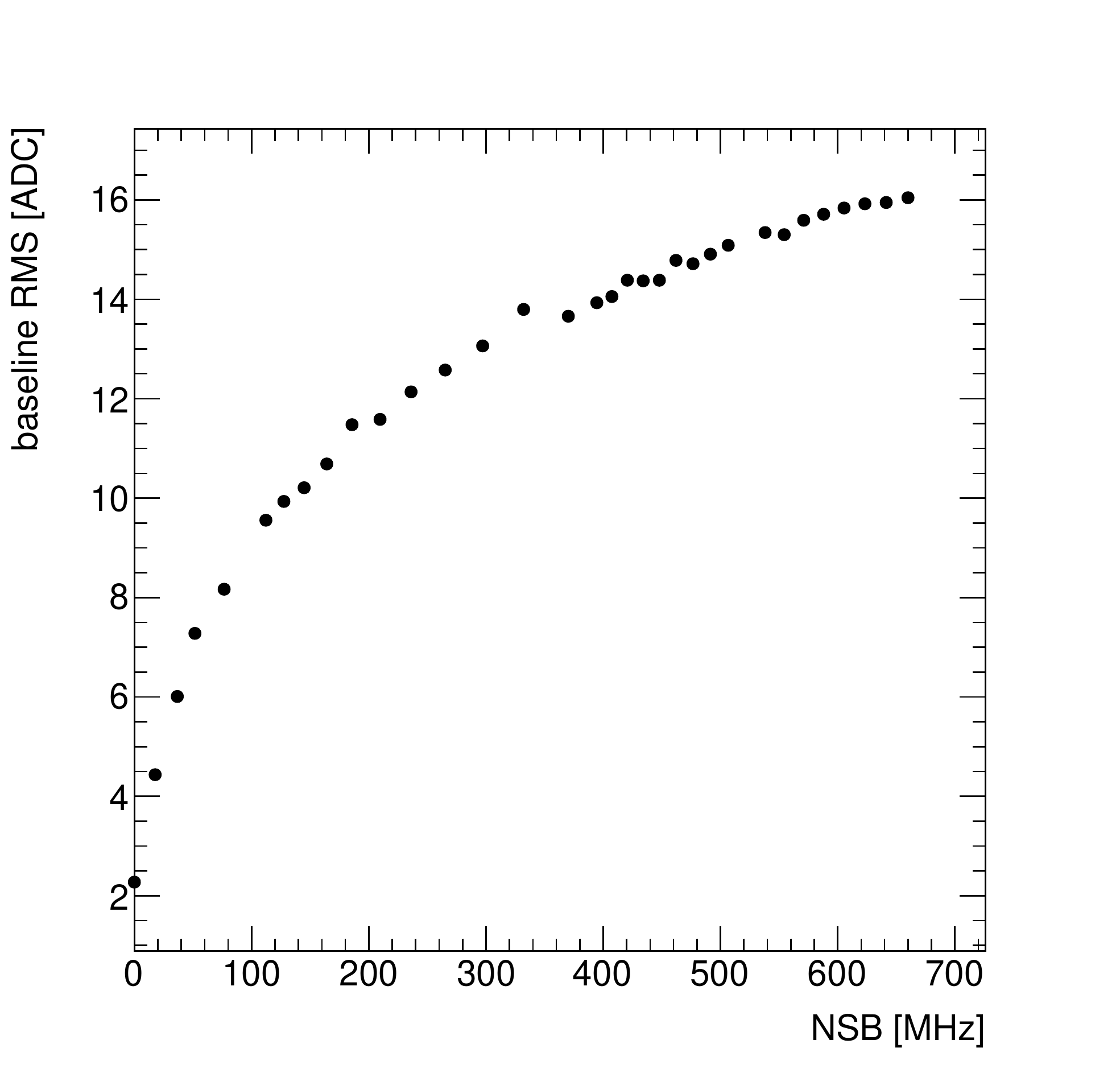}
	\caption{Dependency of the baseline noise (RMS) on the \acl{NSB} level.}
	\label{baseline noise vs NSB}
\end{figure}

The baseline level for a sensor DC coupled to the preamplifier is correlated to the \ac{NSB}. The determination of the baseline level can be done on an event-by-event basis by extending the signal acquisition window before the signal peak arrival time. This is possible thanks to the ring buffer structure implemented in DigiCam and to the relatively low trigger rates expected for the \acp{SST}. 
Using the LED driver board described in Sec.~\ref{sec:performances}, different NSB conditions could be reproduced in laboratory. Data from a pixel in a fully assembled module, readout by DigiCam at different emulated NSB levels, are taken and analyzed to characterize the behavior of the baseline.
The results are shown in Fig.s~\ref{baseline level vs NSB} and ~\ref{baseline noise vs NSB}. Here the baseline position (which here we determine as the mean of the counts) and noise (RMS) are calculated for different emulated \ac{NSB} conditions over a large number of DigiCam samples, corresponding to a total time window of 800~$\upmu$s.

\begin{figure}
	\centering
	\includegraphics[width=0.45\textwidth]{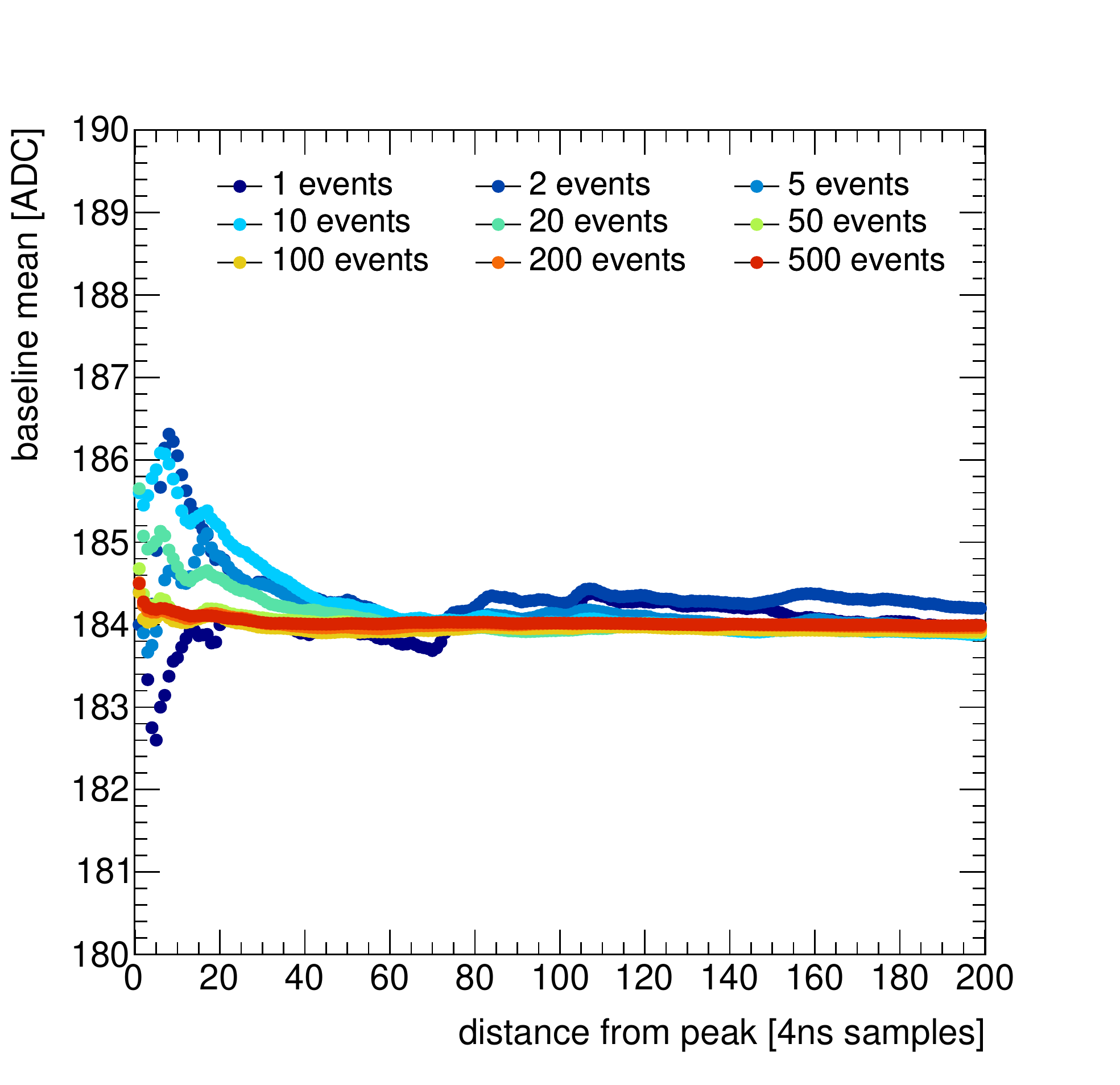}
	\includegraphics[width=0.45\textwidth]{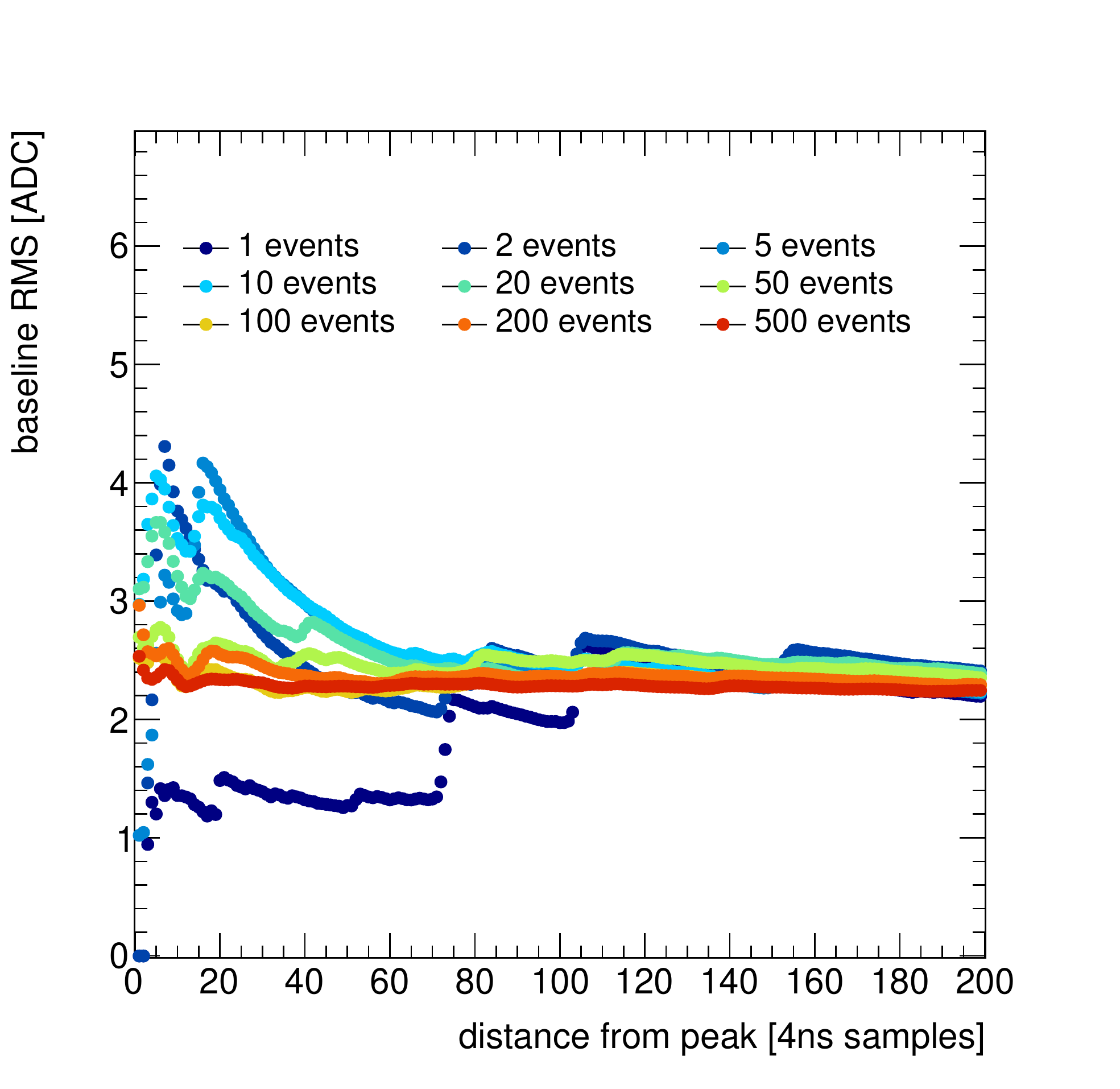}
	\caption{Baseline position (in this case mean, top) and noise (RMS, bottom) measured for different pre-pulse sample sizes (distance from peak, in the horizontal axes), and for a different number of events (colors).}
	\label{baseline window study}
\end{figure}

A detailed study of the baseline level determination was performed and the accuracy of the baseline estimate as a function of the 
number of data samples was estimated. This is studied within the same dataset at different emulated \ac{NSB} levels, and the result is shown in Fig.~\ref{baseline window study}. These plots refer to the dark condition, but similar ones have been made for a number of \ac{NSB} levels between dark and 660~MHz. From these, it can be concluded that sufficiently accurate measurements can be made on a set of around 50 events, when around 30 pre-pulse samples are considered, which in total corresponds to a 10~$\upmu$s window. This means that, even in the case in which the baseline is measured during data taking at the frequency of 1~Hz (which is, most likely, too high), this would add a negligible duty cycle. Thus, in general, the baseline (level and noise) can be measured accurately at a frequency that is high enough to efficiently monitor the \ac{NSB} in real time.

All these studies shown here are done systematically for each pixel of the camera prior to its installation on the telescope structure, by using the cabling test setup as discussed in Sec.~\ref{camera test setup section}. The same setup will also be used to perform a preliminary flat fielding of the camera 
and to test the trigger logic by illuminating the camera with pre-defined light patterns that mimic real Cherenkov events (e.g. elliptical images from gamma-ray and proton showers and rings from muon events). The possibility is also foreseen to reproduce patterns from simulated events.

It is understood that the presence at the sensor bias stage of a 10~kOhm resistor in series~\cite{electronics_paper} with the sensor produces a voltage drop at the sensor cathode when a current (e.g. induced by the NSB photons) flows through the resistor. If the voltage drops, the over-voltage is not anymore the one set by the user and therefore the operation point changes with the following consequences:
\begin{enumerate}
	\item the gain decreases;
	\item the \ac{PDE} decreases;
	\item the optical cross talk decreases;
	\item the dark count rate decreases.
\end{enumerate}

\begin{figure}
	\centering
	\includegraphics[width=0.45\textwidth]{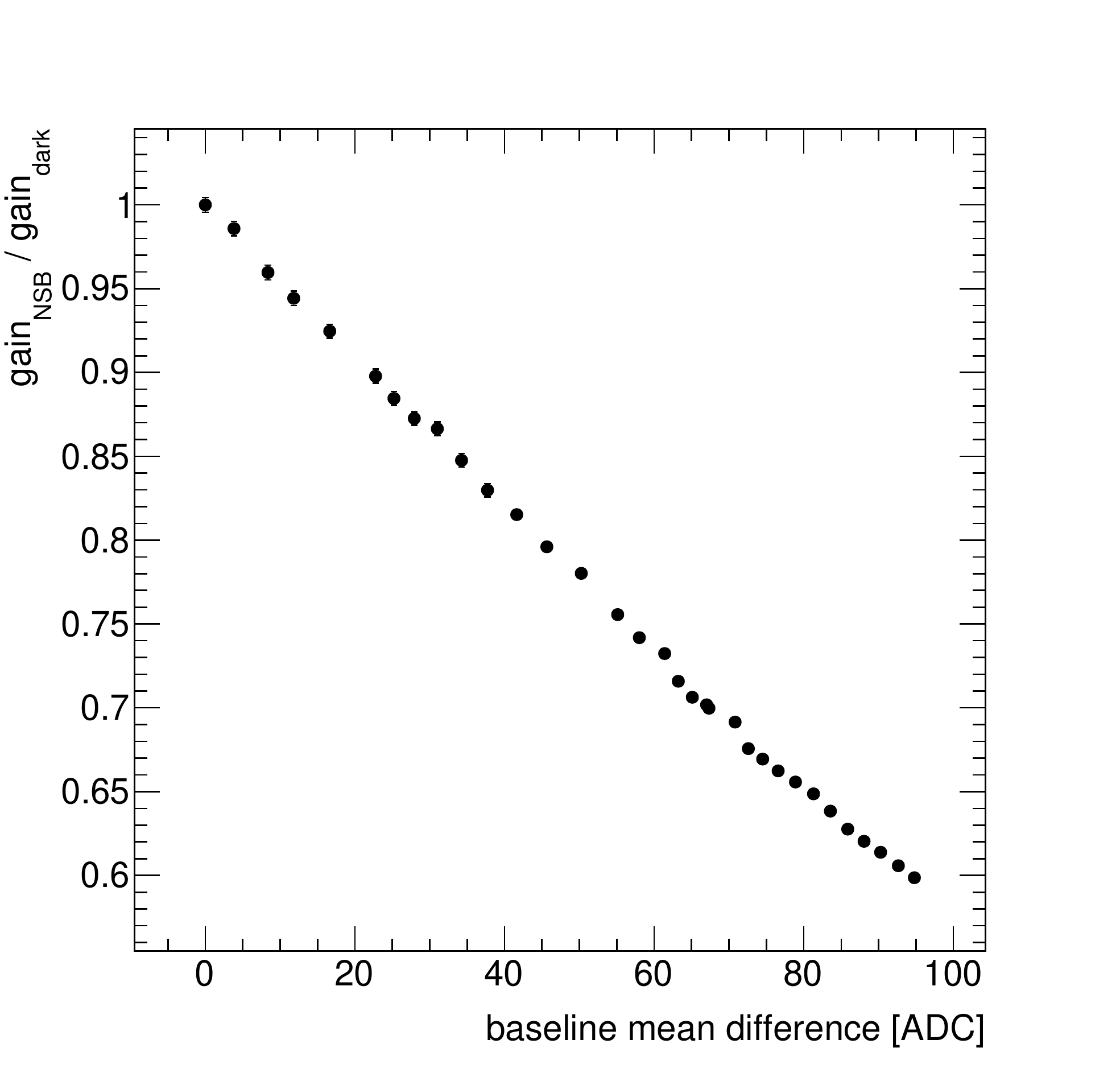}
	\caption{Gain variation, with respect to the nominal gain in the dark, as a function of the baseline shift.}
	\label{gain vs NSB}
\end{figure}

The fact that the gain decreases implies that the conversion factor from \acl{p.e.} to ADC count changes
(see Fig. \ref{gain vs NSB}). For the prototype camera, this effect will not be compensated at the hardware level but will be taken into account during the telescope operation since the baseline shift can be evaluated online and the gain correction can be derived at \ac{FPGA} level or at software level. 

The baseline shift measurement will be part of the data stream and accessible at the data analysis level and will allow one to derive the evolution of the relevant parameters with the operation point, such as the \acl{PDE}. 
As a matter of fact, as shown in Fig.~\ref{PDE_XT}, the \ac{PDE} variation with the operation point with the NSB level has been measured. We will operate
the sensor slightly before the region where the \ac{PDE} becomes independent on the over-voltage. Hence the \ac{PDE} variations have to be monitored since
they could affect the trigger threshold (expressed in terms of \acs{p.e.}), on which the efficiency of the data taking depends. The trigger threshold is set according to Monte Carlo simulations (see Sec. \ref{sec:performances}), which will be benchmarked against real data. The setting point is approximately in the region where cosmic rays begin to emerge on top of the noise. 

\begin{figure}
	\centering
        \includegraphics[width=0.45\textwidth]{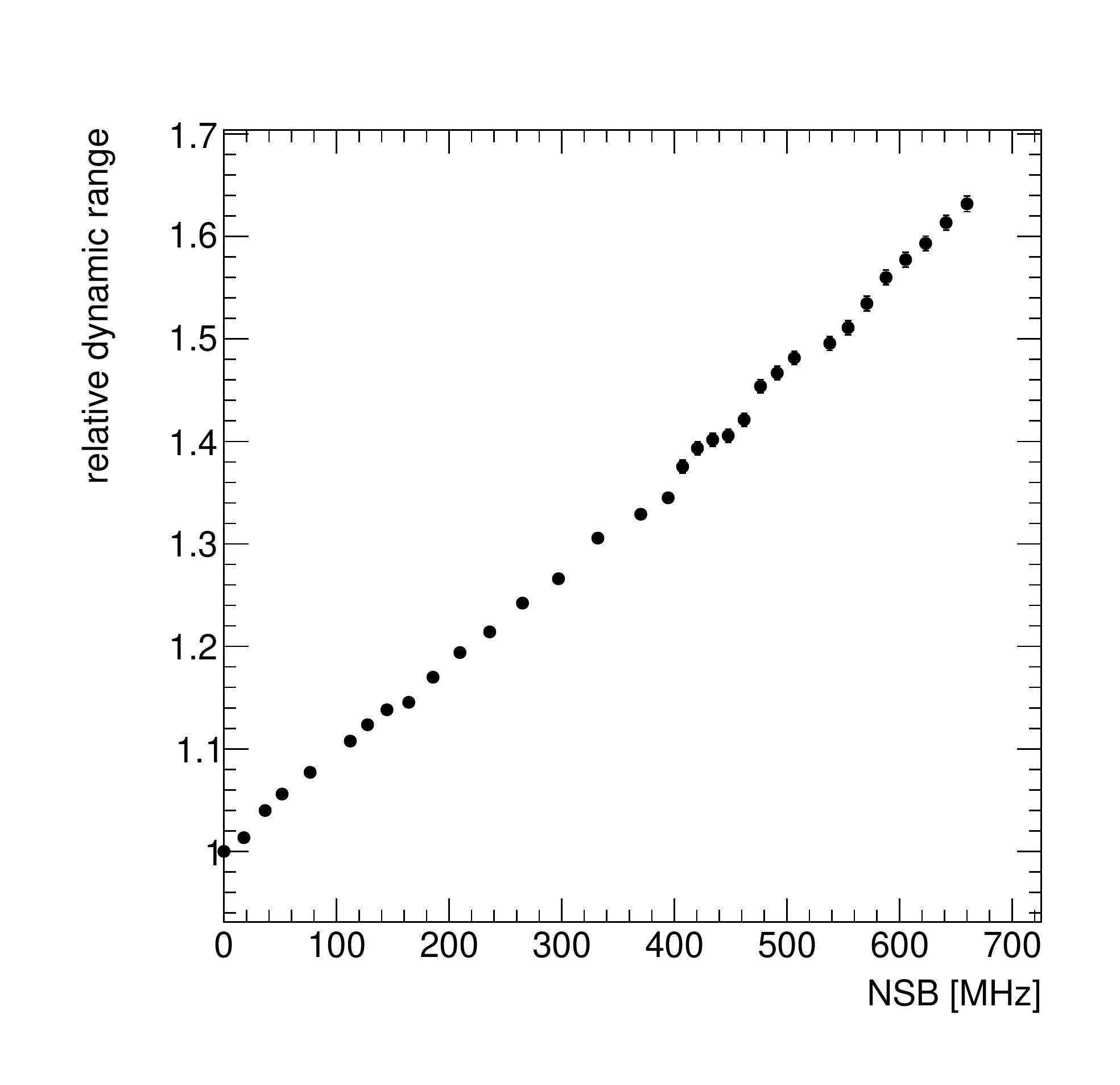}
	\caption{Dynamic range variation (relative to the dynamic range in dark conditions) as a function of the NSB level.}
	\label{fig:dynamic range}
\end{figure}

Incidentally, it is possible that the \ac{PDE} vs wavelength does not scale identically for different over-voltages, therefore different \ac{NSB} levels. 
This effect will be characterized in the laboratory with a Xenon lamp by measuring the PDE vs wavelength for different over-voltages.
Moreover, the contribution of the dark count rate and optical cross talk variations as a function of the over-voltage (see Fig. \ref{PDE_XT}) have to be subtracted to the measured signal to avoid problems with different over-voltages, and hence \ac{NSB} levels. 
These effects require a well defined camera calibration, which can be properly set up as we demonstrate here. Moreover there is a positive counter part. Opposite to what one would expect from a DC coupled front-end electronics, the voltage drop feature leads to an increase in dynamic range even when the baseline or noise increases, as visible in Fig.~\ref{fig:dynamic range}, due to the fact that the gain decreases.

\section{Expected performances through simulations}
\label{sec:performances}
The measurements on the characterization of the camera performance (window transmittance, cone reflectivity, sensor \ac{PDE}, charge resolution, and so on) have been used to reproduce the camera response in Monte Carlo simulations. Through these simulations we can estimate the performance parameters  and compare them to the \ac{CTA} requirements.

Different simulation tools have been used for this study. Atmospheric showers induced by gamma-rays and/or cosmic rays have been simulated with CORSIKA
up to 100~EeV~\cite{corsika}. The simulation of the telescope was done using two different tools which produce comparable results: \emph{sim\_telarray}~\cite{simtelarray} and the combination of GrOptics and CARE~\cite{groptics+care}. \emph{sim\_telarray} is widely used in \ac{CTA} to study the preliminary performance of the array of telescopes (sensitivity, array layout, array trigger, etc.). It simulates the telescope optics and the camera, but it does not account for the shadowing of the elements (such as the masts and the camera box) in an exact way, nor does it simulate the camera with a great deal of detail. 
Hence, a more detailed simulation of the \ac{SST-1M} telescope and its camera was implemented with GrOptics and CARE. GrOptics is a package for ray tracing that considers the mirror transmission in detail and the telescope structure. CARE simulates the camera down to more fundamental properties of the detector, such as the microcells of the \acp{SiPM} and the saturation of the signals, the trigger system and the backgrounds (such as the electronics noise and the \ac{NSB}).

\subsection{Single telescope performance as a function of energy}\label{Eth}
\begin{figure}
	\centering
	\includegraphics[width=0.5\textwidth]{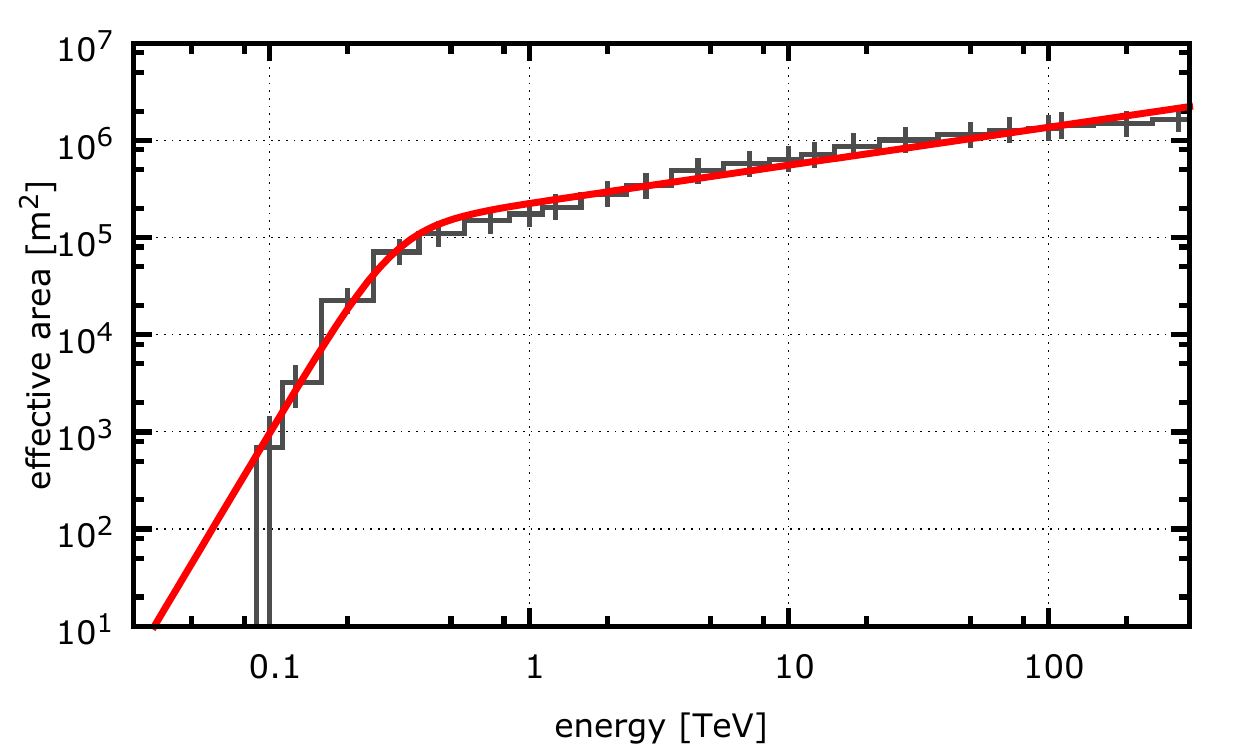}
	\caption{Estimated effective area of a single SST-1M as a function of energy. The red line is a fit via an empirical function of the form $f(x) = \log_{10}(A) + B\cdot x - \log_{10}\left[1+\left(\frac{10^x}{C}\right)^D\right]$.}
	\label{effarea}
\end{figure}
The studies in this section concern the sensitivity as a function of energy of single SST-1M telescopes and they strongly depend on the kind of trigger logic chosen.
CORSIKA simulated gamma-ray showers have been fed to \emph{sim\_telarray} to estimate the differential trigger rate $\mathrm{d}R/\mathrm{d}E$ as a function of the energy of the primaries. The rate is estimated as the Crab flux unit~\cite{CrabHESS} detected over the effective area. The effective area at a given energy is the integral of the distribution of triggered events over the distance between the core of the shower and the telescope, and is shown in Fig.~\ref{effarea}. 

\begin{figure}
	\centering
	\includegraphics[width=0.5\textwidth]{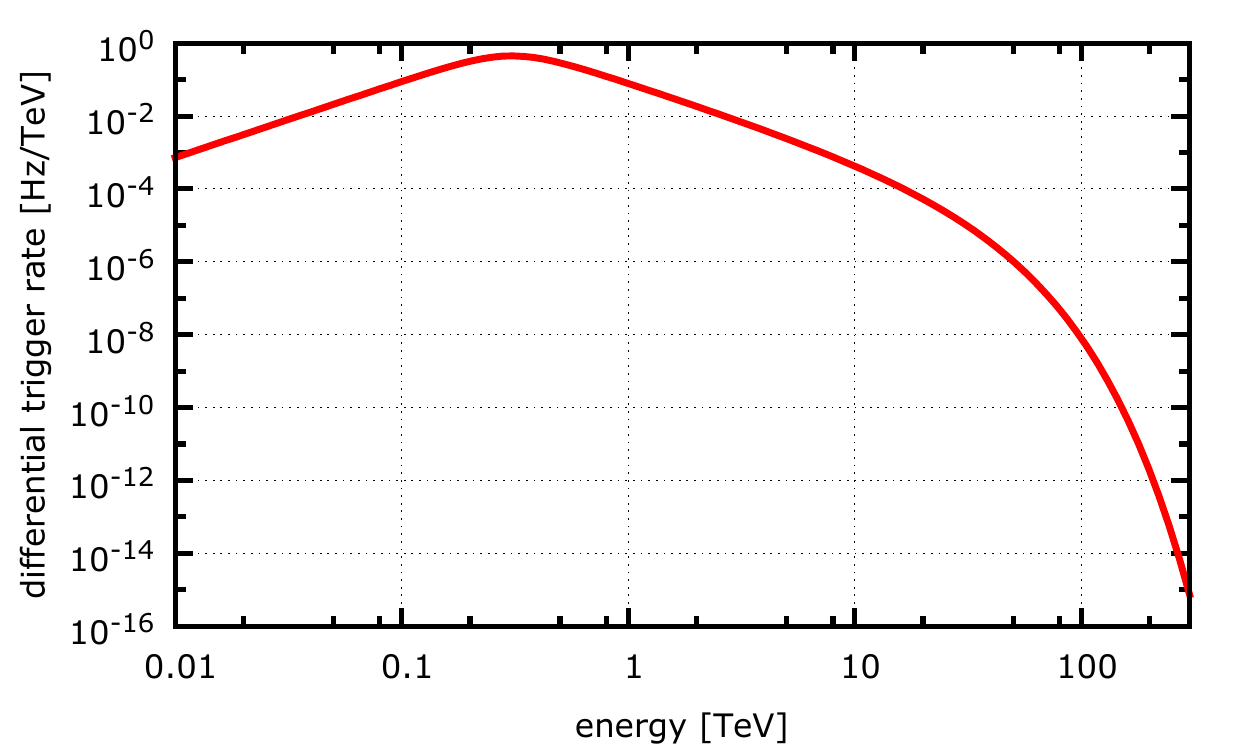}
	\caption{Differential trigger rate $\mathrm{d}R/\mathrm{d}E$ as a function of energy. The maximum of the curve marks the energy threshold of the telescope.}
	\label{diffTriggRate}
\end{figure}

The differential trigger rate is shown in Fig.~\ref{diffTriggRate}. The maximum of the curve, named ``energy threshold'', marks the point above which the telescope becomes most effective. The simulation shows that the threshold for the \ac{SST-1M} telescope is at around 300~GeV, one order magnitude lower than the requirement specified by \ac{CTA}. 
Currently, only a simple majority trigger has been implemented requiring that the trigger is fired if the digitized pulse of the signal in a hexagonal patch of 7 pixels, with a readout window of 200~ns, is above a threshold of 145 ADC counts (28.7 p.e.s summed up in the patch, for a simulated gain of 5~ADC/p.e.). 

\subsection{Estimated average camera efficiency}
\label{averageEfficiency}
\ac{CTA} requires an average camera efficiency above 17\% for the \acp{SST}. This has been estimated as the average of the \ac{PDE} filtered in wavelength by the Fresnel losses due to the entrance window and the funnel transmittance (Fig.~\ref{SNRwindow} on top, red dashed line), and weighed by the Cherenkov spectrum (Fig.~\ref{SNRwindow} on top, blue solid line) in the 300~nm to 550~nm wavelength range, yielding 32.73\%. However, the average efficiency due to the angular dependence of the incoming photons must also be taken into account. Calculating the integral average of the cone angular transmittance (Fig.~\ref{cone measurement} on top, red line) weighed by the probability distribution of the incoming angle (taken from~\cite{CONES}, simulated with Zemax) gives an efficiency of 0.88. Some of the photons are lost due to the dead zones between pixels in the \ac{PDP}. Since the side-to-side size of a pixel active area is 2.32~cm and the side-to-side size of the full \ac{PDP} is 88~cm, the ratio of the active area of the full 1296 pixels matrix to the physical area of the \ac{PDP} is $1296\cdot(2.32/88)^2=0.90$. Hence, the average camera efficiency can be estimated to be $32.73\cdot0.88\cdot0.90=25.94\%$, larger than the requirement.

\subsection{Expected number of photoelectrons}\label{expectedPH}
\begin{figure}
	\centering
	\includegraphics[width=0.5\textwidth, trim = 0mm 0mm 0mm 10mm, clip]{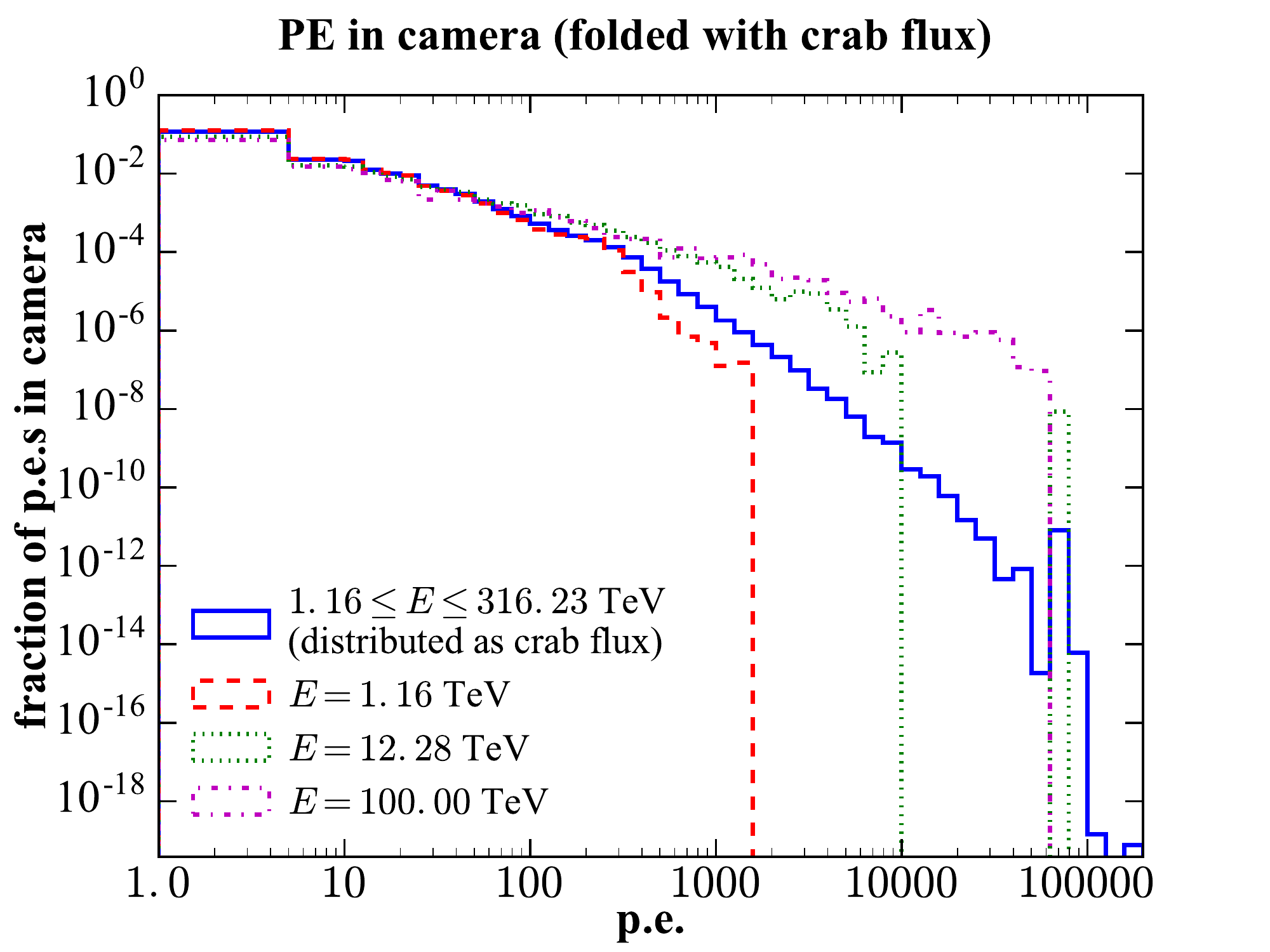}
	\caption{Simulated \ac{p.e.} distribution in the camera for on-axis events with a Crab-like flux from 1~TeV to 316~TeV of energy at 2000~m of altitude (blue solid line). Examples of  mono-energetic events are also shown: 1.16~TeV (red dashed line), 12.28~TeV (green dotted line) and 100 (magenta dash-dotted line). Areas are normalized to one.}
	\label{peCamera}
\end{figure}
\begin{figure}
	\centering
	\includegraphics[width=0.5\textwidth, trim = 0mm 0mm 0mm 10mm, clip]{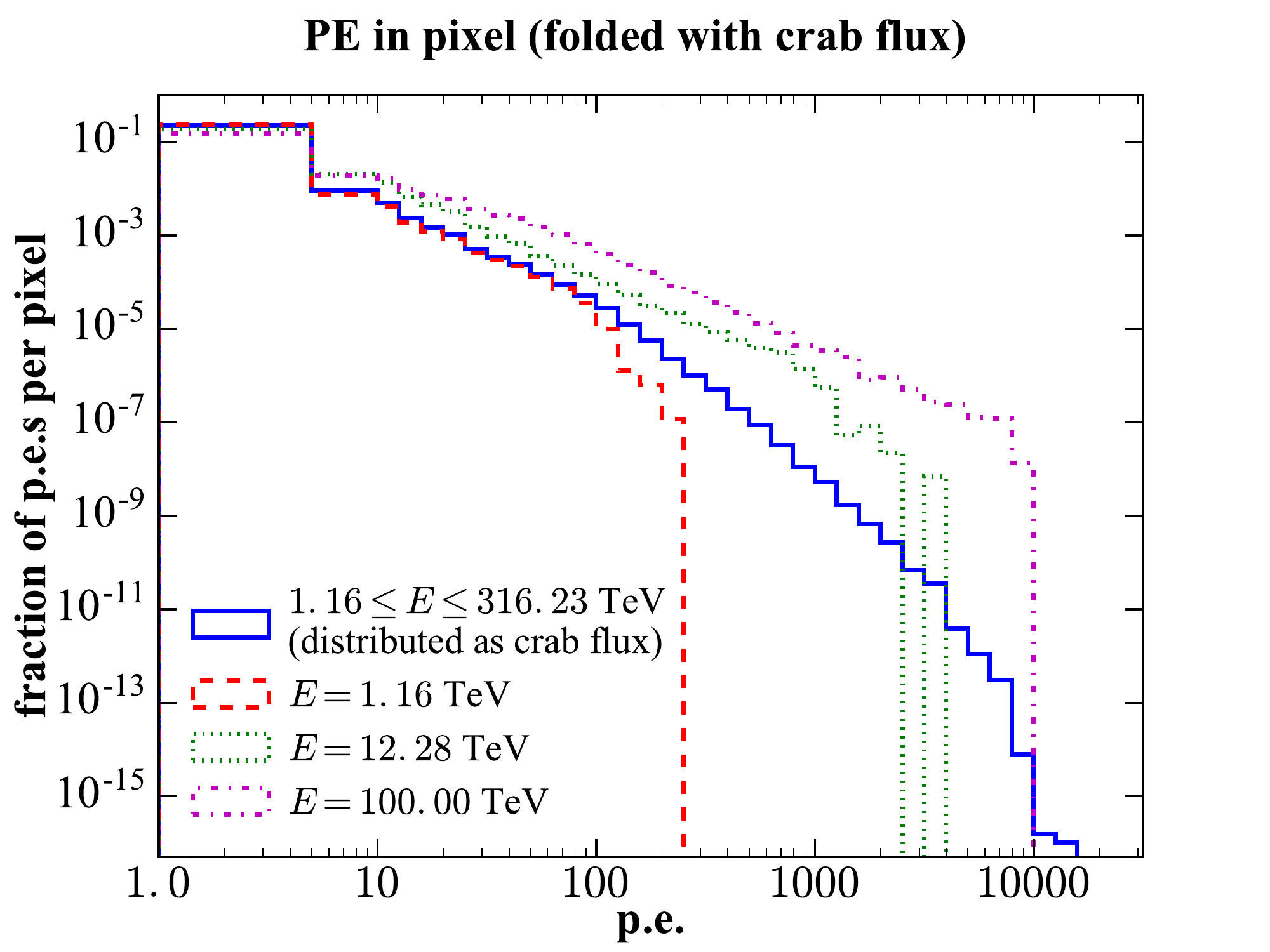}
	\caption{Simulated \ac{p.e.} distribution per pixel for on-axis events with a Crab-like flux from 1~TeV to 316~TeV of energy at 2000~m of altitude (blue solid line). Examples of mono-energetic events are also shown: 1.16~TeV (red dashed line), 12.28~TeV (green dotted line) and 100 (magenta dash-dotted line). Areas are normalized to one.}
	\label{pePixel}
\end{figure}
\begin{figure}
	\centering
	\includegraphics[width=0.5\textwidth, trim = 0mm 0mm 0mm 10mm, clip]{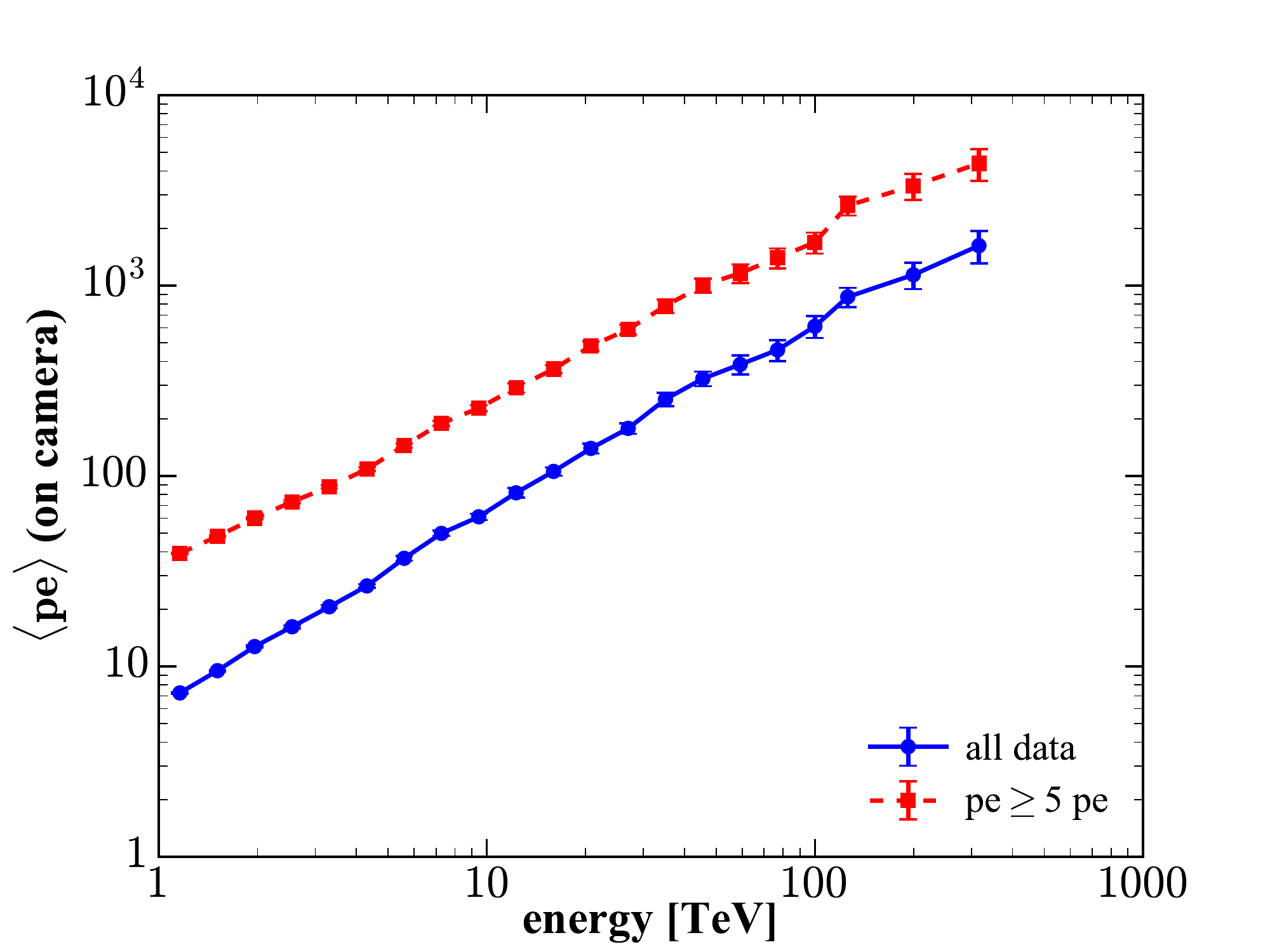}
	\caption{Averages of the simulated \ac{p.e.} distributions induced by mono-energetic photons in the camera.}
	\label{peCamera_mean}
\end{figure}
\begin{figure}
	\centering
	\includegraphics[width=0.5\textwidth, trim = 0mm 0mm 0mm 10mm, clip]{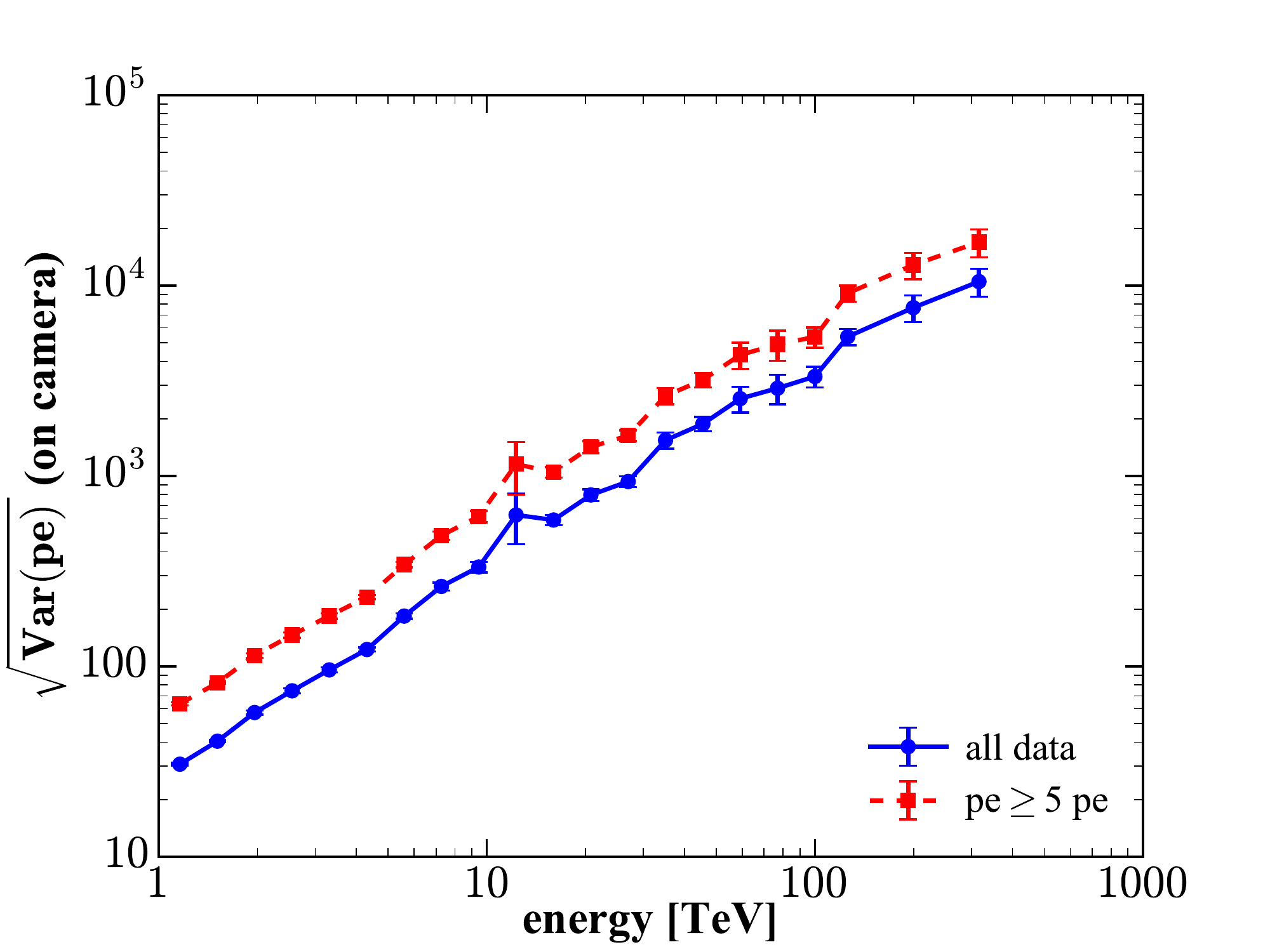}
	\caption{Standard deviations of the simulated \ac{p.e.} distributions induced by mono-energetic photons in the camera.}
	\label{peCamera_std}
\end{figure}
\begin{figure}
	\centering
	\includegraphics[width=0.5\textwidth, trim = 0mm 0mm 0mm 10mm, clip]{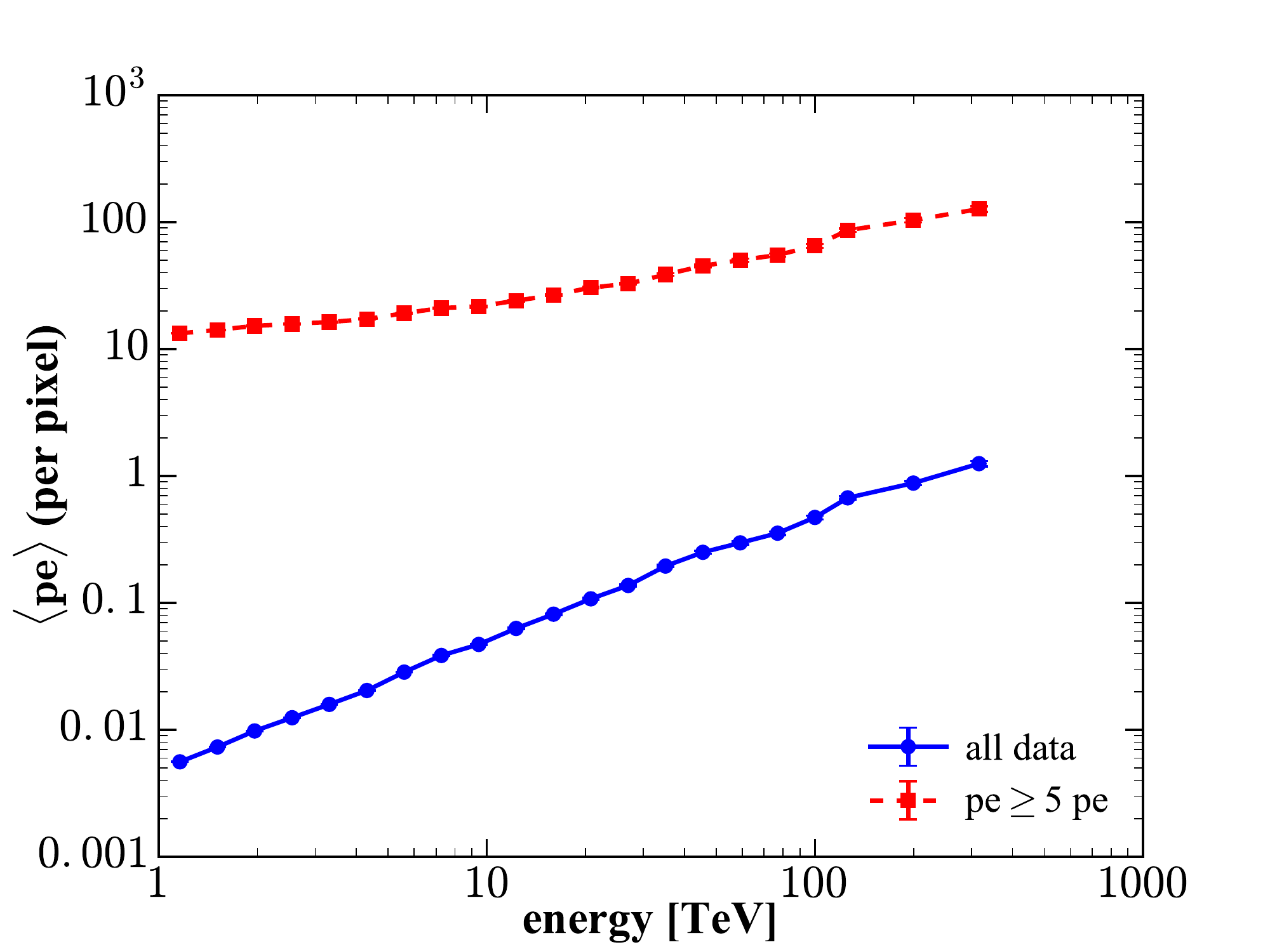}
	\caption{Averages of the simulated \ac{p.e.} distributions per pixel.}
	\label{pePixel_mean}
\end{figure}
\begin{figure}
	\centering
	\includegraphics[width=0.5\textwidth, trim = 0mm 0mm 0mm 10mm, clip]{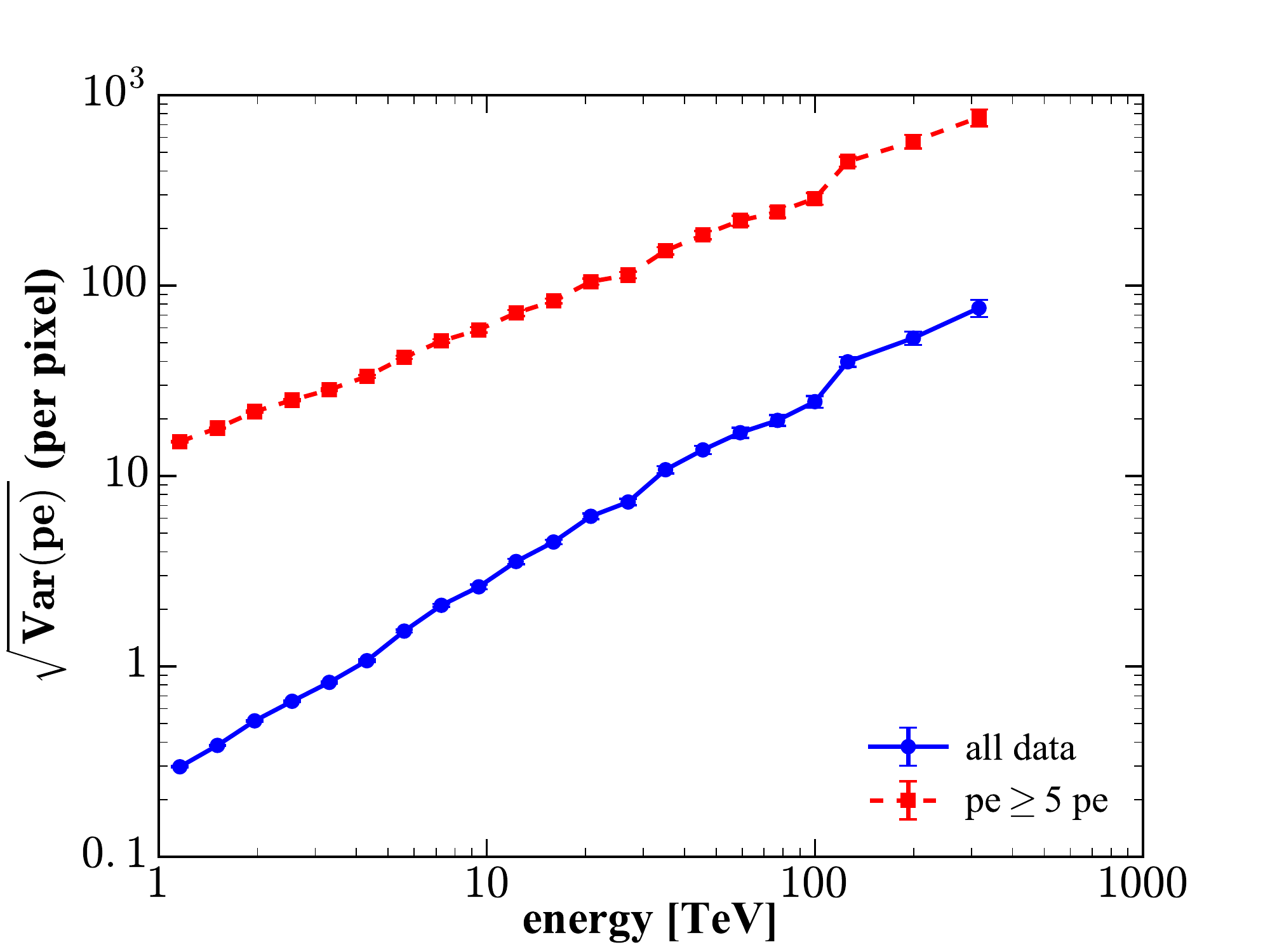}
	\caption{Standard deviations of the simulated \ac{p.e.} distributions per pixel.}
	\label{pePixel_std}
\end{figure}
To estimate the expected number of \acp{p.e.} reaching the camera (full \ac{PDP} and single pixels), on-axis fixed energy gamma events from 1~TeV to 316~TeV have been simulated in CORSIKA, locating the telescope at 2000~m of altitude. This is the typical altitude of the chosen southern site for the installation of \ac{CTA} telescopes.
Since for this study it is essential to estimate all the photons reaching the camera, the trigger is set to require at least 1 \ac{p.e.}/pixel, and the \ac{NSB} is ignored assuming that with the reconstruction of the baseline it can be corrected for. The resulting number of \acp{p.e.} at each gamma-ray energy is weighted with the Crab flux~\cite{CrabHESS}. 

The results are shown in Fig.s~\ref{peCamera} and~\ref{pePixel} for the whole camera and for single pixels, respectively, as blue solid lines. Examples of distributions of events with single energy are also shown: 1.16~TeV (red dashed lines), 12.28~TeV (green dotted lines) and 100~TeV (magenta dash-dotted lines). All the areas are normalized to one. To better understand the behavior of the expected \acs{p.e.}, energy by energy, the averages and standard deviations of monochromatic distributions are shown in Fig.s~\ref{peCamera_mean} and~\ref{peCamera_std} for the entire PDE and Fig.s~\ref{pePixel_mean} and~\ref{pePixel_std} for the pixels (blue solid lines with circle markers). As expected, the number of \ac{p.e.} on average and its standard deviation increase exponentially with the energy and the distributions have a large spread around the mean.

Notice that when a minimal realistic trigger condition is introduced (\ac{p.e.}$\geq5$ per pixel), the events with the smallest number of \acp{p.e.} per pixel will be suppressed (first bin in Fig.~\ref{pePixel}); therefore also the averages and standard deviations change accordingly, as shown in Fig.s from \ref{peCamera_mean} to \ref{pePixel_std} as red dashed lines with square markers. In this case both averages and standard deviations increase: the spread around the mean is still large, but more comparable with the average value. The cut is meant to discard the pixels with less light detected and this is particularly evident looking at the change from the blue to red data in Fig.s~\ref{pePixel_mean} and~\ref{pePixel_std}. At small energies there are too many pixels poorly illuminated and discarded by this selection; therefore, the average increases at $\simeq$1~TeV from $\simeq$0.005 \ac{p.e.} to $\simeq$10 \ac{p.e.} and the spread from the mean goes from $\simeq$0.3 \ac{p.e.} to $\simeq$10 \ac{p.e.}.

\begin{figure}
	\centering
	\includegraphics[width=0.5\textwidth, trim = 0mm 0mm 0mm 10mm, clip]{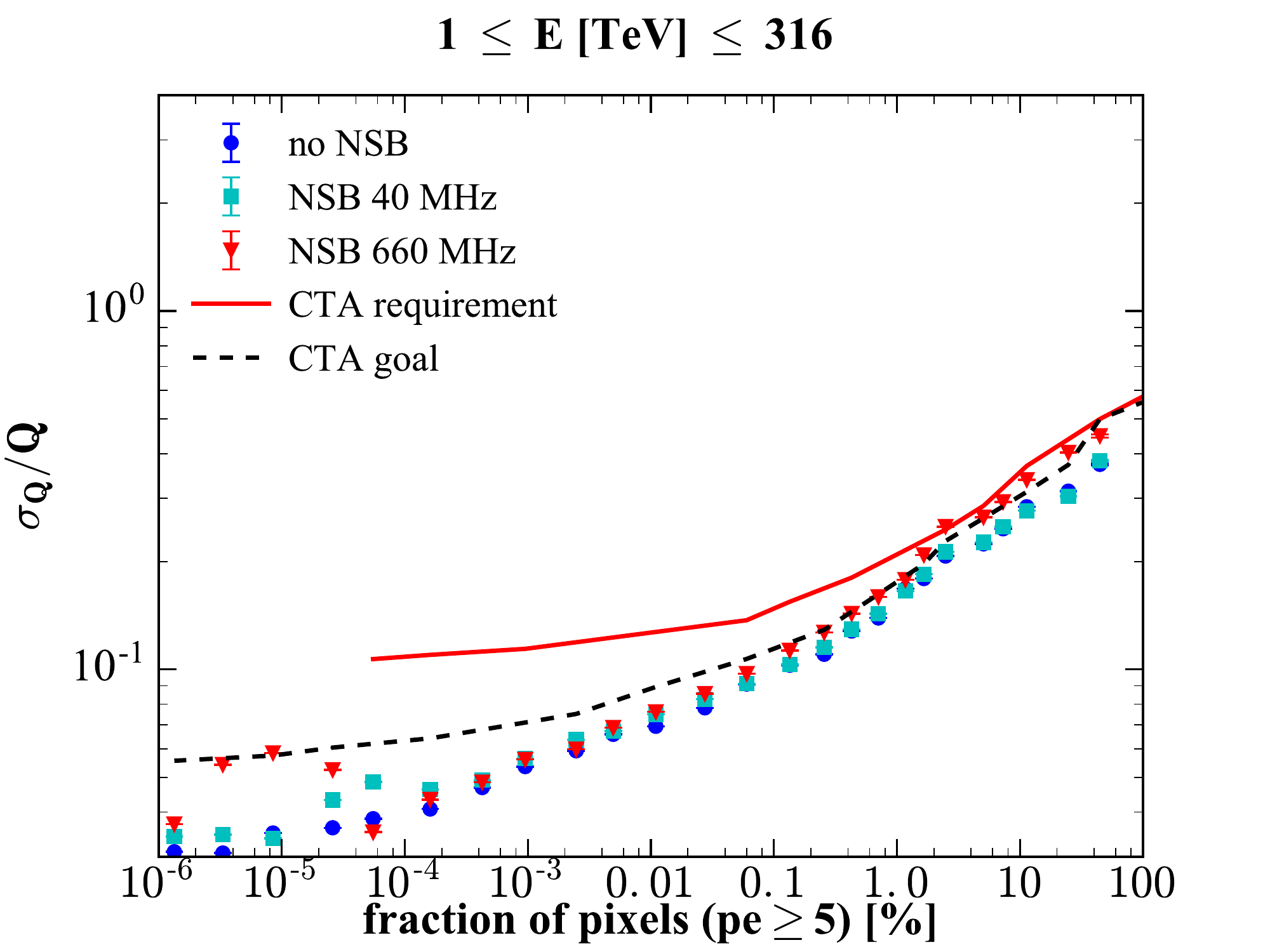}
	\caption{Charge resolution as a function of the fraction of expected events simulated per pixel, expressed in percentage, for the preferred gain configuration (low gain).}
	\label{chargeXpePixel}
\end{figure}

Correlating the measurement of the charge resolution of the preferred gain configuration at different \ac{NSB} (Fig.~\ref{charge resolution} on top, the low gain) with the expected \acp{p.e.} per pixel (Fig.~\ref{pePixel}) gives the distribution of the charge resolution per sensor, at each \ac{NSB} level, as a function of the fraction of expected events. This is shown in Fig.~\ref{chargeXpePixel}, where the \ac{CTA} requirement and goal curves are shown as well. The $x$-axis is normalized to the total number of simulated events and expressed in percentage. Data with at least 5 \acp{p.e.} are considered and each point in Fig.~\ref{chargeXpePixel} is the average of the $\sigma_\text{Q}/\text{Q}$ values belonging to the same \ac{p.e.} bin in Fig.~\ref{pePixel}, with the uncertainties summed in quadrature. The plot shows that with low NSB levels, all the events will be detected with a charge resolution better than the CTA goal. Even in half moon nights (\ac{NSB} 660~MHz), where the \ac{CTA} goal is reached by just 1-2\% of the events (those at higher energies), still the remaining events present a charge resolution below the requirement. In particular, for a fraction of pixels below $10^{-4}$\%, when the transition between the non-saturated and the saturated regimes occurs (see Fig.~\ref{charge resolution} on top, between 1000-2000 \ac{p.e.}), the charge resolution on average is always below the goal; this means that, with a dedicated data analysis, this few but important events, might be recovered with an energy resolution below the goal even for half moon nights.
This study is the starting point for the evaluation of the energy reconstruction performance and energy resolution of the camera.

\section{Conclusions}
\label{sec:concl}
The prototype camera proposed for the \ac{SST-1M} telescope of the \ac{CTA} project adopts several innovative solutions conceived to provide high performance and reliability on long time scales, as well as being cost effective in view of a possible production of up to 20 units. The challenges encountered during the design phase (such as the realization and operation of large area hexagonal \acp{SiPM} and the hollow light concentrators, the stabilization of the working point of the sensors and the cooling strategy) have been all successfully addressed, and the camera is now being assembled at the University of Geneva, were it will be fully tested and characterized. Preliminary measurements and simulations have shown that the camera fully complies with the CTA requirements. Installation on the prototype telescope structure hosted at the H. Niewodnicza\'nski institute of Nuclear Physics in Krakow is foreseen in fall 2016.


\section{Acknowledgments}
We gratefully acknowledge support from the University of Geneva, the Swiss National Foundation, the Ernest Boninchi Foundation and the agencies and organizations listed under Funding Agencies at this website: http://www.cta-observatory.org/. In particular we are grateful for support from the NCN grant DEC-2011/01/M/ST9/01891 and the MNiSW grant 498/1/FNiTP/FNiTP/2010 in Poland. The authors gratefully acknowledge the
support by the projects LE13012 and LG14019 of the Ministry of Education, Youth and Sports of the Czech Republic.
This paper has gone through internal review by the CTA Consortium. 


\bibliographystyle{spphys}
\bibliography{mybibfile}

\begin{thebibliography}{10}
\providecommand{\url}[1]{{#1}}
\providecommand{\urlprefix}{URL }
\expandafter\ifx\csname urlstyle\endcsname\relax
  \providecommand{\doi}[1]{DOI \discretionary{}{}{}#1}\else
  \providecommand{\doi}{DOI \discretionary{}{}{}\begingroup
  \urlstyle{rm}\Url}\fi

\bibitem{CTAconcept}
{B.S. Acharya \emph{et al.} (CTA consortium)}, Introducing the CTA concept
  \textbf{43}, 3 (2013).
\newblock \doi{10.1016/j.astropartphys.2013.01.007}

\bibitem{ASTRI}
{S.~Vercellone for the ASTRI and CTA Consortium}, in \emph{{Proceedings, 34th
  International Cosmic Ray Conference (ICRC 2015)}} (2015)

\bibitem{GCT}
{A.~De~Franco for the CTA Consortium}, in \emph{{Proceedings, 34th
  International Cosmic Ray Conference (ICRC 2015)}} (2015)

\bibitem{CONES}
J.A. Aguilar, A.~Basili, V.~Boccone, F.~Cadoux, A.~Christov, D.~della Volpe,
  T.~Montaruli, L.~Platos, {M.~Rameez for the SST-1M sub-Consortium},
  Astropart. Phys. \textbf{60}, 32 (2015).
\newblock \doi{10.1016/j.astropartphys.2014.05.010}

\bibitem{ICRC_FlashCam}
{G.~P\"uhlhofer \emph{et al.} for the CTA consortium}, in \emph{{Proceedings,
  34th International Cosmic Ray Conference (ICRC 2015)}} (2015)

\bibitem{FACT}
H.~Anderhub, M.~Backes, A.~Biland, V.~Boccone, I.~Braun, {T.~Bretz \emph{et
  al.}}, JINST \textbf{8}, P06008 (2013).
\newblock \doi{10.1088/1748-0221/8/06/P06008}

\bibitem{electronics_paper}
{J.~A.~Aguilar \emph{et al.} for the SST-1M sub-Consortium}, The front-end
  electronics and slow control of large area sipm for the sst-1m camera for the
  cta experiment (2015).
\newblock To be submitted to Nuclear Physics B

\bibitem{NSB_1}
C.~Benn, S.~Ellison, New Astronomy Reviews \textbf{42}(6-8), 503 (1998).
\newblock \doi{10.1016/S1387-6473(98)00062-1}.
\newblock \urlprefix\url{http://arxiv.org/abs/astro-ph/9909153}

\bibitem{NSB_2}
S.~Preu{\ss}, G.~Hermann, W.~Hofmann, A.~Kohnle, Nuclear Instruments and
  Methods in Physics Research Section A: Accelerators, Spectrometers, Detectors
  and Associated Equipment \textbf{481}(1-3), 229 (2002).
\newblock \doi{10.1016/S0168-9002(01)01264-5}.
\newblock \urlprefix\url{http://arxiv.org/abs/astro-ph/0107120}

\bibitem{NSB_3}
D.~Britzger, E.~Carmona, P.~Majumdar, O.~Blanch, J.~Rico, J.~Sitarek,
  R.~Wagner, f.t.M. Collaboration, in \emph{{Proceedings, 31st International
  Cosmic Ray Conference}} (2009).
\newblock \urlprefix\url{http://arxiv.org/abs/0907.0973}

\bibitem{zemax}
Zemax.
\newblock \urlprefix\url{http://www.zemax.com}

\bibitem{Okumura}
A.~Okumura, Astropart. Phys. \textbf{38}, 18 (2012).
\newblock \doi{10.1016/j.astropartphys.2012.08.008}

\bibitem{RoughDavies}
{H.E Bennett and J. O. Porteus}, Journal of the Optimal Society of America
  \textbf{51}(2) (1961)

\bibitem{MPPC}
V.~Boccone, A.~Basili, J.A. Aguilar, A.~Christov, D.~della Volpe, T.~Montaruli,
  M.~Rameez, IEEE Trans. Nucl. Sc. \textbf{61}(3), 1474 (2014).
\newblock \doi{10.1109/TNS.2014.2321339}

\bibitem{FACT2}
{T.~Bretz \emph{et al.}}, in \emph{{Proceedings, 2013 IEEE Nuclear Science
  Symposium and Medical Imaging Conference (NSS/MIC 2013)}} (Proc. of the
  Nuclear Science Symp. and Medical Imaging Conf. (IEEE-NSS/MIC), 2013).
\newblock \doi{10.1109/NSSMIC.2013.6829590}

\bibitem{ICRC_DigiCam}
{P.~Rajda \emph{et al.} for the SST-1M sub-Consortium}, in \emph{{Proceedings,
  34th International Cosmic Ray Conference (ICRC 2015)}} (2015)

\bibitem{1748-0221-9-10-P10012}
{A.~Biland \emph{et al.}}, JINST \textbf{9}(10), P10012 (2014).
\newblock \doi{10.1088/1748-0221/9/10/P10012}

\bibitem{ICRC_Elisa}
{E.~Prandini \emph{et al.} for the SST-1M sub-Consortium}, in
  \emph{{Proceedings, 34th International Cosmic Ray Conference (ICRC 2015)}}
  (2015)

\bibitem{corsika}
{D. Heck \emph{et al.}},
  \href{https://inspirehep.net/record/469835/files/FZKA6019.pdf}{CORSIKA: A
  Monte Carlo Code to Simulate Extensive Air Showers}.
\newblock {Report FZKA-6019}, Forschungszentrum Karlsruhe (1998)

\bibitem{simtelarray}
K.~Bernl\"ohr, Astropart. Phys. \textbf{30}, 149 (2008).
\newblock \doi{10.1016/j.astropartphys.2008.07.009}

\bibitem{groptics+care}
{GrOptics \& CARE}.
\newblock \urlprefix\url{http://otte.gatech.edu/care/tutorial/}

\bibitem{CrabHESS}
{Aharonia \emph{et al.} (H.E.S.S.\ collaboration)}, Astron. Astrophys.
  \textbf{457}, 899 (2006).
\newblock \doi{10.1051/0004-6361:20065351}

\end{thebibliography}


\begin{acronym}[OPC-UA]
	\acro{ACS}{ALMA Common Software}
	\acro{ACTL}{central array control system}
	\acro{AMC}{active mirror control}
	\acro{CCD}{CCD camera}
	\acro{CDTS}{clock distribution and trigger time stamping}
	\acro{CS}{camera server}
	\acro{CSC}{camera slow control}
	\acro{CTA}{Cherenkov Telescope Array}
	\acro{dPLC}{drive programmable logic controller}
	\acro{DAQ}{data acquisition system}
	\acro{EAS}{extensive air shower}
	\acro{FADC}{fast analog to digital converter}
	\acro{FPGA}{field programmable gate array}
	\acro{FEA}{finite element analysis}
	\acro{FoV}{field-of-view}
	\acro{IACT}{imaging atmospheric Cherenkov telescope}
	\acro{LST}{large-size telescope}
	\acro{MST}{medium-size telescope}
	\acro{MPE}{multiple photoelectron}
	\acro{NSB}{night sky background}
	\acro{NTC}{negative temperature coefficient}
	\acro{OPC-UA}{Open Platform Communications Unified Architecture}
	\acro{PCB}{printed circuit board}
	\acro{PDE}{photo-detection efficiency}
	\acro{PDP}{photo-detection plane}
	\acro{PLC}{programmable logic controller}
	\acro{PMT}{photo-multiplier tube}
	\acro{PreAmp}{preamplifier}
	\acro{PSF}{point spread function}
	\acro{SCB}{slow control board}
	\acro{SiPM}{silicon photo-multiplier}
	\acro{SNR}{signal to noise ratio}
	\acro{SPE}{single photoelectron}
	\acro{sPLC}{safety programmable logic controller}
	\acro{SST}{small-size telescope}
	\acro{SST-1M}{single-mirror small-size telescope}
	\acro{SWAT}{software array trigger}
	\acro{VHE}{very high-energy}
	\acro{WR}{WhiteRabbit}
	 \acro{FFT}{Fast Fourier Transform}
	 \acro{p.e.}{photoelectron}
\end{acronym}

\end{document}